\documentclass[12pt,a4paper]{article}
\usepackage{amsmath,amsfonts,amsthm}
\allowdisplaybreaks[3]
\numberwithin{equation}{section} 
\newtheorem{theorem}{Theorem}[section]
\newtheorem{lemma}[theorem]{Lemma}
\newtheorem{proposition}[theorem]{Proposition}

\def\gbf{{\mathbf g}}
\def\ubf{{\mathbf u}}
\def\Rbf{{\mathbf R}}
\def\Rsf{{\mathsf R}}
\def\Rbb{{\mathbb R}}

\def\rsf{{\mathsf r}}

\def\Lbf{{\mathbf L}}
\def\Lbb{{\mathbb L}}
\def\Ebf{{\mathbf E}}
\def\Hbf{{\mathbf H}}
\def\bsf{{\mathsf b}}
\def\xbf{{\mathbf x}}
\def\Xbf{{\mathbf X}}
\def\ybf{{\mathbf y}}
\def\zbf{{\mathbf z}}
\def\Jbf{{\mathbf J}}
\def\Jsf{{\mathsf J}}
\def\Bcl{{\mathcal B}}
\newcommand{\Rcal}{{\mathcal R}}
\newcommand{\Rbar}{\overline{\Rcal}}

\def\A{{\mathcal A}}
\def\R{\mathcal{R}}
\def\Rb{{\mathbf R}}
\usepackage[body={15.5cm,22cm}]{geometry}
\begin{document}
\title{Quantum groups, 
Yang-Baxter maps and quasi-determinants
}
\author{
Zengo Tsuboi
\vspace{.5cm}
\\ 
Laboratoire de physique th\'eorique, 
D\'epartement de physique  \\
de l'ENS, \'Ecole normale sup\'erieure, 
PSL Research 
University, 
\\
Sorbonne Universit\'es, 
UPMC Univ. Paris 06, CNRS, 
\\
75005 Paris, France
}

\date{}

\maketitle
\begin{abstract}
For any quasi-triangular Hopf algebra, there exists the universal R-matrix, 
which satisfies the Yang-Baxter equation. 
It is known that the adjoint action of the universal R-matrix 
on the elements of the tensor square of the algebra 
constitutes  a quantum Yang-Baxter map, which 
 satisfies the set-theoretic Yang-Baxter equation. 
  The map has a zero curvature representation 
  among L-operators defined as images of the universal R-matrix. 
  We find that the zero curvature representation can be solved by 
the Gauss decomposition of a product of L-operators. 
Thereby obtained a quasi-determinant
 expression of the quantum Yang-Baxter map  
 associated with the quantum algebra $U_{q}(gl(n))$. 
 Moreover, the map is identified with products of quasi-Pl\"{u}cker coordinates over a matrix 
composed of the L-operators. 
We also consider the 
quasi-classical limit, where the underlying quantum algebra 
reduces to a Poisson algebra. 
The quasi-determinant expression of the quantum Yang-Baxter map   
reduces to ratios of determinants, which give a new expression of  
  a classical Yang-Baxter map. 
\end{abstract}
Keywords: universal R-matrix, L-operator, 
quasi-determinant, Yang-Baxter map, 
Gauss decomposition, quantum group
\\\\
arXiv:1708.06323 [math-ph]
\\
Nuclear Physics B 926 (2018) 200-238
\\
https://doi.org/10.1016/j.nuclphysb.2017.11.005

\newpage 
\section{Introduction}
Let $\Xbf$ be a set.  
The Yang-Baxter maps \cite{Veselov:2003} are the maps 
\begin{align}
\Rcal : \qquad  \Xbf \times \Xbf  \mapsto  \Xbf \times \Xbf ,
 \label{map-set}
\end{align}
which satisfy the set-theoretic Yang-Baxter equation \cite{Drinfeld:1992}
\begin{align}
\Rcal_{12} \circ \Rcal_{13} \circ \Rcal_{23}=
\Rcal_{23} \circ \Rcal_{13} \circ \Rcal_{12}
 , \label{YBE-set}
\end{align}
 where this equation is defined on the direct product of 
 three sets $ \Xbf \times \Xbf  \times \Xbf $,
 and $\Rcal_{ij}$ acts non-trivially on the $i$-th and $j$-th components. 
 
 The Yang-Baxter maps attract interest mainly in the context of 
  discrete integrable evolution equations
\footnote{A Yang-Baxter map also appeared in a different context: see 
section 3.4 in \cite{BHH17}.} 
 \cite{Adler:1993,Hietarinta:1997,Etingof:1999,Quispel:1999,
Kajiwara:2002,Odesskii:2003,Goncharenko:2004,Adler:2004,AdlerBobenkoSuris,KNW09,Joshi:2010,Doliwa14}. 
However, the Hamiltonian structures for the Yang-Baxter maps were not very clear, 
and systematic procedures for 
quantization of them or their applications to 
Hamiltonian evolution systems did not exist.
To overcome this difficulty, Bazhanov and Sergeev proposed  \cite{BS15}
 a new approach to the Yang-Baxter maps. It is based on the theory of quantum groups
\cite{Dri87,Jim86}, in particular on the universal R-matirx \cite{Dri87}. 
They presented their scheme on the example of the quantum algebra $U_{q}(sl(2))$. 
In this paper, we will review their paper \cite{BS15} by 
generalizing a part of their results to 
 the higher rank algebra $U_{q}(gl(n))$ for any $n$, and 
 propose some algebraic relations not addressed in \cite{BS15}. 

For any quasi-triangular Hopf algebra $\A$, 
 there exists the universal R-matrix $\Rb\in\A\otimes\A$, 
 which satisfies the (universal) quantum Yang-Baxter equation
\begin{align}
\Rbf_{12} \Rbf_{13} \Rbf_{23}=
\Rbf_{23} \Rbf_{13} \Rbf_{12} . 
\label{YBE-int}
\end{align}
By using the universal R-matrix, 
one can define \cite{Sklyanin:1988,Kashaev:2004} 
a map 
\begin{align}
\Rcal: 
\qquad \A \otimes \A \mapsto
 \A \otimes \A ,
\end{align}
acting on the generators of the algebra 
$ x^{(1)} = x \otimes 1$, $ y^{(2)}=1 \otimes y$, 
$x,y \in \A$ as 
\begin{align}
\Rcal:
\quad  x^{(1)}  \mapsto 
\widetilde{x}^{(1)} =\Rbf  x^{(1)}  \Rbf^{-1}, 
\qquad 
 y^{(2)}  \mapsto 
\widetilde{y}^{(2)} =\Rbf y^{(2)} \Rbf^{-1} 
\label{Rmap}. 
\end{align}
This satisfies \eqref{YBE-set} due to \eqref{YBE-int}. 
In addition, it keeps the defining relations of the algebra unchanged, and 
thus is an automorphism of  $\A \otimes \A$.  
The set $\Xbf$ in \eqref{map-set} is not just a set but the algebra $\A$ 
(or its generators)
in this case. 
They showed \cite{BS15} that the map \eqref{Rmap}
defines Hamiltonian evolution equations for a two dimensional discrete quantum  
integrable system with an algebra of observables formed by a tensor power of $\A$. 
In addition,  the map \eqref{Rmap} has a discrete analogue of the 
zero curvature representation \cite{BS15}. 
For $U_{q}(gl(n))$, it has the following from (cf.\ \eqref{ZC-an}): 
\begin{align}
\Lbf^{+(1)}\Lbf^{+(2)}
=\widetilde{\Lbf}^{+(2)}\widetilde{\Lbf}^{+(1)},
\quad
\Lbf^{-(1)}\Lbf^{+(2)}
=\widetilde{\Lbf}^{+(2)}\widetilde{\Lbf}^{-(1)},
\quad 
\Lbf^{-(1)}\Lbf^{-(2)}
=\widetilde{\Lbf}^{-(2)}\widetilde{\Lbf}^{-(1)}, 
\label{ZC-an-into}
\end{align}
where $\Lbf^{+(i)}$ 
(resp.\ $\Lbf^{-(i)}$) are 
upper triangular  (resp.\ lower triangular) $n \times n$-matrices, called L-operators, which are  
given as images of the universal R-matrix (cf.\ \eqref{L-gauge1}, \eqref{L-gauge2}), 
and $ \widetilde{\Lbf}^{\pm (i)}=\Rbf  \Lbf^{\pm (i)} \Rbf^{-1}$. The indices $i=1,2$ 
of these L-operators 
denote the spaces where the matrix elements of them non-trivially act, and 
correspond to the ones in \eqref{Rmap}. 
In principal, one can obtain an explicit expression of the 
quantum Yang-Baxter map \eqref{Rmap} by using the explicit expression 
of the universal R-matrix. However, this is a cumbersome task in practice. 
The zero curvature representation \eqref{ZC-an-into} gives us a shortcut for this \cite{BS15}.  
We find that the zero curvature representation can be explicitly 
solved based on the Gauss decomposition of a product of L-operators 
$\Jbf=\Lbf^{-(1)} \Lbf^{+(2)}$. 
The resultant expression can be written in terms of the so-called 
quasi-determinants (Theorem \ref{sol-zero1}). 
The quasi-determinants \cite{GR91,GR92,GGRW05} are defined for matrices with  
non-commutative matrix elements, and are related to inverse of the matrices. 
They reduce to ratios of determinants if all the matrix elements are commutative. 
Moreover, the map is identified with products of quasi-Pl\"{u}cker coordinates over a matrix 
\eqref{singlmat-q} 
composed of the L-operators. This may imply an underlying quantum Grassmannian of the system. 
We also consider the inverse map (Theorem \ref{sol-zero1-o})
 and the map associated with $\Rbf_{21}^{-1}$.

In the quasi-classical limit,  the quasi-triangular algebra $\A$ reduces 
 to a Poisson algebra. The Yang-Baxter maps associated with $\A$ reduce to 
classical Yang-Baxter maps, which preserve the structure of a Hamiltonian map \cite{BS15}. 
Our quasi-determinant expressions of the quantum Yang-Baxter maps reduce to 
 ratios of minor determinants of a 
single matrix \eqref{singlmat-cl}, 
which give new expressions of classical Yang-Baxter maps.

 We consider the quantum case in section 2 and the classical case in section 3. 
 We will follow the style of presentation of \cite{BS15} and generalize their results 
on $U_{q}(sl(2))$ to  $U_{q}(gl(n))$. 
The quasi-determinant expressions of the quantum Yang-Baxter maps in section \ref{subsec-YBQD} 
and the determinant formulas for their classical counterparts in section \ref{subsec-CYBD} 
 are our new results, which are not mentioned in \cite{BS15} even for $U_{q}(sl(2))$ case, 
although the map itself  for $U_{q}(sl(2))$ case is known in \cite{Kashaev:2004}. 
The appendices are devoted to technical details. 

\section{Quantum Yang-Baxter map for the algebra $U_{q}(gl(n))$}
In this section, we will generalize results of Bazhanov and Sergeev \cite{BS15} 
to the higher rank case, and propose new quasi-determinant expressions of 
 quantum Yang-Baxter maps for $U_{q}(gl(n))$. 
Subsections \ref{subsec-QA}-\ref{section-Rmatform0} are higher rank analogues of 
 subsections 2.1-2.3, 2.5 and 2.8 of \cite{BS15}. 
The twisting of L-operators in subsection \ref{subsec-twist}, 
a review on quasi-determinants in subsection \ref{subsec-QD} and 
the quasi-determinant formulas in subsection \ref{subsec-YBQD} are 
not addressed in \cite{BS15}. 
\subsection{Quantum algebra $U_{q}(gl(n))$}
\label{subsec-QA}
In this subsection, we give a short digest on the definition of the quantum algebra
\footnote{The notation $U_{h}(gl(n))$  is often used in literatures for 
this form of presentation of the quantum algebra.
The deformation parameter $q$ is related to $h$ as $q=e^{h}$. 
We assume that $q$ is generic.}
 $U_{q}(gl(n))$.
The quantum algebra 
${\mathcal A}=U_{q}(gl(n))$ is generated by 
the generators $\Ebf_{i,i+1}, \Ebf_{j+1,j},\Ebf_{k,k}$ 
($i,j \in \{1,2,\dots, n-1\}$, 
$k \in \{1,2,\dots, n\}$), with the following relations
\footnote{
The normalization commonly used in literatures 
can be obtained by the replacement: $
\Ebf_{ij} \to (q-q^{-1})\Ebf_{ij}$, for $i \ne j$, and 
$\Ebf_{ii} \to  \Ebf_{ii} $.}
: 
\begin{align}
\begin{split}
&[ \Ebf_{kk}, \Ebf_{ij}]=
(\delta_{ik}-\delta_{jk})\Ebf_{ij}, 
\\[5pt]
&[\Ebf_{i,i+1},\Ebf_{j+1,j}]=\delta_{ij}(q-q^{-1})
 (q^{\Ebf_{ii}-\Ebf_{i+1,i+1}}-q^{-\Ebf_{ii}+\Ebf_{i+1,i+1}}),
\\[5pt]
&[\Ebf_{i,i+1},\Ebf_{j,j+1}]=[\Ebf_{i+1,i},\Ebf_{j+1,j}]=0
\quad \text{for} \quad |i-j| \ge 2,
\end{split}
\label{def-gln}
\end{align}
and,  for $i \in \{1,2,\dots,n-2\}$, the Serre relations
\begin{align}
\begin{split}
&
\Ebf_{i,i+1}^2\Ebf_{i+1,i+2}-(q+q^{-1})\Ebf_{i,i+1}\Ebf_{i+1,i+2}\Ebf_{i,i+1}
+ \Ebf_{i+1,i+2}\Ebf_{i,i+1}^2=0,
\\[5pt]
&
\Ebf_{i+1,i+2}^2\Ebf_{i,i+1}-(q+q^{-1})\Ebf_{i+1,i+2}\Ebf_{i,i+1}\Ebf_{i+1,i+2}
+ \Ebf_{i,i+1}\Ebf_{i+1,i+2}^2=0,
\\[5pt]
&
\Ebf_{i+1,i}^2\Ebf_{i+2,i+1}-(q+q^{-1})\Ebf_{i+1,i}\Ebf_{i+2,i+1}\Ebf_{i+1,i}
+ \Ebf_{i+2,i+1}\Ebf_{i+1,i}^2=0,
\\[5pt]
&
\Ebf_{i+2,i+1}^2\Ebf_{i+1,i}-(q+q^{-1})\Ebf_{i+2,i+1}\Ebf_{i+1,i}\Ebf_{i+2,i+1}
+ \Ebf_{i+1,i}\Ebf_{i+2,i+1}^2=0. 
\end{split}
\label{Serre}
\end{align}
For $i,j \in \{1,2,\dots, n\}$ with $j-i \ge 2 $ and $k \in \{i+1,i+2,\dots, j-1 \}$, we define 
\begin{align}
\begin{split}
\Ebf_{ij}&=(q-q^{-1})^{-1}(\Ebf_{ik}\Ebf_{kj}-q\Ebf_{kj}\Ebf_{ik}),
\\[5pt]
\Ebf_{ji}&=(q-q^{-1})^{-1}(\Ebf_{jk}\Ebf_{ki}-q^{-1} \Ebf_{ki}\Ebf_{jk}).
\end{split}
 \label{def-eij}
\end{align}
Here \eqref{def-eij} does not depend on the choice of $k$. 
%
Let us introduce elements 
$\Hbf_{i}=q^{\Ebf_{ii}-\Ebf_{i+1,i+1}}$, $i \in \{1,2,\dots, n-1\}$. 
Then the first two relations in \eqref{def-gln} become 
\begin{align}
&\Hbf_{k} \Ebf_{ij} \Hbf_{k}^{-1}=
q^{\delta_{ik}-\delta_{jk}}\Ebf_{ij} 
, \qquad 
[\Ebf_{i,i+1},\Ebf_{j+1,j}]=\delta_{ij}(q-q^{-1})
 (\Hbf_{k}- \Hbf_{k}^{-1}) . 
\end{align}
The restriction of  ${\mathcal A}$ 
to the algebra generated by 
the generators $\Ebf_{i,i+1}, \Ebf_{j+1,j},\Ebf_{ii}-\Ebf_{i+1,i+1}$ 
($i,j \in \{1,2,\dots, n-1\}$) is isomorphic to  
$U_{q}(sl(n))$. 
Note that 
${\mathbf c}=\sum_{j=1}^{n}\Ebf_{jj}$ 
is a central element of ${\mathcal A}$. 
The algebra  ${\mathcal A}$ becomes isomorphic 
to an enlarged version of $U_{q}(sl(n))$ if the
 condition ${\mathbf c}=0 $ 
is imposed. 
As a Hopf algebra, ${\mathcal A}$ has 
 the co-multiplication 
 $\Delta$, the co-unit $\epsilon$ and the 
antipode $S$. 
The co-multiplication  $\Delta$ is an algebra homomorphism
 from the algebra $ {\mathcal A}  $ to its tensor square 
\begin{align}
\Delta : \qquad  {\mathcal A} \to {\mathcal A} \otimes {\mathcal A}, 
\end{align}
defined by 
\begin{align}
\begin{split}
\Delta(\Ebf_{i,i+1})&=\Ebf_{i,i+1} \otimes q^{\Ebf_{ii}-\Ebf_{i+1,i+1}}
+1 \otimes \Ebf_{i,i+1},
\\[5pt]
\Delta(\Ebf_{i+1,i})&=\Ebf_{i+1,i} \otimes 1 + q^{-\Ebf_{ii}+\Ebf_{i+1,i+1}}
 \otimes \Ebf_{i+1,i} \qquad \text{for}
\quad i \in \{ 1,2,\dots, n-1 \}, 
\\[5pt]
\Delta(\Ebf_{kk})&=\Ebf_{kk} \otimes 1
+1 \otimes \Ebf_{kk}
\qquad \text{for} 
\quad k \in \{ 1,2,\dots, n\}. 
\end{split}
\label{co-pro}
\end{align}
We will also use 
the opposite co-multiplication $\Delta^{\prime}$, defined 
by
\begin{align}
\Delta^{\prime} =\sigma \circ \Delta,
\end{align}
where $\sigma ({\mathbf a} \otimes {\mathbf b} )=
{\mathbf b} \otimes {\mathbf a} $ 
for any ${\mathbf a} ,{\mathbf b} \in {\mathcal A}  $.
The co-unit is an algebra homomorphism
 from the algebra $ {\mathcal A}  $ to complex numbers, defined by 
\begin{align}
 \epsilon (\Ebf_{ij})=0, 
 \qquad \epsilon (1)=1.   
 \label{counit}
\end{align}
The antipode $S$ is an algebra anti-homomorphism 
on the algebra $ {\mathcal A}  $, defined by 
\begin{multline}
S(\Ebf_{kk})=-\Ebf_{kk}, 
\qquad 
S(\Ebf_{i,i+1})=-\Ebf_{i,i+1} q^{-\Ebf_{ii} +\Ebf_{i+1,i+1}}, 
\\
S(\Ebf_{i+1,i})=-q^{\Ebf_{ii} -\Ebf_{i+1,i+1}} \Ebf_{i+1,i} . 
\label{antipode1}
\end{multline}
It satisfies 
\begin{align}
(S \otimes S) \circ \Delta 
 = \Delta^{\prime } \circ S, 
\qquad 
 S(1)=1, 
 \qquad 
 \epsilon \circ S= \epsilon 
. 
\label{antipode2}
\end{align}
We also remark relations for the element $\Hbf_{i}$: 
\begin{align}
\Delta (\Hbf_{i})=\Hbf_{i} \otimes \Hbf_{i}, \qquad 
\epsilon(\Hbf_{i})=1, \qquad 
S(\Hbf_{i})=\Hbf_{i}^{-1}.  
 \label{relationsH}
\end{align}
\subsection{Universal R-matrix}
The quantum algebra ${\mathcal A}$ is a quasi-triangular Hopf algebra. 
Then there exists an element $\Rbb$, called the universal R-matrix, in the completion of 
the tensor product 
$ {\mathcal A} \otimes {\mathcal A}$, which obeys 
\begin{align}
\begin{split}
\Delta^{\prime } ({\mathbf a})  \, \Rbb &=\Rbb  \,
\Delta({\mathbf a}) \qquad 
\text{for all } \, {\mathbf a} \in {\mathcal A},
\\[5pt]
(\Delta \otimes 1) \, \Rbb &=
\Rbb_{13} \, \Rbb_{23}, 
\\[5pt]
(1 \otimes \Delta ) \, \Rbb &=
\Rbb_{13} \, \Rbb_{12}, 
\end{split}
\label{UR-def}
\end{align}
where
\footnote{More generally, we will embed an element of the form 
${\mathbf A}=\sum_{k} {\mathbf a}_{1}^{(k)} \otimes    {\mathbf a}_{2}^{(k)} \in {\mathcal A} \otimes {\mathcal A} $ into ${\mathcal A}^{\otimes L}$  (for $L \in {\mathbb Z}_{\ge 2}$) and use the notation 
${\mathbf A}_{ij}=\sum_{k} 
 1^{\otimes (i-1)} \otimes {\mathbf a}_{1}^{(k)} \otimes 1^{\otimes (j-i-1)} \otimes  
 {\mathbf a}_{2}^{(k)}  \otimes  1^{\otimes (L-j)}$ for $1 \le i<j \le L$. 
 One should not confuse this notation with the ones for matrix elements of a matrix, or 
 generators of algebras, which will also be 
 used in this paper. 
}
 $\Rbb_{12}=\Rbb \otimes 1 $, 
$\Rbb_{23}=1 \otimes \Rbb  $ and 
$\Rbb_{13}=(\sigma \otimes 1)\Rbb_{23}$. 
The so-called (universal) quantum Yang-Baxter equation
\begin{align}
\Rbb_{12}\Rbb_{13}\Rbb_{23}=
\Rbb_{23}\Rbb_{13}\Rbb_{12}
 \label{YBE}
\end{align}
follows from these relations \eqref{UR-def}.  
In addition,  \eqref{antipode1}, \eqref{antipode2} and \eqref{UR-def} produce 
the following relations 
\begin{align}
\begin{split}
(\epsilon \otimes 1) \Rbb& =(1 \otimes \epsilon ) \Rbb =1 \otimes 1, 
\\[5pt]
(S \otimes 1) \Rbb & =(1 \otimes S^{-1}) \Rbb =\Rbb^{-1}. 
\end{split} 
 \label{UR-relations}
\end{align}
Thus $\Rbb$ and $\Rbb^{-1}$ are invariant under $S \otimes S$.
\begin{align}
(S \otimes S) \Rbb & = \Rbb , \qquad (S \otimes S) \Rbb^{-1} =\Rbb^{-1}. 
 \label{UR-relations2}
\end{align}
Let us introduce a q-analogue of the exponential function defined 
by 
\begin{align}
\exp_{q}(x)&=1 + \sum_{k=1}^{\infty} \frac{x^{k}}{(k)_{q}!},
\\ & (k)_{q}!=(1)_{q}(2)_{q}\cdots (k)_{q}, 
\quad (k)_{q}=(1-q^{k})/(1-q).
\end{align}
We will also use the relation $\exp_{q}(x)^{-1}=\exp_{q^{-1}}(-x)$. 
In general, the universal R-matrix is not unique as an element of 
$ {\mathcal A} \otimes {\mathcal A}$. However, 
if we assume that the universal R-matrix has the form 
$\Rbb=q^{\sum_{i=1}^{n}\Ebf_{ii} \otimes \Ebf_{ii}}\overline{\Rbb} $, where $\overline{\Rbb} $ 
is an element of 
${\mathcal N}_{+}\otimes {\mathcal N}_{-}$:
 ${\mathcal N}_{+}$ and ${\mathcal N}_{-}$ 
are nilpotent sub-algebras 
generated by $\{ \Ebf_{ij} \}$ and 
$\{ \Ebf_{ji} \}$ for $i<j$, $i,j \in \{1,2,\dots, n\}$ 
respectively, then it is uniquely
fixed by \cite{Rosso89} 
\begin{align}
\Rbb=q^{\sum_{i=1}^{n}\Ebf_{ii} \otimes \Ebf_{ii}}
\overrightarrow{\prod}_{i<j}
\exp_{q^{-2}}
\left(
(q-q^{-1})^{-1}\Ebf_{ij}\otimes \Ebf_{ji}
\right),
 \label{UR-exp}
\end{align}
where the product is taken over 
 the reverse lexicographical order on $(i,j)$: 
$(i_{1},j_{1}) \prec  (i_{2},j_{2}) $ if $i_{1}> i_{2}$,  
or $ i_{1}=i_{2}$ and $j_{1}> j_{2} $.
The algebras 
generated by $\{ \Ebf_{ij} \}$ and 
$\{ \Ebf_{ji} \}$ for $i\le j$, $i,j \in \{1,2,\dots, n\}$  
are called Borel subalgebras and denoted as 
$\Bcl_{+}$ and $\Bcl_{-}$, respectively. The universal R-matrix \eqref{UR-exp} is 
in the completion of $\Bcl_{+} \otimes \Bcl_{-}$.

Let $E_{ij}$ be the $n\times n$ matrix unit whose $(a,b)$-matrix elements are given by 
$(E_{ij})_{ab}=\delta_{ia}\delta_{jb}$, $i,j,a,b \in \{1,2,\dots,n\}$. 
Then we define 
the $n$-dimensional fundamental representation $\pi $ of ${\mathcal A}$ 
 by 
%
%
$\pi(\Ebf_{kk})=E_{kk}$, 
$\pi (\Ebf_{ij})=(q-q^{-1})E_{ij}$, 
for $i \ne j$. 

\subsection{Quantum Yang-Baxter map}
Let ${\mathbf X}=\{ \Ebf_{ij}, q^{\Ebf_{kk}}, q^{-\Ebf_{kk}} | i,j,k \in \{1,2,\dots , n\}, i \ne j \}$ 
 be the set of generators of 
${\mathcal A}$, and 
${\mathbf X}^{(a)}$ be the corresponding 
components in ${\mathcal A} \otimes {\mathcal A}$: 
\begin{align}
{\mathbf X}^{(1)}=
\{ {\mathbf x}^{(1)}:={\mathbf x} \otimes 1 | {\mathbf x} \in {\mathbf X} \}
, \qquad 
{\mathbf X}^{(2)}=
\{{\mathbf x}^{(2)}:=1\otimes {\mathbf x} | {\mathbf x} \in {\mathbf X} \}
. 
\end{align}
The quantum Yang-Baxter map ${\mathcal R}$ is defined by the 
adjoint action of the universal R-matrix,
\begin{multline}
{\mathcal R} : 
({\mathbf X}^{(1)}, {\mathbf X}^{(2)})
\mapsto 
(\widetilde{\mathbf X}^{(1)}, \widetilde{\mathbf X}^{(2)}), 
\\
\widetilde{\mathbf X}^{(a)}=\Rbb{\mathbf X}^{(a)}
\Rbb^{-1}
=
\left\{ \widetilde{\mathbf x}^{(a)}:=
\Rbb{\mathbf x}^{(a)}
\Rbb^{-1} | {\mathbf x}^{(a)} \in {\mathbf X}^{(a)} 
\right\}, 
\qquad a=1,2. 
\label{defYB-map}
\end{multline}
Note that any elements of $\widetilde{\mathbf X}^{(1)}$ 
commute with those of $\widetilde{\mathbf X}^{(2)}$. 
In addition, the algebra generated by the elements of  the set 
$\widetilde{\mathbf X}^{(a)}$ is isomorphic to the algebra $\A$. 
Both ${\mathbf X}^{(a)}$ and $\widetilde{\mathbf X}^{(a)}$ can 
 be regarded as copies of ${\mathbf X}$. 
The map can  be naturally  extended to 
 an automorphism of $\A \otimes \A$.
 
As was discussed in \cite{BS15},  \eqref{defYB-map} 
defines a rational map among generators for  the case 
$U_{q}(sl(2))$. However, we find that square roots 
appear in the map for $U_{q}(sl(n))$, $n \ge 3$ case. 
To overcome this difficulty, we will change the gauge of the universal R-matrix afterward.

\subsection{R-matrix form of the $U_{q}(gl(n))$ 
defining relations}
\label{section-Rmatform0}
One can prove that if  $\Rbb_{12} \in {\mathcal B}_{+} \otimes  {\mathcal B}_{-}$ 
satisfies \eqref{UR-def}, then 
\begin{align}
\Rbb^{*}_{12}=\Rbb^{-1}_{21} 
\in {\mathcal B}_{-} \otimes  {\mathcal B}_{+}, 
 \label{inv-R}
\end{align}
which is different 
from $\Rbb_{12} $, also satisfies \eqref{UR-def}.  
The explicit expression of it can be obtained from \eqref{UR-exp}.
\begin{align}
\Rbb^{*}_{12} 
=
\overleftarrow{\prod}_{i<j}
\exp_{q^{2}}
\left(-
(q-q^{-1})^{-1}\Ebf_{ji}\otimes \Ebf_{ij}
\right)
q^{-\sum_{i=1}^{n}\Ebf_{ii} \otimes \Ebf_{ii}} .
 \label{UR-expi}
\end{align}
In addition, from the definition \eqref{inv-R} and 
\eqref{YBE}, one can show that 
these satisfy the following Yang-Baxter type equations. 
\begin{align}
\Rbb_{12}\Rbb_{13}\Rbb^{*}_{23}&=
\Rbb^{*}_{23}\Rbb_{13}\Rbb_{12}, 
\label{YBE00}
\\[5pt]
\Rbb^{*}_{12}\Rbb_{13}\Rbb_{23}&=
\Rbb_{23}\Rbb_{13}\Rbb^{*}_{12}, 
\label{YBE01}
\\[5pt]
\Rbb^{*}_{12}\Rbb^{*}_{13}\Rbb_{23}&=
\Rbb_{23}\Rbb^{*}_{13}\Rbb^{*}_{12} ,
\label{YBE02}
\\[5pt]
\Rbb_{12}\Rbb^{*}_{13}\Rbb^{*}_{23}&=
\Rbb^{*}_{23}\Rbb^{*}_{13}\Rbb_{12} ,
\label{YBE03}
\\[5pt]
\Rbb^{*}_{12}\Rbb^{*}_{13}\Rbb^{*}_{23}&=
\Rbb^{*}_{23}\Rbb^{*}_{13}\Rbb^{*}_{12} . 
\label{YBE04}
\end{align}

Evaluating the universal R-matrices  
\eqref{UR-exp} and \eqref{UR-expi} in the fundamental representation 
in the first space (called an auxiliary space), we introduce two types of 
L-operators: 
\begin{align}
\begin{split}
\Lbb^{-}&=(\pi \otimes 1) \Rbb^{*}
=\sum_{k=1}^{n}E_{kk}\otimes q^{-\Ebf_{kk}}
 -\sum_{i<j}E_{ji} \otimes   \Ebf_{ij} q^{-\Ebf_{ii}},
\\[5pt]
\Lbb^{+}&=(\pi \otimes 1)\Rbb
=\sum_{k=1}^{n}E_{kk}\otimes q^{\Ebf_{kk}}
 +\sum_{i<j}E_{ij} \otimes  q^{\Ebf_{ii}} \Ebf_{ji}. 
\end{split}
\label{Lmp0}
\end{align} 
The matrix elements of the L-operators are elements of $U_{q}(gl(n))$ and 
act in the quantum space. 
Evaluating further the second space (quantum space) of 
these L-operators in the fundamental representation $\pi$, we obtain the 
R-matrices
\begin{align}
\begin{split}
R^{-}&=(1\otimes \pi )
\Lbb^{-} 
=(\pi \otimes \pi)
\Rbb^{*}  
\\[5pt]
&=\sum_{i,j}q^{-\delta_{ij}}E_{ii}\otimes E_{jj}
 -(q-q^{-1})\sum_{i<j}E_{ji} \otimes  E_{ij} ,
\\[8pt]
R^{+}&=(1\otimes \pi )
\Lbb^{+} =(\pi \otimes \pi) \Rbb 
\\[5pt]
&=\sum_{i,j}q^{\delta_{ij}}E_{ii}\otimes E_{jj}
 +(q-q^{-1})\sum_{i<j}E_{ij} \otimes  E_{ji}. 
\end{split}
\label{Rmp0}
\end{align}
Then we define the spectral parameter $\lambda \in {\mathbb C}$ dependent L-operator
\begin{align}
\Lbb(\lambda)= \lambda \Lbb^{+} - \lambda^{-1} \Lbb^{-},
\label{L-sp0}
\end{align}
and the R-matrix 
\begin{align}
R(\lambda)= \lambda R^{+} - \lambda^{-1} R^{-}. 
\label{Rs0}
\end{align}
The latter coincides with the R-matrix of Cherednik 
\cite{Cherednik80}. 
The permutation matrix 
 $P=\sum_{i,j}E_{ij}\otimes E_{ji}$ appears at 
a special value of the spectral parameter of the R-matrix: 
\begin{align}
R(1)= (q-q^{-1})P.
\label{R01}
\end{align}
We will also use
\begin{align}
\Check{R}^{\pm}= P {R}^{\pm}, 
\qquad 
\Check{R}(\lambda)= P {R}(\lambda ). 
\label{Rch-sp0}
\end{align}
Evaluating the first and the second space of \eqref{YBE} and  \eqref{YBE00}-\eqref{YBE04} for  
$(\pi \otimes \pi  \otimes 1)$, 
and multiplying the permutation matrix $P_{12}$ from the left, 
 we obtain the R-matrix form \cite{Faddeev:1987ih} 
of the defining relations 
\eqref{def-gln}-\eqref{def-eij} of $U_{q}(gl(n))$: 
\begin{align}
\begin{split}
\check{R}^{\pm}_{12} \Lbb^{-}_{13}
 \Lbb^{-}_{23} 
&=
\Lbb^{-}_{13} \Lbb^{-}_{23} \check{R}^{\pm }_{12}, 
\qquad 
\check{R}^{\pm}_{12} \Lbb^{+}_{13}
 \Lbb^{+}_{23} 
=
\Lbb^{+}_{13} \Lbb^{+}_{23} \check{R}^{\pm }_{12},
\\[5pt]
\check{R}^{-}_{12} \Lbb^{-}_{13}
 \Lbb^{+}_{23} 
&=
\Lbb^{+}_{13} \Lbb^{-}_{23} \check{R}^{- }_{12}, 
\qquad 
\check{R}^{+}_{12} \Lbb^{+}_{13}
 \Lbb^{-}_{23} 
=
\Lbb^{-}_{13} \Lbb^{+}_{23} \check{R}^{+ }_{12}.
\end{split}
\label{FRT0}
\end{align}
In terms of \eqref{L-sp0} and \eqref{Rch-sp0},
 \eqref{FRT0} can also be rewritten as 
 a spectral parameter dependent Yang-Baxter equation
\footnote{\eqref{YBE-sp0} also contains 
a term 
$\check{R}^{-}_{12} \Lbb^{+}_{13} \Lbb^{-}_{23} -
\check{R}^{+}_{12} \Lbb^{-}_{13} \Lbb^{+}_{23} 
=
\Lbb^{-}_{13} \Lbb^{+}_{23} \check{R}^{- }_{12}-
\Lbb^{+}_{13} \Lbb^{-}_{23} \check{R}^{+ }_{12}
 $, 
whose relation to \eqref{FRT0} 
can be shown with the help of \eqref{R01}, namely  $\check{R}^{+}_{12}-\check{R}^{-}_{12}=q-q^{-1}$. 
(cf. \cite{Faddeev:1987ih}).} 
\begin{align}
\check{R}_{12}(\lambda / \mu ) \Lbb_{13}(\lambda)
 \Lbb_{23}(\mu)
&=
\Lbb_{13}(\mu) \Lbb_{23}(\lambda) \check{R}_{12}(\lambda / \mu ).
\label{YBE-sp0}
\end{align}
Let us evaluate the first space of the third relation in \eqref{UR-def} 
(and the corresponding relation for \eqref{inv-R}) 
in the fundamental representation $ \pi $.
\begin{align}
(1 \otimes \Delta) (\Lbb^{+} )=\Lbb^{+}_{13} \Lbb^{+}_{12}, 
\qquad
(1 \otimes \Delta) (\Lbb^{-} )=\Lbb^{-}_{13} \Lbb^{-}_{12}. 
 \label{coproL0}
\end{align}
Matrix elements of these give the co-multiplications for the generators
\footnote{To be precise, 
the co-multiplications for the Cartan elements defined by these relations are 
the ones for $q^{\pm \Ebf_{kk}}$ rather than the one for $ \Ebf_{kk}$. 
}
Let us rewrite
\footnote{Here we relabel the indices as $(1,2,3) \to (0,1,2)$.}
\eqref{YBE} and \eqref{YBE01}-\eqref{YBE02} in the following form 
\begin{align}
\begin{split}
\Rbb_{01}\Rbb_{02}&=
(\Rbb_{12}\Rbb_{02}\Rbb^{-1}_{12})
(\Rbb_{12}\Rbb_{01}\Rbb^{-1}_{12}).
\\[5pt]
\Rbb^{*}_{01}\Rbb^{*}_{02}&=
(\Rbb_{12}\Rbb^{*}_{02}\Rbb^{-1}_{12})
(\Rbb_{12}\Rbb^{*}_{01}\Rbb^{-1}_{12}), 
\\[5pt]
\Rbb^{*}_{01}\Rbb_{02}&=
(\Rbb_{12}\Rbb_{02}\Rbb^{-1}_{12})
(\Rbb_{12}\Rbb^{*}_{01}\Rbb^{-1}_{12})
\end{split}
\label{RR-rel0}
\end{align}
Evaluating the first space of these (labeled by $0$) 
in the fundamental representation $\pi$, we obtain 
the so-called zero curvature representation 
\begin{align}
\Lbb^{+(1)}\Lbb^{+(2)}
=\widetilde{\Lbb}^{+(2)}\widetilde{\Lbb}^{+(1)},
\quad
\Lbb^{-(1)}\Lbb^{+(2)}
=\widetilde{\Lbb}^{+(2)}\widetilde{\Lbb}^{-(1)},
\quad 
\Lbb^{-(1)}\Lbb^{-(2)}
=\widetilde{\Lbb}^{-(2)}\widetilde{\Lbb}^{-(1)}, 
\label{ZC-an0}
\end{align}
where $\Lbb^{\pm (a)}=\Lbb^{\pm}_{0a}$, 
$\widetilde{\Lbb}^{\pm (a)}=
\Rbb_{12} \Lbb^{\pm (a)} \Rbb^{-1}_{12} $, $a=1,2$. 
As remarked in \cite{BS15}, \eqref{ZC-an0} defines 
 a quantum Yang-Baxter map independent of 
the notion of the universal R-matrix. 
It gives a rational map among generators for the case 
$U_{q}(sl(2))$. However, we find that square roots 
appear in the map for $U_{q}(sl(n))$, $n \ge 3$ case. 
To overcome this difficulty, we will 
change the gauge of the universal R-matrix in the next 
subsection.
\subsection{Twisting L-operators}
\label{subsec-twist}
Here we reformulate the contents of the previous subsection in terms of 
twisted version of R-matrices. 
Let us introduce an element
\begin{align}
{\mathbf F}=q^{\sum_{i=1}^{n}\omega_{i-1} \otimes \Ebf_{ii}
},
\end{align}
where $\omega_{i}=\Ebf_{11}+\cdots + \Ebf_{ii}$ and $\omega_{0}=0$. 
This satisfies the following relations 
\footnote{We do not impose the condition 
${\mathbf F}_{12}{\mathbf F}_{21}=1$ 
(We thank S.\ Khoroshkin for a comment on this condition). 
This holds true if 
we modify the twist as 
${\mathbf F}_{12}=q^{\sum_{i=1}^{n}\omega_{i-1} \otimes \Ebf_{ii}
-\frac{1}{2}({\mathbf c}\otimes {\mathbf c} -
\sum_{i=1}^{n} \Ebf_{ii} \otimes \Ebf_{ii} )}$. 
} 
\begin{align}
\begin{split}
(\Delta \otimes 1) \, {\mathbf F} &=
{\mathbf F}_{13} \, {\mathbf F}_{23}, 
\\[5pt]
(1 \otimes \Delta ) \, {\mathbf F} &=
{\mathbf F}_{13} \, {\mathbf F}_{12}, 
\\[5pt]
{\mathbf F}_{12}{\mathbf F}_{13}{\mathbf F}_{23}&=
{\mathbf F}_{23}{\mathbf F}_{13}{\mathbf F}_{12}.
\end{split}
\end{align} 
We also have 
\begin{align}
{\mathbf F}_{21}=q^{\sum_{i=1}^{n} \Ebf_{ii} \otimes \omega_{i-1} }, 
\qquad 
{\mathbf F}_{12}^{-1}=q^{\sum_{i=1}^{n} \Ebf_{ii} \otimes \omega_{i} -{\mathbf c}\otimes {\mathbf c} 
}.
\label{F-inv}
\end{align}
It is known  \cite{Reshetikhin:1990ep} that 
 the gauge transformed universal R-matrices
\footnote{The central element $q^{-{\mathbf c}\otimes {\mathbf c} }$
in \eqref{F-inv} plays no role in our discussion below. 
Then we remove this from the universal R-matrices.} 
\begin{align}
\Rbf &={\mathbf F}_{21}\Rbb{\mathbf F}_{12}^{-1}
q^{{\mathbf c}\otimes {\mathbf c} }
=
q^{\sum_{i=1}^{n} \Ebf_{ii} \otimes \omega_{i}}
\overrightarrow{\prod}_{i<j}
\exp_{q^{-2}}
\left(
(q-q^{-1})^{-1}\Ebf_{ij}\otimes \Ebf_{ji}
\right)
q^{\sum_{i=1}^{n} \Ebf_{ii} \otimes \omega_{i}  
}
, 
\label{UR-expi-ren}
\\[5pt]
\Rbf^{*} &={\mathbf F}_{21}\Rbb^{*}{\mathbf F}_{12}^{-1}q^{{\mathbf c}\otimes {\mathbf c} }
=
q^{\sum_{i=1}^{n} \Ebf_{ii} \otimes \omega_{i-1} }
\overleftarrow{\prod}_{i<j}
\exp_{q^{2}}
\left(-
(q-q^{-1})^{-1}\Ebf_{ji}\otimes \Ebf_{ij}
\right)
q^{\sum_{i=1}^{n} \Ebf_{ii} \otimes \omega_{i-1} 
}
\label{UR-expi-ren2}
\end{align}
satisfy the defining relations of the universal R-matrix \eqref{UR-def} 
for the gauge transformed co-multiplication
 $\Delta^{F}(a)={\mathbf F} \Delta(a) {\mathbf F}^{-1}, a \in \A $: 
\begin{align}
\begin{split}
\Delta^{F}(\Ebf_{i,i+1})&=\Ebf_{i,i+1} \otimes q^{\Ebf_{ii}}
+q^{-\Ebf_{ii}} \otimes \Ebf_{i,i+1},
\\[5pt]
\Delta^{F}(\Ebf_{i+1,i})&=\Ebf_{i+1,i} \otimes q^{-\Ebf_{i+1,i+1}} + q^{\Ebf_{i+1,i+1}}
 \otimes \Ebf_{i+1,i} \qquad \text{for}
\quad i \in \{ 1,2,\dots, n-1 \}, 
\\[5pt]
\Delta^{F}(\Ebf_{kk})&=\Ebf_{kk} \otimes 1
+1 \otimes \Ebf_{kk}
\qquad \text{for} 
\quad k \in \{ 1,2,\dots, n\}. 
\end{split}
\label{co-pro-gauge}
\end{align}
Consequently,  
\eqref{UR-expi-ren} and \eqref{UR-expi-ren2} 
also satisfy the same equations as \eqref{YBE} and \eqref{YBE00}-\eqref{YBE04}: 
\begin{align}
\Rbf_{12}\Rbf_{13}\Rbf_{23}&=
\Rbf_{23}\Rbf_{13}\Rbf_{12}, 
\label{YBEt}
\\[5pt]
\Rbf_{12}\Rbf_{13}\Rbf^{*}_{23}&=
\Rbf^{*}_{23}\Rbf_{13}\Rbf_{12}, 
\label{YBE0}
\\[5pt]
\Rbf^{*}_{12}\Rbf_{13}\Rbf_{23}&=
\Rbf_{23}\Rbf_{13}\Rbf^{*}_{12}, 
\label{YBE1}
\\[5pt]
\Rbf^{*}_{12}\Rbf^{*}_{13}\Rbf_{23}&=
\Rbf_{23}\Rbf^{*}_{13}\Rbf^{*}_{12} . 
\label{YBE2}
\\[5pt]
\Rbf_{12}\Rbf^{*}_{13}\Rbf^{*}_{23}&=
\Rbf^{*}_{23}\Rbf^{*}_{13}\Rbf_{12} . 
\label{YBE3}
\\[5pt]
\Rbf^{*}_{12}\Rbf^{*}_{13}\Rbf^{*}_{23}&=
\Rbf^{*}_{23}\Rbf^{*}_{13}\Rbf^{*}_{12} . 
\label{YBE4}
\end{align}

We introduce images of the twist operators 
\begin{align}
\Psi=(\pi \otimes 1)({\mathbf F}_{21})=
\sum_{i=1}^{n} E_{ii} \otimes q^{\omega_{i-1}},
\qquad 
\Phi=(\pi \otimes 1)({\mathbf F}_{12}^{-1}
q^{{\mathbf c}\otimes {\mathbf c} })=
\sum_{i=1}^{n} E_{ii} \otimes q^{\omega_{i}}, 
\end{align}
and define the gauge transformed $\Lbf$-operators
\begin{align}
\begin{split}
\Lbf^{-}&=
(\pi \otimes 1)(\Rbf^{*})
=\Psi \Lbb^{-} \Phi =
 \\
&=\sum_{k=1}^{n}E_{kk}\otimes q^{2\omega_{k-1}}
 -\sum_{i<j}E_{ji} \otimes  q^{\omega_{i-1}+\omega_{j-1}}\Ebf_{ij} ,
\end{split}
\label{L-gauge1}
\\
\begin{split}
\Lbf^{+}&=(\pi \otimes 1) (\Rbf)
=\Psi \Lbb^{+} \Phi =
 \\
&=\sum_{k=1}^{n}E_{kk}\otimes q^{2\omega_{k}}
 +\sum_{i<j}E_{ij} \otimes  q^{\omega_{i}+\omega_{j}}\Ebf_{ji}.
\end{split}
\label{L-gauge2}
\end{align}
%
Evaluating further the second space (quantum space) of 
these L-operators in the fundamental representation $\pi$, we obtain the block 
R-matrices
%
%
\begin{align}
\begin{split}
\Rsf^{-}&=(1\otimes \pi )
\Lbf^{-} 
=(\pi \otimes \pi)
\Rbf^{*} 
\\[5pt]
&=\sum_{i,j}q^{2\theta(i > j)}E_{ii}\otimes E_{jj}
 -(q^{2}-1)\sum_{i<j}E_{ji} \otimes  E_{ij} ,
\\[8pt]
\Rsf^{+}&=(1\otimes \pi )
\Lbf^{+} =(\pi \otimes \pi) \Rbf
\\[5pt]
&=\sum_{i,j}q^{2\theta(i \ge j)}E_{ii}\otimes E_{jj}
 +(q^{2}-1)\sum_{i<j}E_{ij} \otimes  E_{ji},
\end{split}
\label{Rmp}
\end{align}
where $\theta(\text{true})=1$ and $\theta(\text{false})=0$.
Then we define the spectral parameter dependent L-operator
\begin{align}
\Lbf(\lambda)= \lambda \Lbf^{+} - \lambda^{-1} \Lbf^{-}
\label{L-sp}
\end{align}
and the R-matrix 
\begin{align}
\Rsf(\lambda)= \lambda \Rsf^{+} - \lambda^{-1} \Rsf^{-}. 
\label{Rs}
\end{align}
It has a property
\begin{align}
\Rsf(1)= (q^{2}-1)P.
\label{R1}
\end{align}
We will also use
\begin{align}
\Check{\Rsf}^{\pm}= P {\Rsf}^{\pm}, 
\qquad 
\Check{\Rsf}(\lambda)= P {\Rsf}(\lambda ). 
\label{Rch-sp}
\end{align}
Evaluating the first and the second space of \eqref{YBEt}-\eqref{YBE4} for 
$(\pi \otimes \pi  \otimes 1)$, 
and multiplying $P_{12}$ from the left, 
 we obtain the twisted version of the R-matrix form \cite{Faddeev:1987ih} 
of the defining relations 
\eqref{def-gln}-\eqref{def-eij}
\begin{align}
\begin{split}
\check{\Rsf}^{\pm}_{12} \Lbf^{-}_{13}
 \Lbf^{-}_{23} 
&=
\Lbf^{-}_{13} \Lbf^{-}_{23} \check{\Rsf}^{\pm }_{12}, 
\qquad 
\check{\Rsf}^{\pm}_{12} \Lbf^{+}_{13}
 \Lbf^{+}_{23} 
=
\Lbf^{+}_{13} \Lbf^{+}_{23} \check{\Rsf}^{\pm }_{12},
\\[5pt]
\check{\Rsf}^{-}_{12} \Lbf^{-}_{13}
 \Lbf^{+}_{23} 
&=
\Lbf^{+}_{13} \Lbf^{-}_{23} \check{\Rsf}^{- }_{12}, 
\qquad 
\check{\Rsf}^{+}_{12} \Lbf^{+}_{13}
 \Lbf^{-}_{23} 
=
\Lbf^{-}_{13} \Lbf^{+}_{23} \check{\Rsf}^{+ }_{12}.
\end{split}
\label{FRT}
\end{align}
The algebra 
generated by the matrix elements of $\Lbf^{-}$ (resp.\ $\Lbf^{+}$)
with the first (resp.\ second) relation in \eqref{FRT}
is the Borel subalgebra 
$\Bcl_{+}$ (resp.\ $\Bcl_{-}$). 
In terms of \eqref{L-sp} and \eqref{Rch-sp},
 \eqref{FRT} can also be rewritten as 
 a spectral parameter dependent Yang-Baxter equation
\footnote{\eqref{YBE-sp} also contains 
a term 
$\check{\Rsf}^{-}_{12} \Lbf^{+}_{13} \Lbf^{-}_{23} -
\check{\Rsf}^{+}_{12} \Lbf^{-}_{13} \Lbf^{+}_{23} 
=
\Lbf^{-}_{13} \Lbf^{+}_{23} \check{\Rsf}^{- }_{12}-
\Lbf^{+}_{13} \Lbf^{-}_{23} \check{\Rsf}^{+ }_{12}
 $, 
whose relation to \eqref{FRT} 
can be shown with the help of \eqref{R1}, namely  $\check{R}^{+}_{12}-\check{R}^{-}_{12}=q^{2}-1$. 
(cf. \cite{Faddeev:1987ih}).} 
\begin{align}
\check{\Rsf}_{12}(\lambda / \mu ) \Lbf_{13}(\lambda)
 \Lbf_{23}(\mu)
&=
\Lbf_{13}(\mu) \Lbf_{23}(\lambda) \check{\Rsf}_{12}(\lambda / \mu ).
\label{YBE-sp}
\end{align}
One can derive the co-multiplication for the L-operators in the same way as 
\eqref{coproL0}.
\begin{align}
(1 \otimes \Delta) (\Lbf^{+} )=\Lbf^{+}_{13} \Lbf^{+}_{12}, 
\qquad
(1 \otimes \Delta) (\Lbf^{-} )=\Lbf^{-}_{13} \Lbf^{-}_{12}. 
 \label{coproL}
\end{align}

We can rewrite 
\eqref{YBEt} and \eqref{YBE1}-\eqref{YBE2} in the same 
form as \eqref{RR-rel0}, 
from which 
the zero curvature representation 
follows 
\begin{align}
\Lbf^{+(1)}\Lbf^{+(2)}
=\widetilde{\Lbf}^{+(2)}\widetilde{\Lbf}^{+(1)},
\quad
\Lbf^{-(1)}\Lbf^{+(2)}
=\widetilde{\Lbf}^{+(2)}\widetilde{\Lbf}^{-(1)},
\quad 
\Lbf^{-(1)}\Lbf^{-(2)}
=\widetilde{\Lbf}^{-(2)}\widetilde{\Lbf}^{-(1)}, 
\label{ZC-an}
\end{align}
where 
$\Lbf^{\pm (a)}=\Lbf^{\pm}_{0a}$, 
$\widetilde{\Lbf}^{\pm (a)}=
\Rbf_{12} \Lbf^{\pm (a)} \Rbf^{-1}_{12} $, $a=1,2$. 
Now the set of generators ${\mathbf X}$ corresponding to the one in 
 \eqref{defYB-map} becomes  
$ {\mathbf X}=\{ \Lbf^{+}_{ij}, \ \Lbf^{-}_{ji} \}_{i\le j}$, 
where these are related to the matrix elements
\footnote{Here we abuse notation. The subscripts are used for the indices of matrix elements 
 $\Lbf^{\pm}_{ij}$ 
of $(n \times n)$-matrices $\Lbf^{\pm}=\sum_{ij}E_{ij}\otimes \Lbf^{\pm}_{ij}$, 
which should not be confused with the space operators
 $\Lbf^{\pm}$ (in \eqref{FRT}-\eqref{coproL})
 are acting on.} of the L-operators 
 \eqref{L-gauge1} and \eqref{L-gauge2} as:
\begin{align}
\begin{split}
\Lbf^{+}_{ij}&=q^{\omega_{i}+\omega_{j}}\Ebf_{ji}, 
\qquad 
\Lbf^{-}_{ji}=-q^{\omega_{i-1}+\omega_{j-1}} \Ebf_{ij}, 
\qquad 
\Lbf^{+}_{ji}=\Lbf^{-}_{ij}=0
\quad \text{for} \quad i<j, 
\\[5pt]
\Lbf^{+}_{kk}&=q^{2\omega_{k}}, 
\qquad 
\Lbf^{-}_{kk}=q^{2\omega_{k-1}}. 
\end{split}
\label{L-Chv}
\end{align}
We will also use the notation $\ubf_{k}=q^{2\omega_{k}}$. 
In these variables, the zero-curvature relation \eqref{ZC-an} 
realizes a rational (without square roots) quantum Yang-Baxter map:  
\begin{multline}
{\mathcal R} : 
({\mathbf X}^{(1)}, {\mathbf X}^{(2)})
\mapsto 
(\widetilde{\mathbf X}^{(1)}, \widetilde{\mathbf X}^{(2)}), 
\\[6pt]
\widetilde{\mathbf X}^{(a)}=\Rbf{\mathbf X}^{(a)}
\Rbf^{-1}
=
\left\{ \widetilde{\mathbf x}^{(a)}:=
\Rbf{\mathbf x}^{(a)}
\Rbf^{-1} | {\mathbf x}^{(a)} \in {\mathbf X}^{(a)} 
\right\}, 
\qquad a=1,2, 
\\[6pt]
{\mathbf X}^{(1)}=
\{ {\mathbf x}^{(1)}:={\mathbf x} \otimes 1 | {\mathbf x} \in {\mathbf X} \}
, \qquad 
{\mathbf X}^{(2)}=
\{{\mathbf x}^{(2)}:=1\otimes {\mathbf x} | {\mathbf x} \in {\mathbf X} \} ,
\label{defYB-map-t}
\end{multline}
as will be shown explicitly 
in subsection \ref{subsec-YBQD}. 
That the L-operators  \eqref{L-gauge1}-\eqref{L-gauge2} for  $U_{q}(sl(3))$ case 
give a rational quantum Yang-Baxter map through \eqref{ZC-an} 
is a suggestion by S.\ Sergeev \cite{Sergeev15}.

We remark that one can derive another zero curvature representation from 
\eqref{YBE0}, \eqref{YBE3} and \eqref{YBE4}: 
\begin{align}
\Lbf^{+(1)}\Lbf^{+(2)}
=\widetilde{\Lbf}^{+(2)}\widetilde{\Lbf}^{+(1)},
\quad
\Lbf^{+(1)}\Lbf^{-(2)}
=\widetilde{\Lbf}^{-(2)}\widetilde{\Lbf}^{+(1)},
\quad 
\Lbf^{-(1)}\Lbf^{-(2)}
=\widetilde{\Lbf}^{-(2)}\widetilde{\Lbf}^{-(1)}, 
\label{ZC-an*}
\end{align}
where 
$\widetilde{\Lbf}^{\pm (a)}=
\Rbf^{*}_{12} \Lbf^{\pm (a)} \Rbf^{*-1}_{12} $, $a=1,2$. 
This defines another Yang-Baxter map: 
\begin{multline}
{\mathcal R}^{*} : 
({\mathbf X}^{(1)}, {\mathbf X}^{(2)})
\mapsto 
(\widetilde{\mathbf X}^{(1)}, \widetilde{\mathbf X}^{(2)}), 
\\[6pt]
\widetilde{\mathbf X}^{(a)}=\Rbf^{*} {\mathbf X}^{(a)}
\Rbf^{*-1}
=
\left\{ \widetilde{\mathbf x}^{(a)}:=
\Rbf^{*} {\mathbf x}^{(a)}
\Rbf^{*-1} | {\mathbf x}^{(a)} \in {\mathbf X}^{(a)} 
\right\}, 
\qquad a=1,2, 
\\[6pt]
{\mathbf X}^{(1)}=
\{ {\mathbf x}^{(1)}:={\mathbf x} \otimes 1 | {\mathbf x} \in {\mathbf X} \}
, \qquad 
{\mathbf X}^{(2)}=
\{{\mathbf x}^{(2)}:=1\otimes {\mathbf x} | {\mathbf x} \in {\mathbf X} \} ,
\label{defYB-map-t*}
\end{multline}
which is different from \eqref{defYB-map-t}. 
The zero-curvature relation \eqref{ZC-an*} 
is mapped to \eqref{ZC-an} by the replacement: 
\begin{align}
(\Lbf^{\pm (1)} , \Lbf^{\pm (2)}, 
\widetilde{\Lbf}^{\pm (1)} ,\widetilde{\Lbf}^{\pm (2)}) 
\mapsto
 (\widetilde{\Lbf}^{\pm (2)} ,\widetilde{\Lbf}^{\pm (1)},
\Lbf^{\pm (2)} , \Lbf^{\pm (1)}) . 
\label{*trans}
\end{align}
Thus 
the map ${\mathcal R}^{*}$ can be obtained 
by swapping the superscripts `$^{(1)}$' and `$^{(2)}$' of the variables  in 
the inverse map ${\mathcal R}^{-1}$. 
\subsection{Review on quasi-determinants}
\label{subsec-QD}

In this subsection, we summarize formulas on quasi-determinants 
based on  \cite{GR91,GR92,GGRW05}
 (see also \cite{CFR09,CFRS12,Lauve04,Molev-book}). 

Let $A=(a_{ij})_{1 \le i,j \le N}$ be a $N \times N$ matrix 
whose matrix elements $a_{ij}$ are elements of an 
associative algebra. Here we do not require the 
commutativity of the algebra. 
For $i_{1},i_{2},\dots, i_{m} , 
j_{1},j_{2},\dots, j_{n} \in \mathcal{I}=\{1,2,\dots, N\}$
we introduce the following notation for a matrix:
\begin{align}
A^{i_{1},i_{2},\dots, i_{m}}_{j_{1},j_{1},\dots, j_{n}}
=
\begin{pmatrix}
a_{i_{1},j_{1}} & a_{i_{1},j_{2}} & \cdots & a_{i_{1},j_{n}} 
\\
a_{i_{2},j_{1}} & a_{i_{2},j_{2}} & \cdots & a_{i_{2},j_{n}} 
\\
\vdots & \vdots & \ddots & \vdots \\
a_{i_{m},j_{1}} & a_{i_{m},j_{2}} & \cdots & a_{i_{m},j_{n}} 
\end{pmatrix}.
\end{align}
We will also use the notation 
$A^{ij}=A^{1,2,\dots, \widehat{i}, \dots, N}_{1,2,\dots,\widehat{j}, \dots, N}$, 
where ``$1,2,\dots, \widehat{i}, \dots, N $'' means that 
$i$ is removed from the sequence of numbers ``$1,2, \dots, N$''. 
In general, there are $N^2$ quasi-determinants for a $N \times N$ matrix. 
For a $1 \times 1$-sub-matrix $A^{i}_{j}=(a_{ij})$ of 
$A=A^{1,2,\dots,N}_{1,2,\dots,N}$, 
{\em $(i,j)$-th quasi-determinant}
\footnote{Here $(i,j)$ denotes the subscript of the matrix element 
$a_{ij}$ of the original matrix $A$ 
rather than $i$-th row and 
$j$-th row of the matrix $A^{i}_{j}$. 
This is also the case with more general sub-matrices. 
In particular, the quasi-determinant does not depend on permutations 
on rows and columns in the matrix.} is defined by 
$|A^{i}_{j}|_{ij}= a_{ij}$. 
Let us consider the case where 
 subsets $ \{i_{1},i_{2},\dots,i_{m}\}, \{j_{1},j_{2},\dots,j_{m}\}$ 
 of $\mathcal{I}$ satisfy the following conditions: $m \ge 2$; 
$ i_{1},i_{2},\dots,i_{m}$ (resp.\ $\{j_{1},j_{2},\dots,j_{m}\}$) 
are pairwise distinct; 
$ i \in \{i_{1},i_{2},\dots,i_{m}\}$, $  j \in \{j_{1},j_{2},\dots,j_{m}\}$. 
Then the $(i,j)$-th quasi-determinant of the submatrix 
$\tilde{A}=A^{i_{1},i_{2},\dots,i_{m}}_{j_{1},j_{2},\dots,j_{m}}$ of 
$A$ is recursively defined 
 by 
\begin{align}
|\tilde{A}|_{ij}= a_{ij}- 
\sum_{k \in \{j_{1},j_{2},\dots,j_{m}\} \setminus \{j \}, \atop 
l \in \{i_{1},i_{2},\dots,i_{m}\} \setminus \{i \} }
a_{ik}(|\tilde{A}^{ij}|_{lk})^{-1}a_{lj} .
\label{def-quasi-det}
\end{align}
In case all the quasi-determinants of $A$ are not zero, 
the inverse matrix of $A$ can be expressed in terms of them:
\begin{align}
A^{-1}=(|A|_{ji}^{-1})_{1 \le i,j \le N}. 
 \label{invdet}
\end{align}
If $A^{ij}$ is invertible, then 
the $(i,j)$-th quasi-determinant can also be defined as 
\begin{align}
|A|_{ij}= a_{ij}- 
\sum_{k \in \mathcal{I} \setminus \{j \}, \atop 
l \in \mathcal{I}\setminus \{i \} }
a_{ik}((A^{ij})^{-1})_{kl}a_{lj} ,
\label{def-quasi-det2}
\end{align}
where $((A^{ij})^{-1})_{kl}$ is the matrix element
 at $k$-th row and $l$-th column of the inverse matrix $(A^{ij})^{-1}$. 
In case all the matrix elements of $A$ are commutative, 
the $(i,j)$-th quasi-determinant of $A$ reduces to 
$|A|_{ij}=(-1)^{i+j}\det A / \det A^{ij}$.
We also use the following the notation for the $(i,j)$-th quasi-determinant of $A$,
\begin{align}
|A|_{ij}
=
\begin{array}{|ccccc|}
a_{11} & \cdots & a_{1j} & \cdots & a_{1 N} \\
\vdots   &           & \vdots  &           & \vdots \\
a_{i1}  & \cdots  & \fbox{$a_{ij}$} & \cdots & a_{iN} \\
\vdots   &           & \vdots  &           & \vdots \\
a_{N1}  & \cdots  & a_{Nj} & \cdots & a_{NN} 
\end{array}
\ .
\end{align}
Let us give examples for $N=2$ case:
\begin{align}
|A^{i_{1}i_{2}}_{j_{1}j_{2}}|_{i_{2}j_{1}}
&=
\begin{array}{|cc|}
a_{i_{1}j_{1}} & a_{i_{1}j_{2}}  \\
\fbox{$a_{i_{2}j_{1}}$}  &  a_{i_{2}j_{2}}
\end{array}
=a_{i_{2}j_{1}}-a_{i_{2}j_{2}} a^{-1}_{i_{1}j_{2}} a_{i_{1}j_{1}},
\end{align}
and for $N=3$ case:
\begin{align}
\begin{split}
|A|_{33}
&=
\begin{array}{|ccc|}
a_{11} & a_{1 2} & a_{1 3}\\
a_{21}  &  a_{22} & a_{2 3} \\
a_{31}  &  a_{32} &\fbox{$a_{33}$}
\end{array}
=a_{33}
-
a_{31}
\,
\begin{array}{|cc|}
\fbox{$a_{11}$} & a_{1 2} \\
a_{21}  &  a_{22} 
\end{array}
^{\ -1}
a_{13}
-
a_{32}
\,
\begin{array}{|cc|}
a_{11} &\fbox{$a_{12}$} \\
a_{21}  &  a_{22} 
\end{array}
^{\ -1}
a_{13}
\\[6pt]
& 
\hspace{130pt}
-
a_{31}
\,
\begin{array}{|cc|}
a_{11} & a_{1 2} \\
\fbox{$a_{21}$}  &  a_{22} 
\end{array}
^{\ -1}
a_{23}
-
a_{32}
\,
\begin{array}{|cc|}
a_{11} & a_{1 2} \\
a_{21}  &  \fbox{$a_{22}$} 
\end{array}
^{\ -1}
a_{23}
\\[6pt]
&= 
a_{33}
-a_{31}(a_{11}-a_{12} (a_{22})^{-1}a_{21})^{-1}a_{13}
-a_{32}(a_{12}-a_{11} (a_{21})^{-1}a_{22})^{-1}a_{13}
\\[6pt]
&
\hspace{31pt}
-a_{31}(a_{21}-a_{22} (a_{12})^{-1}a_{11})^{-1}a_{23}
-a_{32}(a_{22}-a_{21} (a_{11})^{-1}a_{12})^{-1}a_{23} .
\end{split}
\end{align}
In case, some of the matrix elements of a block matrix have the same subscript, 
we use an overline `$\Bar{\quad}$' to remove ambiguity. 
For example, for a block matrix 
$
(B^{ijk}_{i}\ C^{ijk}_{ij})=
\begin{pmatrix}
b_{ii} & c_{ii} & c_{ij} \\
b_{ji} & c_{ji} & c_{jj} \\
b_{ki} & c_{ki} & c_{kj}
\end{pmatrix}
$
, we use the notation 
$
|B^{ijk}_{i}\ C^{ijk}_{\Bar{i}j}|_{j\Bar{i}}
=
\begin{vmatrix}
b_{ii} & c_{ii} & c_{ij} \\
b_{ji} & \fbox{$c_{ji}$} & c_{jj} \\
b_{ki} & c_{ki} & c_{kj}
\end{vmatrix}
$, 
$
|B^{ijk}_{\Bar{i}}\ C^{ijk}_{ij}|_{j\Bar{i}}
=
\begin{vmatrix}
b_{ii} & c_{ii} & c_{ij} \\
\fbox{$b_{ji}$} & c_{ji} & c_{jj} \\
b_{ki} & c_{ki} & c_{kj}
\end{vmatrix}
$
 since both $b_{ji}$ and $c_{ji}$ have the same 
subscript `$ji$'.

There are relations among the quasi-determinants of the matrix $A$ and 
those of its sub-matrices, namely, 
\begin{proposition}
The row homological relations:
\begin{align}
-|A|_{ij} \, |A^{il}|^{-1}_{sj}=|A|_{il}  \,  |A^{ij}|^{-1}_{sl} 
\quad \text{for} \quad 
s \ne i, \quad l \ne j,
\label{row-hom}
\end{align}
and the column homological relations:
\begin{align}
- |A^{kj}|^{-1}_{it} \, |A|_{ij}= |A^{ij}|^{-1}_{kt} \,  |A|_{kj}
\quad \text{for} \quad 
k \ne i, \quad 
t \ne j.
\label{col-hom}
\end{align}
\end{proposition}
There are analogues of the {\em Laplace expansion formulas} for quasi-determinants. 
\begin{proposition}
\begin{align}
|A|_{ij}&=a_{ij}-\sum_{k \ne j} a_{ik} (|A^{ij}|_{sk})^{-1}|A^{ik}|_{sj} 
\quad \text{for} \quad s \ne i, 
\label{Lap-row}
\\[6pt]
|A|_{ij}&=a_{ij}-\sum_{k \ne i} |A^{kj}|_{is}  (|A^{ij}|_{ks})^{-1} a_{kj}
\quad \text{for} \quad s \ne j.
\label{Lap-col}
\end{align}
\end{proposition}
Let us take subsets $P$ and $Q$ of $\mathcal{I}$ with the same size $|P|=|Q|<N$.
Let $B=A^{-1}$. Then the following relation holds
\footnote{This relation is formulated for more general sets ``$I,J$'' in \cite{GGRW05}. 
Here we swapped the sets ``$I \setminus P$'' and ``$J \setminus Q$'' 
in Theorem 1.5.4 in \cite{GGRW05}.}
\begin{proposition}
For $k \notin P$ and $l \notin Q$, 
\begin{align}
\left|A^{P \cup \{k \}}_{Q \cup \{l \}}\right|_{kl} \,  
\left|B^{\mathcal{I} \setminus Q}_{\mathcal{I} \setminus P}\right|_{lk}=1 . 
 \label{inversion}
\end{align}
\end{proposition}
Note that 
this reduces to \eqref{invdet} for $P=\mathcal{I}\setminus \{k \}$ 
and $Q=\mathcal{I}\setminus \{l \}$.

Fix the numbers
 $m<N$, $i,j,j_{1},j_{2},\dots, j_{m-1} \in \mathcal{I}$, and 
$ i \notin \{ j_{1},j_{2},\dots, j_{m-1} \}$. 
Then the {\em left quasi-Pl\"{u}cker coordinates} of 
the $m \times N$  matrix $A^{1,2,\dots, m}_{1,2,\dots, N}$ are defined by 
\begin{align}
q^{j_{1},j_{2},\dots , j_{m-1}}_{ij}
(A^{1,2,\dots, m}_{1,2,\dots, N})=
(|A^{1,2,\dots , m}_{i,j_{1},\dots, j_{m-1}}|_{si})^{-1}
|A^{1,2,\dots , m}_{j,j_{1},\dots, j_{m-1}}|_{sj} 
\label{left-qP}
\end{align}
for any $s \in \{1,2,\dots , m\}$. Here \eqref{left-qP} 
does not depend on $s$. 
Fix numbers
 $m<N$, $i,j,i_{1},i_{2},\dots, i_{m-1} \in \mathcal{I}$, and 
$ j \notin \{ i_{1},i_{2},\dots, i_{m-1} \}$. 
Then the {\em right quasi-Pl\"{u}cker coordinates} of 
the $N \times m$ matrix $A^{1,2,\dots, N}_{1,2,\dots, m}$ are defined by 
\begin{align}
r^{i_{1},i_{2},\dots , i_{m-1}}_{ij}
(A^{1,2,\dots, N}_{1,2,\dots, m})=
|A^{i,i_{1},\dots, i_{m-1}}_{1,2,\dots , m}|_{it}
(|A^{j,i_{1},\dots, i_{m-1}}_{1,2,\dots , m}|_{jt})^{-1}
\label{right-qP}
\end{align}
for any $t \in \{1,2,\dots , m\}$. Here \eqref{right-qP} 
does not depend on $t$. 
These quasi-Pl\"{u}cker coordinates reduce to $1$ at $i=j$. 
They also satisfy 
quasi-Pl\"{u}cker relations, which reduce to 
Pl\"{u}cker relations in case all the matrix elements are commutative. 
In addition, they reduce to ratios of Pl\"{u}cker coordinates:
\begin{align}
\begin{split}
q^{j_{1},j_{2},\dots , j_{m-1}}_{ij}
(A^{1,2,\dots, m}_{1,2,\dots, N}) &=
\det (A^{1,2,\dots , m}_{i,j_{1},\dots, j_{m-1}})^{-1}
\det(A^{1,2,\dots , m}_{j,j_{1},\dots, j_{m-1}})
\, \,  \text{[commutative case]},
\\[6pt]
r^{i_{1},i_{2},\dots , i_{m-1}}_{ij}
(A^{1,2,\dots, N}_{1,2,\dots, m}) &=
\det (A^{i,i_{1},\dots, i_{m-1}}_{1,2,\dots , m})
\det (A^{j,i_{1},\dots, i_{m-1}}_{1,2,\dots , m})^{-1} 
\, \,  \text{[commutative case]}.
\end{split}
\label{qP-cl}
\end{align}

There are two types of 
Gauss decomposition formulas for the matrix $A$. 
\begin{theorem}
\label{Gauss1}
Suppose that the quasi-determinants 
\begin{align}
{\mathbb H}_{k}=|A^{k,k+1,\dots, N}_{k,k+1,\dots, N}|_{kk}, 
\qquad k \in \{1,2,\dots N \}
\end{align}
are well-defined and invertible. Then we have 
\begin{align}
A& =
\begin{pmatrix}
1 & {\mathbb E}_{12} & \dots & {\mathbb E}_{1N}\\
0 & 1 & \dots & {\mathbb E}_{2N} \\
\vdots & \vdots &\ddots & \vdots \\
0 & 0  & \dots & 1
\end{pmatrix}
\begin{pmatrix}
{\mathbb H}_{1} & 0 & \dots & 0\\
0 & {\mathbb H}_{2} & \dots & 0 \\
\vdots & \vdots &\ddots & \vdots \\
0 & 0  & \dots & {\mathbb H}_{N}
\end{pmatrix}
\begin{pmatrix}
1 & 0 & \dots & 0\\
{\mathbb F}_{21} & 1 & \dots & 0 \\
\vdots & \vdots &\ddots & \vdots \\
{\mathbb F}_{N1} & {\mathbb F}_{N2}  & \dots & 1
\end{pmatrix},
\end{align}
where 
\begin{align}
{\mathbb E}_{ij}&=
 r^{j+1,j+2,\dots, N}_{ij}(A^{1,2,\dots, N}_{j,j+1,\dots, N}),
\\[5pt]
{\mathbb F}_{ji}&=
 q^{j+1,j+2,\dots, N}_{ji}(A^{j,j+1,\dots, N}_{1,2,\dots, N}), 
\qquad 1 \le i < j \le N.
\end{align}
\end{theorem}
\begin{theorem}
\label{Gauss2}
Suppose that the quasi-determinants 
\begin{align}
\widetilde{\mathbb H}_{k}=|A^{1,2,\dots, k}_{1,2,\dots, k}|_{kk}, 
\qquad k \in \{1,2,\dots N \}
\end{align}
are well-defined and invertible. Then we have 
\begin{align}
A& =
\begin{pmatrix}
1 & 0 & \dots & 0\\
\widetilde{\mathbb F}_{21} & 1 & \dots & 0 \\
\vdots & \vdots &\ddots & \vdots \\
\widetilde{\mathbb F}_{N1} & \widetilde{\mathbb F}_{N2}  
 & \dots & 1
\end{pmatrix}
\begin{pmatrix}
\widetilde{\mathbb H}_{1} & 0 & \dots & 0\\
0 & \widetilde{\mathbb H}_{2} & \dots & 0 \\
\vdots & \vdots &\ddots & \vdots \\
0 & 0  & \dots & \widetilde{\mathbb H}_{N}
\end{pmatrix}
\begin{pmatrix}
1 & \widetilde{\mathbb E}_{12} & \dots & 
\widetilde{\mathbb E}_{1N}\\
0 & 1 & \dots & \widetilde{\mathbb E}_{2N} \\
\vdots & \vdots &\ddots & \vdots \\
0 & 0  & \dots & 1
\end{pmatrix}
,
\end{align}
where 
\begin{align}
\widetilde{\mathbb E}_{ij}&=
 q^{1,2,\dots, i-1}_{ij}(A^{1,2,\dots, i}_{1,2,\dots, N}),
\\[5pt]
\widetilde{\mathbb F}_{ji}&=
 r^{1,2,\dots, i-1}_{ji}(A^{1,2,\dots, N}_{1,2,\dots, i}), 
\qquad 1 \le i < j \le N.
\end{align}
\end{theorem}
A detailed explanation of Theorem \ref{Gauss2} can also 
be found in \cite{Molev-book}.

\subsection{Solution of the zero-curvature representation}
\label{subsec-YBQD}
Let us solve the system of equations \eqref{ZC-an} 
for the matrices  with non-commutative entries 
$\Lbf^{\pm (a)}=(\Lbf^{\pm (a)}_{ij})_{1 \le i,j \le n}$ and 
$\tilde{\Lbf}^{\pm (a)}=
(\tilde{\Lbf}^{\pm (a)}_{ij})_{1 \le i,j \le n}$ 
for $a=1,2$ 
under the condition 
$\Lbf^{+(a)}_{ij}=\widetilde{\Lbf}^{+(a)}_{ij}=
\widetilde{\Lbf}^{-(a)}_{ji}=\Lbf^{-(a)}_{ji}=0$ for $i>j$, 
$\Lbf^{+(a)}_{ii}=\ubf_{i}^{(a)}$, 
$\widetilde{\Lbf}^{+(a)}_{ii}=\widetilde{\ubf}_{i}^{(a)}$, 
$\Lbf^{-(a)}_{ii}=\ubf_{i-1}^{(a)}$, 
$\widetilde{\Lbf}^{-(a)}_{ii}=\widetilde{\ubf}_{i-1}^{(a)}$ 
and 
  $\ubf_{0}^{(a)}=\widetilde{\ubf}_{0}^{(a)}=1$. 
  In addition, we will impose the condition $\ubf_{n}^{(a)}=
\widetilde{\ubf}_{n}^{(a)}=1$ 
if we consider $U_{q}(sl(n))$. 
These variables come from \eqref{L-Chv} and \eqref{defYB-map-t}. 
However, 
at this stage, we do not a priori assume the defining relations of $U_{q}(gl(n))$ explicitly
\footnote{However, structures on $U_{q}(gl(n))$ are incorporated in the above conditions
 on the matrix elements and the fact that we are solving the zero-curvature relation, which 
 comes from Yang-Baxter relations. 
 The variables of the form ${\mathbf A}^{(1)}$ and  ${\mathbf B}^{(2)}$ 
 (resp.\  $\widetilde{\mathbf A}^{(1)}$ and  $\widetilde{\mathbf B}^{(2)}$) 
 commute each other. However, this commutativity is not explicitly assumed 
 in Theorems \ref{sol-zero1} and \ref{sol-zero1-o}.}. 

Let us denote the left hand side of 
the second equation in \eqref{ZC-an} as 
$\Jbf=(\Jbf_{a,b})_{1 \le a,b \le n}
:=\Lbf^{-(1)} \Lbf^{+(2)}$, where 
\begin{align}
\Jbf_{a,b}=
\sum_{k=1}^{\min\{a,b\}} \Lbf^{-(1)}_{ak}\Lbf^{+(2)}_{kb}.
\end{align} 
Then we find the following solution. 
\begin{theorem}
\label{sol-zero1}
For $ 1 \le i \le j \le n $,
\begin{align}
\widetilde{\Lbf}_{ij}^{+(1)}&=
\left(
\overrightarrow{\prod}_{k=1}^{i-1} 
|\Jbf^{k,\dots,n}_{k,\dots,n}|^{-1}_{kk}
\ubf_{k}^{(1)}\ubf_{k}^{(2)}
\right)
|\Jbf^{i,\dots,n}_{i,\dots,n}|^{-1}_{ii}
\begin{array}{|cc|}
(\Lbf^{+(1)}\Lbf^{+(2)})^{i,i+1,\dots,n}_{\Bar{j}}
&
 \Jbf^{i,i+1,\dots , n}_{i+1,i+2,\dots, n} 
\end{array}
_{\, i\Bar{j}} ,
 \label{Lp1-th} 
\\[6pt]
\widetilde{\Lbf}_{ji}^{-(1)}&=
\left(
\overrightarrow{\prod}_{k=1}^{j-1} 
|\Jbf^{k,\dots,n}_{k,\dots,n}|^{-1}_{kk}
\ubf_{k}^{(1)}\ubf_{k}^{(2)}
\right)
|\Jbf^{j,\dots,n}_{j,\dots,n}|_{jj}^{-1}
 |\Jbf^{j,j+1,\dots,n}_{i,j+1,\dots,n}|_{ji} ,
\label{Lm1-th}
\\[6pt]
\widetilde{\Lbf}_{ij}^{+(2)}&=
 |\Jbf^{i,j+1,\dots,n}_{j,j+1,\dots,n}|_{ij}
\overleftarrow{\prod}_{k=1}^{j-1}
 (\ubf_{k}^{(1)}\ubf_{k}^{(2)})^{-1}
|\Jbf^{k,\dots,n}_{k,\dots,n}|_{kk} ,
\label{Lp2-th}
\\[6pt]
\widetilde{\Lbf}_{ji}^{-(2)}&=
\begin{array}{|c|}
(\Lbf^{-(1)}\Lbf^{-(2)})^{\Bar{j}}_{i,i+1,\dots,n}
\\
 \Jbf^{i+1,i+2,\dots , n}_{i,i+1,\dots, n} 
\end{array}
_{\, \Bar{j}i} 
\, 
\overleftarrow{\prod}_{k=1}^{i-1} 
(\ubf_{k}^{(1)}\ubf_{k}^{(2)})^{-1}
| \Jbf^{k,\dots,n}_{k,\dots,n} |_{kk} 
 \label{Lm2-th}
\end{align}
solve the zero-curvature relation \eqref{ZC-an}.
\end{theorem}
%
Before we go into the proof of this theorem, we give an example for $n=3$ case.
All the formulas can be written in terms of matrices: 
\begin{align}
 \Lbf^{-(a)} &=\begin{pmatrix}
1 & 0 & 0 \\
 \Lbf^{-(a)}_{21} & \ubf^{(a)}_{1} & 0 \\ 
 \Lbf^{-(a)}_{31} &  \Lbf^{-(a)}_{32} & \ubf^{(a)}_{2} 
\end{pmatrix},
\qquad 
 \Lbf^{+(a)} =\begin{pmatrix}
\ubf^{(a)}_{1}  & \Lbf^{+(a)}_{12} &  \Lbf^{+(a)}_{13} \\
0 & \ubf^{(a)}_{2} &  \Lbf^{+(a)}_{23} \\ 
0 & 0 & \ubf^{(a)}_{3}
\end{pmatrix},
\qquad 
a=1,2, 
\nonumber 
\\[5pt]
 \Lbf^{-(1)} \Lbf^{+(2)} &=\Jbf =
\begin{pmatrix}
\Jbf_{11} & \Jbf_{12} & \Jbf_{13} \\
\Jbf_{21} & \Jbf_{22} & \Jbf_{23} \\ 
\Jbf_{31} & \Jbf_{32} & \Jbf_{33} 
\end{pmatrix}
\nonumber \\[6pt]
&=
\begin{pmatrix}
\ubf^{(2)}_{1} & \Lbf^{+(2)}_{12} & \Lbf^{+(2)}_{13} \\
\Lbf^{-(1)}_{21} \ubf^{(2)}_{1} &
   \Lbf^{-(1)}_{21}\Lbf^{+(2)}_{12} + \ubf^{(1)}_{1} \ubf^{(2)}_{2} 
   & \Lbf^{-(1)}_{21}\Lbf^{+(2)}_{13} + \ubf^{(1)}_{1} \Lbf^{+(2)}_{23} \\
\Lbf^{-(1)}_{31} \ubf^{(2)}_{1} &\Lbf^{-(1)}_{31}\Lbf^{+(2)}_{12} 
 + \Lbf^{-(1)}_{32}\ubf^{(2)}_{2} 
   & \Lbf^{-(1)}_{31}\Lbf^{+(2)}_{13} +\Lbf^{-(1)}_{32} \Lbf^{+(2)}_{23}
     +\ubf^{(1)}_{2}\ubf^{(2)}_{3}
\end{pmatrix}
.
\end{align}
Then the solution is given by
\begin{align}
\widetilde{\ubf}_{0}^{(1)}&=\widetilde{\Lbf}_{11}^{-(1)}=1,
\nonumber 
\\[6pt]
\widetilde{\ubf}_{1}^{(1)}&=\widetilde{\Lbf}_{11}^{+(1)}=
\widetilde{\Lbf}_{22}^{-(1)}=
\begin{array}{|ccc|}
\fbox{$\Jbf_{11}$} & \Jbf_{12} & \Jbf_{13} \\
\Jbf_{21} & \Jbf_{22} & \Jbf_{23} \\
\Jbf_{31} & \Jbf_{32} & \Jbf_{33} 
\end{array}
^{\,  -1}
\ubf^{(1)}_{1} \ubf^{(2)}_{1},
\nonumber 
\\[6pt]
\widetilde{\ubf}_{2}^{(1)}&=
\widetilde{\Lbf}_{22}^{+(1)}=
\widetilde{\Lbf}_{33}^{-(2)}=
\begin{array}{|ccc|}
\fbox{$\Jbf_{11}$} & \Jbf_{12} & \Jbf_{13} \\
\Jbf_{21} & \Jbf_{22} & \Jbf_{23} \\
\Jbf_{31} & \Jbf_{32} & \Jbf_{33} 
\end{array}
^{\,  -1}
\ubf^{(1)}_{1} \ubf^{(2)}_{1}
\begin{array}{|cc|}
 \fbox{$\Jbf_{22}$} & \Jbf_{23} \\
 \Jbf_{32} & \Jbf_{33} 
\end{array}
^{\,  -1}
\ubf^{(1)}_{2} \ubf^{(2)}_{2},
\nonumber 
\\[6pt]
\widetilde{\ubf}_{3}^{(1)}&=
\widetilde{\Lbf}_{33}^{+(1)}=
\begin{array}{|ccc|}
\fbox{$\Jbf_{11}$} & \Jbf_{12} & \Jbf_{13} \\
\Jbf_{21} & \Jbf_{22} & \Jbf_{23} \\
\Jbf_{31} & \Jbf_{32} & \Jbf_{33} 
\end{array}
^{\,  -1}
\ubf^{(1)}_{1} \ubf^{(2)}_{1}
\begin{array}{|cc|}
 \fbox{$\Jbf_{22}$} & \Jbf_{23} \\
 \Jbf_{32} & \Jbf_{33} 
\end{array}
^{\,  -1}
\ubf^{(1)}_{2} \ubf^{(2)}_{2}
\Jbf_{33}^{-1}
\ubf^{(1)}_{3} \ubf^{(2)}_{3},
\nonumber 
\\[6pt]
\widetilde{\Lbf}_{12}^{+(1)}&=
\begin{array}{|ccc|}
\fbox{$\Jbf_{11}$} & \Jbf_{12} & \Jbf_{13} \\
\Jbf_{21} & \Jbf_{22} & \Jbf_{23} \\
\Jbf_{31} & \Jbf_{32} & \Jbf_{33} 
\end{array}
^{\,  -1} \,
\begin{array}{|ccc|}
\fbox{$({\Lbf}^{+(1)}{\Lbf}^{+(2)})_{12}$} & \Jbf_{12} & \Jbf_{13} \\
({\Lbf}^{+(1)}{\Lbf}^{+(2)})_{22} & \Jbf_{22} & \Jbf_{23} \\
({\Lbf}^{+(1)}{\Lbf}^{+(2)})_{32} & \Jbf_{32} & \Jbf_{33} 
\end{array}
, \nonumber 
\\[6pt]
\widetilde{\Lbf}_{13}^{+(1)}&=
\begin{array}{|ccc|}
\fbox{$\Jbf_{11}$} & \Jbf_{12} & \Jbf_{13} \\
\Jbf_{21} & \Jbf_{22} & \Jbf_{23} \\
\Jbf_{31} & \Jbf_{32} & \Jbf_{33} 
\end{array}
^{\,  -1} \,
\begin{array}{|ccc|}
\fbox{$({\Lbf}^{+(1)}{\Lbf}^{+(2)})_{13}$} & \Jbf_{12} & \Jbf_{13} \\
({\Lbf}^{+(1)}{\Lbf}^{+(2)})_{23} & \Jbf_{22} & \Jbf_{23} \\
({\Lbf}^{+(1)}{\Lbf}^{+(2)})_{33} & \Jbf_{32} & \Jbf_{33} 
\end{array}
,
\nonumber 
\\[6pt]
\widetilde{\Lbf}_{23}^{+(1)}&=
\begin{array}{|ccc|}
\fbox{$\Jbf_{11}$} & \Jbf_{12} & \Jbf_{13} \\
\Jbf_{21} & \Jbf_{22} & \Jbf_{23} \\
\Jbf_{31} & \Jbf_{32} & \Jbf_{33} 
\end{array}
^{\,  -1} 
\ubf^{(1)}_{1} \ubf^{(2)}_{1}
\begin{array}{|cc|}
 \fbox{$\Jbf_{22}$} & \Jbf_{23} \\
 \Jbf_{32} & \Jbf_{33} 
\end{array}
^{\,  -1} \,
\begin{array}{|cc|}
\fbox{$({\Lbf}^{+(1)}{\Lbf}^{+(2)})_{23}$}  & \Jbf_{23} \\
({\Lbf}^{+(1)}{\Lbf}^{+(2)})_{33}  & \Jbf_{33} 
\end{array}
, \nonumber 
\\[6pt]
\widetilde{\Lbf}_{21}^{-(1)}&=
\begin{array}{|ccc|}
\fbox{$\Jbf_{11}$} & \Jbf_{12} & \Jbf_{13} \\
\Jbf_{21} & \Jbf_{22} & \Jbf_{23} \\
\Jbf_{31} & \Jbf_{32} & \Jbf_{33} 
\end{array}
^{\,  -1} 
\ubf^{(1)}_{1} \ubf^{(2)}_{1}
\begin{array}{|cc|}
 \fbox{$\Jbf_{22}$} & \Jbf_{23} \\
 \Jbf_{32} & \Jbf_{33} 
\end{array}
^{\,  -1} \,
\begin{array}{|cc|}
 \fbox{$\Jbf_{21}$} & \Jbf_{23} \\
 \Jbf_{31} & \Jbf_{33} 
\end{array}
, \nonumber 
\\[6pt]
\widetilde{\Lbf}_{31}^{-(1)}&=
\begin{array}{|ccc|}
\fbox{$\Jbf_{11}$} & \Jbf_{12} & \Jbf_{13} \\
\Jbf_{21} & \Jbf_{22} & \Jbf_{23} \\
\Jbf_{31} & \Jbf_{32} & \Jbf_{33} 
\end{array}
^{\,  -1} 
\ubf^{(1)}_{1} \ubf^{(2)}_{1}
\begin{array}{|cc|}
 \fbox{$\Jbf_{22}$} & \Jbf_{23} \\
 \Jbf_{32} & \Jbf_{33} 
\end{array}
^{\,  -1} 
\ubf^{(1)}_{2} \ubf^{(2)}_{2}
 \Jbf^{-1}_{33}
  \Jbf_{31}
, \nonumber 
\\[6pt]
\widetilde{\Lbf}_{32}^{-(1)}&=
\begin{array}{|ccc|}
\fbox{$\Jbf_{11}$} & \Jbf_{12} & \Jbf_{13} \\
\Jbf_{21} & \Jbf_{22} & \Jbf_{23} \\
\Jbf_{31} & \Jbf_{32} & \Jbf_{33} 
\end{array}
^{\,  -1} 
\ubf^{(1)}_{1} \ubf^{(2)}_{1}
\begin{array}{|cc|}
 \fbox{$\Jbf_{22}$} & \Jbf_{23} \\
 \Jbf_{32} & \Jbf_{33} 
\end{array}
^{\,  -1} 
\ubf^{(1)}_{2} \ubf^{(2)}_{2}
 \Jbf^{-1}_{33}
  \Jbf_{32}
, \nonumber 
\\[6pt]
\widetilde{\ubf}_{0}^{(2)}&=\widetilde{\Lbf}_{11}^{-(2)}=1, 
\nonumber 
\\[6pt]
\widetilde{\ubf}_{1}^{(2)}&=\widetilde{\Lbf}_{11}^{+(2)}=
\widetilde{\Lbf}_{22}^{-(2)}=
\begin{array}{|ccc|}
\fbox{$\Jbf_{11}$} & \Jbf_{12} & \Jbf_{13} \\
\Jbf_{21} & \Jbf_{22} & \Jbf_{23} \\
\Jbf_{31} & \Jbf_{32} & \Jbf_{33} 
\end{array}
, \nonumber 
\\[6pt]
\widetilde{\ubf}_{2}^{(2)}&=\widetilde{\Lbf}_{22}^{+(2)}=
\widetilde{\Lbf}_{33}^{-(2)}=
\begin{array}{|cc|}
 \fbox{$\Jbf_{22}$} & \Jbf_{23} \\
 \Jbf_{32} & \Jbf_{33} 
\end{array}
(\ubf^{(1)}_{1} \ubf^{(2)}_{1})^{\,  -1} 
\begin{array}{|ccc|}
\fbox{$\Jbf_{11}$} & \Jbf_{12} & \Jbf_{13} \\
\Jbf_{21} & \Jbf_{22} & \Jbf_{23} \\
\Jbf_{31} & \Jbf_{32} & \Jbf_{33} 
\end{array}
, \nonumber 
\\[6pt]
\widetilde{\ubf}_{3}^{(2)}&=\widetilde{\Lbf}_{33}^{+(2)}=
\Jbf_{33}
(\ubf^{(1)}_{2} \ubf^{(2)}_{2})^{\,  -1} 
\begin{array}{|cc|}
 \fbox{$\Jbf_{22}$} & \Jbf_{23} \\
 \Jbf_{32} & \Jbf_{33} 
\end{array}
(\ubf^{(1)}_{1} \ubf^{(2)}_{1})^{\,  -1} 
\begin{array}{|ccc|}
\fbox{$\Jbf_{11}$} & \Jbf_{12} & \Jbf_{13} \\
\Jbf_{21} & \Jbf_{22} & \Jbf_{23} \\
\Jbf_{31} & \Jbf_{32} & \Jbf_{33} 
\end{array}
, \nonumber 
\\[6pt]
\widetilde{\Lbf}_{12}^{+(2)}&=
\begin{array}{|cc|}
 \fbox{$\Jbf_{12}$} & \Jbf_{13} \\
 \Jbf_{32} & \Jbf_{33} 
\end{array}
(\ubf^{(1)}_{1} \ubf^{(2)}_{1})^{\,  -1} 
\begin{array}{|ccc|}
\fbox{$\Jbf_{11}$} & \Jbf_{12} & \Jbf_{13} \\
\Jbf_{21} & \Jbf_{22} & \Jbf_{23} \\
\Jbf_{31} & \Jbf_{32} & \Jbf_{33} 
\end{array}
, \nonumber 
\\[6pt]
\widetilde{\Lbf}_{13}^{+(2)}&=
\Jbf_{13}
(\ubf^{(1)}_{2} \ubf^{(2)}_{2})^{\,  -1} 
\begin{array}{|cc|}
 \fbox{$\Jbf_{22}$} & \Jbf_{23} \\
 \Jbf_{32} & \Jbf_{33} 
\end{array}
(\ubf^{(1)}_{1} \ubf^{(2)}_{1})^{\,  -1} 
\begin{array}{|ccc|}
\fbox{$\Jbf_{11}$} & \Jbf_{12} & \Jbf_{13} \\
\Jbf_{21} & \Jbf_{22} & \Jbf_{23} \\
\Jbf_{31} & \Jbf_{32} & \Jbf_{33} 
\end{array}
, \nonumber 
\\[6pt]
\widetilde{\Lbf}_{23}^{+(2)}&=
\Jbf_{23}
(\ubf^{(1)}_{2} \ubf^{(2)}_{2})^{\,  -1} 
\begin{array}{|cc|}
 \fbox{$\Jbf_{22}$} & \Jbf_{23} \\
 \Jbf_{32} & \Jbf_{33} 
\end{array}
(\ubf^{(1)}_{1} \ubf^{(2)}_{1})^{\,  -1} 
\begin{array}{|ccc|}
\fbox{$\Jbf_{11}$} & \Jbf_{12} & \Jbf_{13} \\
\Jbf_{21} & \Jbf_{22} & \Jbf_{23} \\
\Jbf_{31} & \Jbf_{32} & \Jbf_{33} 
\end{array}
, \nonumber 
\\[6pt]
\widetilde{\Lbf}_{21}^{-(2)}&=
\begin{array}{|ccc|}
\fbox{$({\Lbf}^{-(1)}{\Lbf}^{-(2)})_{21}$}  & ({\Lbf}^{-(1)}{\Lbf}^{-(2)})_{22} 
& ({\Lbf}^{-(1)}{\Lbf}^{-(2)})_{23}   \\
\Jbf_{21} & \Jbf_{22} & \Jbf_{23} \\
\Jbf_{31} & \Jbf_{32} & \Jbf_{33} 
\end{array}
, \nonumber 
\\[6pt]
\widetilde{\Lbf}_{31}^{-(2)}&=
\begin{array}{|ccc|}
\fbox{$({\Lbf}^{-(1)}{\Lbf}^{-(2)})_{31}$}  & ({\Lbf}^{-(1)}{\Lbf}^{-(2)})_{32} 
& ({\Lbf}^{-(1)}{\Lbf}^{-(2)})_{33}   \\
\Jbf_{21} & \Jbf_{22} & \Jbf_{23} \\
\Jbf_{31} & \Jbf_{32} & \Jbf_{33} 
\end{array}
, \nonumber 
\\[6pt]
\widetilde{\Lbf}_{32}^{-(2)}&=
\begin{array}{|cc|}
\fbox{$({\Lbf}^{-(1)}{\Lbf}^{-(2)})_{32}$}  & ({\Lbf}^{-(1)}{\Lbf}^{-(2)})_{33} \\
\Jbf_{32}  & \Jbf_{33} 
\end{array}
(\ubf^{(1)}_{1} \ubf^{(2)}_{1})^{\,  -1}  
\begin{array}{|ccc|}
\fbox{$\Jbf_{11}$} & \Jbf_{12} & \Jbf_{13} \\
\Jbf_{21} & \Jbf_{22} & \Jbf_{23} \\
\Jbf_{31} & \Jbf_{32} & \Jbf_{33} 
\end{array}
,
\end{align}
where $({\Lbf}^{+(1)}{\Lbf}^{+(2)})_{33}=\ubf^{(1)}_{3}\ubf^{(2)}_{3},$ 
$({\Lbf}^{+(1)}{\Lbf}^{+(2)})_{32}=({\Lbf}^{-(1)}{\Lbf}^{-(2)})_{23}=0, $ 
$({\Lbf}^{-(1)}{\Lbf}^{-(2)})_{22}=\ubf^{(1)}_{1}\ubf^{(2)}_{1}, $ 
$({\Lbf}^{+(1)}{\Lbf}^{+(2)})_{22}=({\Lbf}^{-(1)}{\Lbf}^{-(2)})_{33}
=\ubf^{(1)}_{2}\ubf^{(2)}_{2}$.
Some of the above expressions simplify when one considers $U_{q}(sl(3))$ case. 
In particular, the quasi-determinant over a $3 \times 3$-matrix can be eliminated based on 
the formula
\begin{align}
\begin{array}{|ccc|}
\fbox{$\Jbf_{11}$} & \Jbf_{12} & \Jbf_{13} \\
\Jbf_{21} & \Jbf_{22} & \Jbf_{23} \\
\Jbf_{31} & \Jbf_{32} & \Jbf_{33} 
\end{array}
=
\ubf^{(1)}_{1} \ubf^{(2)}_{1}
\begin{array}{|cc|}
 \fbox{$\Jbf_{22}$} & \Jbf_{23} \\
 \Jbf_{32} & \Jbf_{33} 
\end{array}
^{\,  -1}
\ubf^{(1)}_{2} \ubf^{(2)}_{2}
\Jbf_{33}^{-1},
\end{align}
which is derived from the condition
$u^{(1)}_{3}=u^{(2)}_{3}=\widetilde{u}^{(1)}_{3}=\widetilde{u}^{(2)}_{3}=1$. 

{\em Proof of Theorem \ref{sol-zero1}:} 
Consider the Gauss decomposition of this matrix 
$\Jbf={\mathbb E}{\mathbb H}{\mathbb F}$, where 
${\mathbb E}=({\mathbb E}_{ij})_{1\le i,j \le n}$, 
${\mathbb H}=(\delta_{ij}{\mathbb H}_{i})_{1\le i,j \le n}$,
${\mathbb F}=({\mathbb F}_{ij})_{1\le i,j \le n}$: 
${\mathbb E}_{kk}={\mathbb F}_{kk}=1$ for $1 \le k \le n$; 
${\mathbb E}_{ji}={\mathbb F}_{ij}=0$ for 
$1 \le i < j \le n$, and 
 apply Theorem \ref{Gauss1} 
 to the matrix $\Jbf$
($A$ and $N$ in Theorem \ref{Gauss1} correspond to $\Jbf$ 
and $n$, respectively). Comparing this 
 with the right hand side 
$\widetilde{\Lbf}^{+(2)} \widetilde{\Lbf}^{-(1)}$ 
of the second equation in \eqref{ZC-an}, 
we obtain the following equations
\begin{align}
\begin{split}
{\mathbb E}_{ij}&=\widetilde{\Lbf}^{+(2)}_{ij}(\widetilde{\Lbf}^{+(2)}_{jj})^{-1} , \qquad 
{\mathbb F}_{ji}= (\widetilde{\Lbf}^{-(1)}_{jj})^{-1}
\widetilde{\Lbf}^{-(1)}_{ji} 
\qquad \text{for} \quad 1 \le i<j \le n, 
\\[5pt]
{\mathbb H}_{i}&=
 \widetilde{\Lbf}^{+(2)}_{ii}\widetilde{\Lbf}^{-(1)}_{ii} 
\qquad \text{for} \quad 1 \le i \le n,
\end{split}
\label{gau1}
\end{align}
where the left hand side of these are written in terms of 
quasi-determinants
\footnote{Here we chose $s=j$ in \eqref{left-qP} and $t=j$ in \eqref{right-qP}.}
\begin{align}
{\mathbb H}_{i}&=|\Jbf^{i,\dots,n}_{i,\dots,n}|_{ii}
=
\Jbf_{ii} - 
\sum_{a,b=i+1}^{n}\Jbf_{ia}
((\Jbf^{i+1,\dots, n}_{i+1,\dots, n})^{-1})_{ab}
\Jbf_{bi} 
\quad \text{for} \quad 1 \le i \le n,
 \label{gauss-h1}
\\[6pt]
{\mathbb E}_{ij}&= 
 |\Jbf^{i,j+1,\dots,n}_{j,j+1,\dots,n}|_{ij}
 |\Jbf^{j,\dots,n}_{j,\dots,n}|_{jj}^{-1}=
\left(\Jbf_{ij} - 
\sum_{a,b=j+1}^{n}\Jbf_{ia}
((\Jbf^{j+1,\dots,n}_{j+1,\dots,n})^{-1})_{ab}
\Jbf_{bj}
\right) {\mathbb H}_{j}^{-1} , 
\label{gauss-e1}
\\[6pt]
{\mathbb F}_{ji}&=
|\Jbf^{j,\dots,n}_{j,\dots,n}|_{jj}^{-1}
 |\Jbf^{j,j+1,\dots,n}_{i,j+1,\dots,n}|_{ji}
\nonumber
\\[5pt]
 &=
{\mathbb H}_{j}^{-1}
\left(\Jbf_{ji} - 
\sum_{a,b=j+1}^{n}\Jbf_{ja}
((\Jbf^{j+1,\dots,n}_{j+1,\dots,n})^{-1})_{ab}
\Jbf_{bi}
\right)  
\quad \text{for} \quad 1 \le i<j \le n. 
\label{gauss-f1} 
\end{align}
%
The diagonal elements of the first and the third equations 
in \eqref{ZC-an} correspond to 
\begin{align}
\Lbf^{+(1)}_{ii} \Lbf^{+(2)}_{ii}
&= \widetilde{\Lbf}^{+(2)}_{ii}\widetilde{\Lbf}^{+(1)}_{ii}, 
\qquad 
\Lbf^{-(1)}_{ii} \Lbf^{-(2)}_{ii} =
 \widetilde{\Lbf}^{-(2)}_{ii}\widetilde{\Lbf}^{-(1)}_{ii} 
\qquad \text{for} \quad 1 \le i \le n. 
\label{gau2}
\end{align}
Taking into account the relations 
$\Lbf^{+(a)}_{ii}=\ubf_{i}^{(a)}$, 
$\widetilde{\Lbf}^{+(a)}_{ii}=\widetilde{\ubf}_{i}^{(a)}$, 
$\Lbf^{-(a)}_{ii}=\ubf_{i-1}^{(a)}$ and 
$\widetilde{\Lbf}^{-(a)}_{ii}=\widetilde{\ubf}_{i-1}^{(a)}$, 
we rewrite \eqref{gau1}-\eqref{gau2} as 
\begin{align}
& {\mathbb E}_{ij} =\widetilde{\Lbf}^{+(2)}_{ij} 
(\widetilde{\ubf}_{j}^{(2)})^{-1}, 
\qquad 
{\mathbb F}_{ji}= ( \widetilde{\ubf}_{j-1}^{(1)})^{-1}
\widetilde{\Lbf}^{-(1)}_{ji} 
\qquad \text{for} \quad 1 \le i<j \le n, 
\label{gau3}
\\[5pt]
& {\mathbb H}_{i}=
\widetilde{\ubf}_{i}^{(2)} \widetilde{\ubf}_{i-1}^{(1)} 
\qquad \text{for} \quad 1 \le i \le n, 
\label{gau4}
\\[5pt]
&\ubf_{i}^{(1)} \ubf_{i}^{(2)} =
\widetilde{\ubf}_{i}^{(2)}\widetilde{\ubf}_{i}^{(1)}
\qquad \text{for} \quad 1 \le i \le n.
\label{gau5}
\end{align}
Solving these equations \eqref{gau4} and \eqref{gau5}, we get 
\begin{align}
\widetilde{\ubf}_{i}^{(1)}&=
\overrightarrow{\prod}_{k=1}^{i}
{\mathbb H}_{k}^{-1}  \ubf_{k}^{(1)}\ubf_{k}^{(2)}
 , 
\label{zca1}
\\
\widetilde{\ubf}_{i}^{(2)}&=
 {\mathbb H}_{i}
\overleftarrow{\prod}_{k=1}^{i-1}
 (\ubf_{k}^{(1)}\ubf_{k}^{(2)})^{-1}
{\mathbb H}_{k} \qquad \text{for} \quad 1 \le i \le n.
\label{zca2}
\end{align}
Substituting \eqref{zca1} and \eqref{zca2} into \eqref{gau3}, we obtain 
\begin{align}
\widetilde{\Lbf}_{ij}^{-(1)}&=
\left(
\overrightarrow{\prod}_{k=1}^{i-1} 
{\mathbb H}_{k}^{-1}
\ubf_{k}^{(1)}\ubf_{k}^{(2)}
\right)
\mathbb{F}_{ij} \qquad \text{for} \quad 1 \le j<i \le n,
\label{zca3}
\\
\widetilde{\Lbf}_{ij}^{+(2)}&=
\mathbb{E}_{ij}
 {\mathbb H}_{j} 
\overleftarrow{\prod}_{k=1}^{j-1}
 (\ubf_{k}^{(1)}\ubf_{k}^{(2)})^{-1}
{\mathbb H}_{k} 
\quad \text{for} \qquad 1 \le i<j \le n. 
\label{zca4}
\end{align}
The other relations can be obtained by substituting these into 
the relations follow from the first and the third relations 
in \eqref{ZC-an}, 
\begin{align}
\widetilde{\Lbf}^{-(2)}&=
\Lbf^{-(1)}\Lbf^{-(2)} (\widetilde{\Lbf}^{-(1)})^{-1},
\label{zca5a}
\\[6pt]
\widetilde{\Lbf}^{+(1)}&=
(\widetilde{\Lbf}^{+(2)})^{-1}\Lbf^{+(1)}\Lbf^{+(2)}. 
 \label{zca5b}
\end{align}
In order to calculate these more explicitly, we need the inverse matrices of 
$\mathbb{E}$, $\mathbb{H}$, and $\mathbb{F}$. 
Applying Theorem \ref{Gauss2} to 
$\Jbf^{-1}=\mathbb{F}^{-1}\mathbb{H}^{-1}\mathbb{E}^{-1}$, 
we obtain
\footnote{$(\mathbb{E}^{-1})_{ij}$ is the $i$-th row and $j$-th column of 
the inverse matrix $\mathbb{E}^{-1}$. 
Here we set $s=i$ for \eqref{left-qP} and $t=i$ for \eqref{right-qP}.}
\begin{align}
(\mathbb{E}^{-1})_{ij}&=
 \left|(\Jbf ^{-1})^{1,2,\dots , i}_{1,2,\dots, i}\right|^{-1}_{ii}
 \, 
 \left|(\Jbf ^{-1})^{1,2,\dots , i-1,i}_{1,2,\dots, i-1,j}\right|_{ij}
 ,
 \label{invEin}
\\[6pt]
(\mathbb{F}^{-1})_{ji}&=\left|(\Jbf ^{-1})^{1,2,\dots , i-1,j}_{1,2,\dots, i-1,i}\right|_{ji}
\, 
 \left|(\Jbf ^{-1})^{1,2,\dots , i}_{1,2,\dots, i}\right|^{-1}_{ii} 
  \quad \text{for} \quad 1 \le i \le j \le n.
 \label{invFin}
\end{align}
Then we use the relations 
\begin{align}
\begin{split}
& \left| \Jbf^{i,i+1,\dots , n}_{i,i+1,\dots, n}\right|_{ii} 
 \,
  \left|(\Jbf ^{-1})^{1,2,\dots , i}_{1,2,\dots, i}\right|_{ii} 
 =1, 
 \quad 
  \left| \Jbf^{i,i+1,\dots , n}_{i,i+1,\dots, n}\right|_{ji} 
 \,
  \left|(\Jbf ^{-1})^{1,2,\dots , i-1,i}_{1,2,\dots, i-1,j}\right|_{ij} 
 =1,
\\[6pt]
& 
  \left| \Jbf^{i,i+1,\dots , n}_{i,i+1,\dots, n}\right|_{ij} 
 \,
  \left|(\Jbf ^{-1})^{1,2,\dots , i-1,j}_{1,2,\dots, i-1,i}\right|_{ji} 
 =1 
    \quad \text{for} \quad 1 \le i \le j \le n.
\end{split}
\end{align}
which follow from the formula \eqref{inversion}, to get 
\begin{align}
(\mathbb{E}^{-1})_{ij}&=
 \left| \Jbf^{i,i+1,\dots , n}_{i,i+1,\dots, n}\right|_{ii} 
 \, 
\left| \Jbf^{i,i+1,\dots , n}_{i,i+1,\dots, n}\right|^{-1}_{ji} 
 ,
 \label{Einv}
\\[6pt]
(\mathbb{F}^{-1})_{ji}&=  \left| \Jbf^{i,i+1,\dots , n}_{i,i+1,\dots, n}\right|^{-1}_{ij} 
\, 
\left| \Jbf^{i,i+1,\dots , n}_{i,i+1,\dots, n}\right|_{ii}
  \quad \text{for} \quad 1 \le i \le j \le n.
 \label{Finv}
\end{align}
Substituting \eqref{zca3} (resp.\ \eqref{zca4}) into  \eqref{zca5a} (resp.\ \eqref{zca5b}),  
and using 
 \eqref{Finv}, \eqref{Lap-row}, \eqref{row-hom} (resp.\ 
  \eqref{Einv}, \eqref{Lap-col}, \eqref{col-hom}), we obtain 
\begin{align}
\widetilde{\Lbf}_{ji}^{-(2)}&=
 \left\{
 \sum_{k=i}^{j} (\Lbf^{-(1)}\Lbf^{-(2)})_{jk}
  |\Jbf^{i,i+1,\dots , n}_{i,i+1,\dots, n}|^{-1}_{ik}
  |\Jbf^{i,i+1,\dots , n}_{i,i+1,\dots, n}|_{ii}
    \right\}
    \overleftarrow{\prod}_{k=1}^{i-1} 
(\ubf_{k}^{(1)}\ubf_{k}^{(2)})^{-1}
{\mathbb H}_{k}
 \nonumber \\
 & \hspace{240pt} 
 \text{[by \eqref{zca5a}, \eqref{zca3}, \eqref{Finv}]}
\nonumber 
\\[6pt]
&=
 \left\{
(\Lbf^{-(1)}\Lbf^{-(2)})_{ji}-
 \sum_{k=i+1}^{j} (\Lbf^{-(1)}\Lbf^{-(2)})_{jk}
  |\Jbf^{i+1,i+2,\dots , n}_{i+1,i+2,\dots, n}|^{-1}_{sk}
    |\Jbf^{i+1,i+2,\dots , n}_{i,i+1,\dots,\hat{k},\dots , n}|_{si}
    \right\}
\nonumber \\
& \hspace{150pt} \times 
    \overleftarrow{\prod}_{k=1}^{i-1} 
(\ubf_{k}^{(1)}\ubf_{k}^{(2)})^{-1}
{\mathbb H}_{k}
\qquad 
 \text{[by \eqref{row-hom}, $s \ne i$]}
\nonumber \\[6pt]
&=
\begin{array}{|c|}
(\Lbf^{-(1)}\Lbf^{-(2)})^{\Bar{j}}_{i,i+1,\dots,n}
\\
 \Jbf^{i+1,i+2,\dots , n}_{i,i+1,\dots, n} 
\end{array}
_{\, \Bar{j}i} 
\, 
\overleftarrow{\prod}_{k=1}^{i-1} 
(\ubf_{k}^{(1)}\ubf_{k}^{(2)})^{-1}
{\mathbb H}_{k}
\qquad \text{[by \eqref{Lap-row}]} ,
 \label{Lm2}
\\[6pt]
\widetilde{\Lbf}_{ij}^{+(1)}
&=
\left(
\overrightarrow{\prod}_{k=1}^{i-1} 
{\mathbb H}_{k}^{-1}
\ubf_{k}^{(1)}\ubf_{k}^{(2)}
\right)
{\mathbb H}_{i}^{-1}
\sum_{k=i}^{j}
  |\Jbf^{i,i+1,\dots , n}_{i,i+1,\dots, n}|_{ii}
  |\Jbf^{i,i+1,\dots , n}_{i,i+1,\dots, n}|^{-1}_{ki}
  (\Lbf^{+(1)}\Lbf^{+(2)})_{kj}
   \nonumber \\
 & \hspace{240pt} 
 \text{[by \eqref{zca5b}, \eqref{zca4}, \eqref{Einv}]}
\nonumber 
\\[6pt]
&=
\left(
\overrightarrow{\prod}_{k=1}^{i-1} 
{\mathbb H}_{k}^{-1}
\ubf_{k}^{(1)}\ubf_{k}^{(2)}
\right)
{\mathbb H}_{i}^{-1}
\nonumber 
\\
& \quad \times 
\left\{
(\Lbf^{+(1)}\Lbf^{+(2)})_{ij}-
 \sum_{k=i+1}^{j} 
  |\Jbf^{i,i+1,\dots ,\hat{k},\dots, n}_{i+1,i+2,\dots, n}|_{it}
    |\Jbf^{i+1,i+2,\dots , n}_{i+1,i+2,\dots , n}|^{-1}_{kt}
    (\Lbf^{+(1)}\Lbf^{+(2)})_{kj}
    \right\}
    \nonumber 
\\ 
 &   \hspace{240pt} \text{[by \eqref{col-hom}, $t \ne i$]} 
\nonumber 
\\[6pt]
&=
\left(
\overrightarrow{\prod}_{k=1}^{i-1} 
{\mathbb H}_{k}^{-1}
\ubf_{k}^{(1)}\ubf_{k}^{(2)}
\right)
{\mathbb H}_{i}^{-1}
\begin{array}{|cc|}
(\Lbf^{+(1)}\Lbf^{+(2)})^{i,i+1,\dots,n}_{\Bar{j}}
&
 \Jbf^{i,i+1,\dots , n}_{i+1,i+2,\dots, n} 
\end{array}
_{\, i\Bar{j}} 
\qquad \text{[by \eqref{Lap-col}]} ,
\nonumber 
\\[6pt]
& \hspace{120pt} \text{for} \quad 1 \le i<j \le n,
 \label{Lp1}
\end{align}
where ``$ i,i+1,\dots, \hat{k}, \dots, n$'' means that the number $k$ is removed from 
the sequence of numbers, ``$ i,i+1, \dots, n$''. 
The fact that $(\Lbf^{-(1)}\Lbf^{-(2)})_{jk}=  (\Lbf^{+(1)}\Lbf^{+(2)})_{kj}=0$ 
for $j<k$ is also used. 
Summarizing 
\eqref{zca1}-\eqref{zca4}, \eqref{Lm2}, \eqref{Lp1} and \eqref{gauss-h1}, 
we arrive at Theorem \ref{sol-zero1}.
$ \square $

We did not explicitly assume the defining relations of $U_{q}(gl(n))$ for 
the matrix elements of the 
L-operators 
to derive \eqref{Lp1-th}-\eqref{Lm2-th}. 
Let us assume the relation \eqref{L-Chv} and 
 the defining relations \eqref{FRT} of $U_{q}(gl(n))$. 
According to \eqref{gau4}, for any $k$, ${\mathbb H}_{k} $  
 is a Cartan element with respect to the generators in 
 $\widetilde{\Xbf}^{(1)}$ and  $\widetilde{\Xbf}^{(2)}$, 
and thus commutes with $\widetilde{\ubf}_{i}^{(a)}$ for any 
$i$ and $a$. 
Then ${\mathbb H}_{k} $ 
commutes
\footnote{This does not mean that ${\mathbb H}_{k} $ commutes 
with $\ubf_{i}^{(a)}$. Namely, ${\mathbb H}_{k} $ is not necessary 
a Cartan element with respect to the  
generators in $\Xbf^{(1)}$ and  $\Xbf^{(2)}$.}
 with 
$\ubf_{i}^{(1)}\ubf_{i}^{(2)}$ for any $i$ 
due to the relation \eqref{gau5}. 
In this sense, we need not mind the ordering of the product 
in \eqref{Lp1-th}-\eqref{Lm2-th}. 
By construction, \eqref{Lp1-th}-\eqref{Lm2-th} give an automorphism of $\A \otimes \A $. 

For $n=2$ case, \eqref{Lp1-th}-\eqref{Lm2-th}  
reduces 
 to eq. (2.17) in \cite{BS15} 
(thus to the map in \cite{Kashaev:2004})
if we assume the relation \eqref{L-Chv}, 
the defining relations \eqref{FRT} of $U_{q}(gl(2))$,  
 the condition $\omega_{2}=0$ 
for $U_{q}(sl(2))$, and 
identify 
$ q^{\Ebf_{11}} \Ebf_{12}=E, \Ebf_{21}q^{-\Ebf_{11}} =F, 
q^{2\Ebf_{11}}=K$, where $E,F$ and $K$ are 
 the generators of $U_{q}(sl(2))$ in the notation of \cite{BS15}. 

Next we want to obtain the inverse transformation. 
Let us denote the right hand side of 
the second equation in \eqref{ZC-an} as 
$\widetilde{\Jbf}=
(\widetilde{\Jbf}_{a,b})_{1 \le a,b \le n}:=\widetilde{\Lbf}^{+(2)} \widetilde{\Lbf}^{-(1)} $, 
where 
\begin{align}
\widetilde{\Jbf}_{a,b}=
\sum_{k=\max\{a,b\}}^{n} \widetilde{\Lbf}^{+(2)}_{ak} \widetilde{\Lbf}^{-(1)}_{kb}.
\end{align} 
\begin{theorem}
\label{sol-zero1-o}
For $1 \le i \le j \le n$, 
\begin{align}
\Lbf_{ij}^{+(1)}&=
\begin{array}{|c|}
\widetilde{\Jbf}^{1,2,\dots , j-1}_{1,2,\dots, j} 
\\[3pt]
(\widetilde{\Lbf}^{+(2)}\widetilde{\Lbf}^{+(1)})^{\Bar{i}}_{1,2,\dots,j}
\end{array}
_{\, \Bar{i}j} 
\, 
 |\widetilde{\Jbf}^{1,\dots,j}_{1,\dots,j}|^{-1}_{jj}
    \overleftarrow{\prod}_{k=1}^{j-1} 
\widetilde{\ubf}_{k}^{(2)}\widetilde{\ubf}_{k}^{(1)}
   |\widetilde{\Jbf}^{1,\dots,k}_{1,\dots,k}|^{-1}_{kk} ,
 \label{Lp1-o-th}
\\[6pt]
\Lbf_{ji}^{-(1)}&=
 |\widetilde{\Jbf}^{1,\dots,i-1,j}_{1,\dots,i-1,i}|_{ji}
|\widetilde{\Jbf}^{1,\dots,i}_{1,\dots,i}|_{ii}^{-1}
\overleftarrow{\prod}_{k=1}^{i-1}
 \widetilde{\ubf}_{k}^{(2)}\widetilde{\ubf}_{k}^{(1)}
 |\widetilde{\Jbf}^{1,\dots,k}_{1,\dots,k}|^{-1}_{kk}
,
 \label{Lm1-o-th}
 \\[6pt]
\Lbf_{ij}^{+(2)}&=
\left(
\overrightarrow{\prod}_{k=1}^{i-1}
 |\widetilde{\Jbf}^{1,\dots,k}_{1,\dots,k}|_{kk}
 (\widetilde{\ubf}_{k}^{(2)}\widetilde{\ubf}_{k}^{(1)})^{-1}
\right)
 |\widetilde{\Jbf}^{1,\dots,i-1,i}_{1,\dots,i-1,j}|_{ij}
,
\label{Lp2-o-th}
\\[6pt]
\Lbf_{ji}^{-(2)}
&=
\left(
\overrightarrow{\prod}_{k=1}^{j-1} 
 |\widetilde{\Jbf}^{1,\dots,k}_{1,\dots,k}|_{kk}
(\widetilde{\ubf}_{k}^{(2)}\widetilde{\ubf}_{k}^{(1)})^{-1}
\right)
\begin{array}{|cc|}
 \widetilde{\Jbf}^{1,2,\dots , j}_{1,2,\dots, j-1} 
 &
(\widetilde{\Lbf}^{-(2)}\widetilde{\Lbf}^{-(1)})^{1,2,\dots,j}_{\Bar{i}}
\end{array}
_{\, j \Bar{i}} 
 \label{Lm2-o-th}
\end{align}
give the inverse map for \eqref{Lp1-th}-\eqref{Lm2-th} 
and solve 
 the zero-curvature relation \eqref{ZC-an}. 
\end{theorem}
{\em Proof}: 
Consider the Gauss decomposition of the matrix 
$\widetilde{\Jbf}=\widetilde{\mathbb F} \widetilde{\mathbb H} \widetilde{\mathbb E}$, where 
$\widetilde{\mathbb E}=(\widetilde{\mathbb E}_{ij})_{1\le i,j \le n}$, 
$\widetilde{\mathbb H}=(\delta_{ij} \widetilde{\mathbb H}_{i})_{1\le i,j \le n}$,
$\widetilde{\mathbb F}=(\widetilde{\mathbb F}_{ij})_{1\le i,j \le n}$: 
$\widetilde{\mathbb E}_{kk}=\widetilde{\mathbb F}_{kk}=1$ for $1 \le k \le n$; 
$\widetilde{\mathbb E}_{ji}=\widetilde{\mathbb F}_{ij}=0$ for 
$1 \le i < j \le n$, and 
 apply Theorem \ref{Gauss2} 
 to the matrix $\widetilde{\Jbf}$
($A$ and $N$ in Theorem \ref{Gauss2} correspond to $\widetilde{\Jbf}$ 
and $n$, respectively). 
Repeating a procedure similar to the one for Theorem \ref{sol-zero1}, we obtain 
the following equations 
\begin{align}
&(\ubf_{i}^{(2)})^{-1} \Lbf^{+(2)}_{ij} 
 = \widetilde{\mathbb E}_{ij} , 
\qquad 
 \Lbf^{-(1)}_{ji} (\ubf_{i-1}^{(1)})^{-1} =
 \widetilde{\mathbb F}_{ji}
\qquad \text{for} \quad 1 \le i<j \le n, 
\label{gau6}
\\[5pt]
&
\ubf_{i-1}^{(1)} \ubf_{i}^{(2)} 
= \widetilde{\mathbb H}_{i}
\qquad \text{for} \quad 1 \le i \le n, 
\label{gau7}
\\[5pt]
&\ubf_{i}^{(1)} \ubf_{i}^{(2)} =
\widetilde{\ubf}_{i}^{(2)}\widetilde{\ubf}_{i}^{(1)}
\qquad \text{for} \quad 1 \le i \le n,
\label{gau8}
\end{align}
where the right hand side of these are written in terms of 
quasi-determinants:
\footnote{Here we set $s=i$ in \eqref{left-qP} and $t=i$ in \eqref{right-qP}.}
\begin{align}
\widetilde{\mathbb H}_{i}& 
 =|\widetilde{\Jbf}^{1,\dots,i}_{1,\dots,i}|_{ii}
=
\widetilde{\Jbf}_{ii} - 
\sum_{a,b=1}^{i-1}\widetilde{\Jbf}_{ia}
((\widetilde{\Jbf}^{1,\dots, i-1}_{1,\dots, i-1})^{-1})_{ab}
\widetilde{\Jbf}_{bi} 
\quad \text{for} \quad 1 \le i \le n,
 \label{gauss-h2}
\\[6pt]
\widetilde{\mathbb E}_{ij}&= 
|\widetilde{\Jbf}^{1,\dots,i}_{1,\dots,i}|_{ii}^{-1}
 |\widetilde{\Jbf}^{1,\dots,i-1,i}_{1,\dots,i-1,j}|_{ij}
 =
\widetilde{\mathbb H}_{j}^{-1}
\left(\widetilde{\Jbf}_{ij} - 
\sum_{a,b=1}^{i-1}\widetilde{\Jbf}_{ia}
((\widetilde{\Jbf}^{1,\dots,i-1}_{1,\dots,i-1})^{-1})_{ab}
\widetilde{\Jbf}_{bj}
\right)  , 
\label{gauss-e2}
\\[6pt]
\widetilde{\mathbb F}_{ji}&=
 |\widetilde{\Jbf}^{1,\dots,i-1,j}_{1,\dots,i-1,i}|_{ji}
|\widetilde{\Jbf}^{1,\dots,i}_{1,\dots,i}|_{ii}^{-1}
\nonumber
\\[5pt]
 &=
\left(\widetilde{\Jbf}_{ji} - 
\sum_{a,b=1}^{i-1}\widetilde{\Jbf}_{ja}
((\widetilde{\Jbf}^{1,\dots,i-1}_{1,\dots,i-1})^{-1})_{ab}
\widetilde{\Jbf}_{bi}
\right)  
\widetilde{\mathbb H}_{i}^{-1}
\quad \text{for} \quad 1 \le i<j \le n. 
\label{gauss-f2} 
\end{align}
Solving equations \eqref{gau6}-\eqref{gau8}, we get 
\begin{align}
\ubf_{i}^{(1)}&=
\overleftarrow{\prod}_{k=1}^{i}
 \widetilde{\ubf}_{k}^{(2)}\widetilde{\ubf}_{k}^{(1)}
\widetilde{\mathbb H}_{k}^{-1} 
 , 
\label{zca6}
\\
\ubf_{i}^{(2)}&=
\left(
\overrightarrow{\prod}_{k=1}^{i-1}
\widetilde{\mathbb H}_{k}
 (\widetilde{\ubf}_{k}^{(2)}\widetilde{\ubf}_{k}^{(1)})^{-1}
\right)
\widetilde{\mathbb H}_{i}
 \qquad \text{for} \quad 1 \le i \le n, 
\\
\Lbf_{ij}^{-(1)}&=
\widetilde{\mathbb{F}}_{ij}
\overleftarrow{\prod}_{k=1}^{j-1}
 \widetilde{\ubf}_{k}^{(2)}\widetilde{\ubf}_{k}^{(1)}
\widetilde{\mathbb H}_{k}^{-1} 
 \qquad \text{for} \quad 1 \le j<i \le n,
 \label{zca8}
\\
\Lbf_{ij}^{+(2)}&=
\left(
\overrightarrow{\prod}_{k=1}^{i-1}
\widetilde{\mathbb H}_{k}
 (\widetilde{\ubf}_{k}^{(2)}\widetilde{\ubf}_{k}^{(1)})^{-1}
\right)
\widetilde{\mathbb H}_{i}
\widetilde{\mathbb{E}}_{ij}
\quad \text{for} \qquad 1 \le i<j \le n. 
\label{zca9}
\end{align}
The other relations can be obtained by substituting these into 
the relations follow from the first and the third relations 
in \eqref{ZC-an}, 
\begin{align}
\Lbf^{+(1)}&=
\widetilde{\Lbf}^{+(2)}\widetilde{\Lbf}^{+(1)}
 (\Lbf^{+(2)})^{-1},
  \label{zca10a}
\\[6pt]
\Lbf^{-(2)}&=
(\Lbf^{-(1)})^{-1}\widetilde{\Lbf}^{-(2)}
\widetilde{\Lbf}^{-(1)}. 
 \label{zca10b}
\end{align}
In order to calculate these more explicitly, we need the inverse matrices of 
$\widetilde{\mathbb{E}}$, $\widetilde{\mathbb{H}}$, and $\widetilde{\mathbb{F}}$. 
Applying Theorem \ref{Gauss1} to 
$\widetilde{\Jbf}^{-1}=
\widetilde{\mathbb{E}}^{-1}\widetilde{\mathbb{H}}^{-1}\widetilde{\mathbb{F}}^{-1}$, 
we obtain
\footnote{Here we set $s=j$ in \eqref{left-qP} and $t=j$ in \eqref{right-qP}.}
\begin{align}
(\widetilde{\mathbb{E}}^{-1})_{ij}&=
 \left|(\widetilde{\Jbf} ^{-1})^{i,j+1,\dots , n}_{j,j+1,\dots, n}\right|_{ij}
 \, 
 \left|(\widetilde{\Jbf }^{-1})^{j,j+1,\dots , n}_{j,j+1,\dots, n}\right|^{-1}_{jj}
 ,
 \label{invEin-o}
\\[6pt]
(\widetilde{\mathbb{F}}^{-1})_{ji}&=
 \left|(\widetilde{\Jbf }^{-1})^{j,j+1,\dots , n}_{j,j+1,\dots, n}\right|^{-1}_{jj}
\, 
 \left|(\widetilde{\Jbf} ^{-1})^{j,j+1,\dots , n}_{i,j+1,\dots, n}\right|_{ji} 
  \quad \text{for} \quad 1 \le i \le j \le n.
 \label{invFin-o}
\end{align}
Then we use the relations 
\begin{align}
\begin{split}
& \left| \widetilde{\Jbf}^{1,2,\dots , j}_{1,2,\dots, j}\right|_{jj} 
 \,
  \left|(\widetilde{\Jbf }^{-1})^{j,j+1,\dots , n}_{j,j+1,\dots, n}\right|_{jj} 
 =1, 
 \quad 
  \left| \widetilde{\Jbf}^{1,2,\dots , j}_{1,2,\dots, j}\right|_{ji} 
 \,
  \left|(\widetilde{\Jbf }^{-1})^{i,j+1,\dots , n}_{j,j+1,\dots, n}\right|_{ij} 
 =1, 
\\[6pt]
& 
  \left| \widetilde{\Jbf}^{1,2,\dots , j}_{1,2,\dots, j}\right|_{ij} 
 \,
  \left|(\widetilde{\Jbf }^{-1})^{j,j+1,\dots , n}_{i,j+1,\dots, n}\right|_{ji} 
 =1 \quad \text{for} \quad 1 \le i \le j \le n, 
 \end{split}
\end{align}
which follow from the formula \eqref{inversion}, to get 
\begin{align}
(\widetilde{\mathbb{E}}^{-1})_{ij}&=
 \left| \widetilde{\Jbf}^{1,2,\dots , j}_{1,2,\dots, j}\right|^{-1}_{ji} 
 \, 
\left| \widetilde{\Jbf}^{1,2,\dots , j}_{1,2,\dots, j}\right|_{jj} 
 ,
 \label{Einv-o}
\\[6pt]
(\widetilde{\mathbb{F}}^{-1})_{ji}&=  \left| \widetilde{\Jbf}^{1,2,\dots , j}_{1,2,\dots, j}\right|_{jj} 
\, 
\left| \widetilde{\Jbf}^{1,2,\dots , j}_{1,2,\dots, j}\right|^{-1}_{ij}
  \quad \text{for} \quad 1 \le i \le j \le n.
 \label{Finv-o}
\end{align}
Substituting \eqref{zca9} (resp.\ \eqref{zca8}) into  \eqref{zca10a} (resp.\ \eqref{zca10b}) 
and using 
 \eqref{Einv-o}, \eqref{Lap-row}, \eqref{row-hom} (resp.\ 
  \eqref{Finv-o}, \eqref{Lap-col}, \eqref{col-hom}), we obtain 
\begin{align}
\Lbf_{ij}^{+(1)}&=
 \left\{
 \sum_{k=i}^{j} (\widetilde{\Lbf}^{+(2)}\widetilde{\Lbf}^{+(1)})_{ik}
  |\widetilde{\Jbf}^{1,2,\dots , j}_{1,2,\dots, j}|^{-1}_{jk}
  |\widetilde{\Jbf}^{1,2,\dots , j}_{1,2,\dots, j}|_{jj}
    \right\}
    \widetilde{\mathbb H}^{-1}_{j}
    \overleftarrow{\prod}_{k=1}^{j-1} 
\widetilde{\ubf}_{k}^{(2)}\widetilde{\ubf}_{k}^{(1)}
   \widetilde{\mathbb H}^{-1}_{k}
 \nonumber \\
 & \hspace{240pt} 
 \text{[by \eqref{zca10a}, \eqref{zca9}, \eqref{Einv-o}]}
\nonumber 
\\[6pt]
&=
 \left\{
(\widetilde{\Lbf}^{+(2)}\widetilde{\Lbf}^{+(1)})_{ij}-
 \sum_{k=1}^{j-1} (\widetilde{\Lbf}^{+(2)}\widetilde{\Lbf}^{+(1)})_{ik}
  |\widetilde{\Jbf}^{1,2,\dots , j-1}_{1,2,\dots, j-1}|^{-1}_{sk}
    |\widetilde{\Jbf}^{1,2,\dots , j-1}_{1,2,\dots,\hat{k},\dots , j}|_{sj}
    \right\}
    \nonumber \\
&  \hspace{130pt} \times   
  \widetilde{\mathbb H}^{-1}_{j}
    \overleftarrow{\prod}_{k=1}^{j-1} 
\widetilde{\ubf}_{k}^{(2)}\widetilde{\ubf}_{k}^{(1)}
   \widetilde{\mathbb H}^{-1}_{k}
\qquad 
 \text{[by \eqref{row-hom}, $s \ne j$]}
\nonumber \\[6pt]
&=
\begin{array}{|c|}
\widetilde{\Jbf}^{1,2,\dots , j-1}_{1,2,\dots, j} 
\\[3pt]
(\widetilde{\Lbf}^{+(2)}\widetilde{\Lbf}^{+(1)})^{\Bar{i}}_{1,2,\dots,j}
\end{array}
_{\, \Bar{i}j} 
\, 
  \widetilde{\mathbb H}^{-1}_{j}
    \overleftarrow{\prod}_{k=1}^{j-1} 
\widetilde{\ubf}_{k}^{(2)}\widetilde{\ubf}_{k}^{(1)}
   \widetilde{\mathbb H}^{-1}_{k}
\qquad \text{[by \eqref{Lap-row}]} ,
 \label{Lp1-o}
\\[6pt]
\Lbf_{ji}^{-(2)}
&=
\left(
\overrightarrow{\prod}_{k=1}^{j-1} 
\widetilde{\mathbb H}_{k}
(\widetilde{\ubf}_{k}^{(2)}\widetilde{\ubf}_{k}^{(1)})^{-1}
\right)
\sum_{k=i}^{j}
  |\widetilde{\Jbf}^{1,2,\dots , j}_{1,2,\dots, j}|_{jj}
  |\widetilde{\Jbf}^{1,2,\dots , j}_{1,2,\dots, j}|^{-1}_{kj}
  (\widetilde{\Lbf}^{-(2)}\widetilde{\Lbf}^{-(1)})_{ki}
   \nonumber \\
 & \hspace{240pt} 
 \text{[by \eqref{zca10b}, \eqref{zca8}, \eqref{Finv-o}]}
\nonumber 
\\[6pt]
&=
\left(
\overrightarrow{\prod}_{k=1}^{j-1} 
\widetilde{\mathbb H}_{k}
(\widetilde{\ubf}_{k}^{(2)}\widetilde{\ubf}_{k}^{(1)})^{-1}
\right)
\nonumber 
\\
& \quad \times 
\left\{
(\widetilde{\Lbf}^{-(2)}\widetilde{\Lbf}^{-(1)})_{ji}-
 \sum_{k=i}^{j-1} 
  |\widetilde{\Jbf}^{1,2,\dots ,\hat{k},\dots, j}_{1,2,\dots, j-1}|_{jt}
    |\widetilde{\Jbf}^{1,2,\dots , j-1}_{1,2,\dots , j-1}|^{-1}_{kt}
    (\widetilde{\Lbf}^{-(2)}\widetilde{\Lbf}^{-(1)})_{ki}
    \right\}
\nonumber \\
&
    \hspace{240pt} \text{[by \eqref{col-hom}, $t \ne j$]}
\nonumber 
\\[6pt]
&=
\left(
\overrightarrow{\prod}_{k=1}^{j-1} 
\widetilde{\mathbb H}_{k}
(\widetilde{\ubf}_{k}^{(2)}\widetilde{\ubf}_{k}^{(1)})^{-1}
\right)
\begin{array}{|cc|}
 \widetilde{\Jbf}^{1,2,\dots , j}_{1,2,\dots, j-1} 
 &
(\widetilde{\Lbf}^{-(2)}\widetilde{\Lbf}^{-(1)})^{1,2,\dots,j}_{\Bar{i}}
\end{array}
_{\, j\Bar{i}} ,
\qquad \text{[by \eqref{Lap-col}]} 
\nonumber 
\\[6pt]
& \hspace{120pt} \text{for} \quad 1 \le i<j \le n. 
 \label{Lm2-o}
\end{align}
Summarizing 
\eqref{zca6}-\eqref{zca9}, \eqref{Lp1-o}, \eqref{Lm2-o} and \eqref{gauss-h2}, 
we arrive at Theorem \ref{sol-zero1-o}.
$ \square $

We did not explicitly assume the defining relations $U_{q}(gl(n))$  for the matrix elements 
of the L-operators 
to derive \eqref{Lp1-o-th}-\eqref{Lm2-o-th}. 
Let us assume the relation \eqref{L-Chv} and 
 the defining relations \eqref{FRT} of $U_{q}(gl(n))$. 
According to \eqref{gau7}, for any $k$, 
$\widetilde{\mathbb H}_{k} $  
 is a Cartan element with respect to the generators in 
 $\Xbf^{(1)}$ and  $\Xbf^{(2)}$, 
and thus commutes with $\ubf_{i}^{(a)}$ for any 
$i$ and $a$. 
Then $\widetilde{\mathbb H}_{k} $ 
commutes
\footnote{This does not mean that 
$\widetilde{\mathbb H}_{k} $ commutes 
with $\widetilde{\ubf}_{i}^{(a)}$. 
Namely, $\widetilde{\mathbb H}_{k} $ is not necessary 
a Cartan element  with respect to the  
generators in $\widetilde{\Xbf}^{(1)}$ and  $\widetilde{\Xbf}^{(2)}$.}
 with 
$\widetilde{\ubf}_{i}^{(2)} \widetilde{\ubf}_{i}^{(1)}$ for any $i$ 
due to the relation \eqref{gau8}. 
In this sense, we need not mind the ordering of the product 
in \eqref{Lp1-o-th}-\eqref{Lm2-o-th}. 

We find that Theorem \ref{sol-zero1} can be rewritten in terms of 
quasi-Pl\"{u}cker coordinates, which are (non-commutative) ratios 
of minor quasi-determinants, of a block matrix:
%
%
\begin{align}
{\mathbf M}&=
\begin{pmatrix}
{\mathbf 0} & \Lbf^{-(1)}\Lbf^{-(2)}  \\
\Lbf^{+(1)}\Lbf^{+(2)} & \Lbf^{-(1)}\Lbf^{+(2)}
\end{pmatrix}
=
\begin{pmatrix}
{\mathbf 0} & \Lbf^{-(1)}\Lbf^{-(2)}  \\
\Lbf^{+(1)}\Lbf^{+(2)} & \Jbf
\end{pmatrix}
, 
\label{singlmat-q}
\end{align}
For $\Bar{i_{1}}, \Bar{i_{2}},\dots , \Bar{i_{a}}, 
k_{1},k_{2},\dots, k_{c}, \Bar{j_{1}}, \Bar{j_{2}},\dots , \Bar{j_{b}}, 
l_{1},l_{2},\dots, l_{d} \in \{1,2,\dots, n \}$, we 
define a block matrix as
\begin{align}
{\mathbf M}^{\Bar{i_{1}}, \Bar{i_{2}},\dots , \Bar{i_{a}}, 
k_{1},k_{2},\dots, k_{c}}_{\Bar{j_{1}}, \Bar{j_{2}},\dots , \Bar{j_{b}}, 
l_{1},l_{2},\dots, l_{d}}&=
\begin{pmatrix}
{\mathbf 0}
 & (\Lbf^{-(1)}\Lbf^{-(2)})^{\Bar{i_{1}}, \Bar{i_{2}},\dots , \Bar{i_{a}}}_ {l_{1},l_{2},\dots, l_{d}} \\[5pt]
(\Lbf^{+(1)}\Lbf^{+(2)})^{k_{1},k_{2},\dots, k_{c}}_{
\Bar{j_{1}}, \Bar{j_{2}},\dots , \Bar{j_{b}}}
 & (\Lbf^{-(1)}\Lbf^{+(2)})^{k_{1},k_{2},\dots, k_{c}}_ {l_{1},l_{2},\dots, l_{d}}
\end{pmatrix}
.
\label{singlsubmat-q}
\end{align}
The notation `$\Bar{\ } $' on indices is used just to remove ambiguity on
 which block matrix we are referring to. Thus it does not appear in 
  subscripts of each matrix element. 
 For example, 
 $(\Lbf^{+(1)}\Lbf^{+(2)})^{k_{1},k_{2},\dots, k_{c}}_{\Bar{j_{1}}, \Bar{j_{2}},\dots , \Bar{j_{b}}}
 =((\Lbf^{+(1)}\Lbf^{+(2)})_{k_{\alpha} j_{\beta}})_{1 \le \alpha \le c, 1 \le \beta \le b}$
 is a $c \times b$ matrix whose $(k_{\alpha}, j_{\beta})$-matrix element is given by
  $(\Lbf^{+(1)}\Lbf^{+(2)})_{k_{\alpha} j_{\beta}}$. 
Taking note on the relations: 
\begin{align}
\begin{split}
\begin{array}{|cc|}
(\Lbf^{+(1)}\Lbf^{+(2)})^{k,k+1,\dots,n}_{\Bar{k}}
&
 \Jbf^{k,k+1,\dots , n}_{k+1,k+2,\dots, n} 
\end{array}
_{\, k \Bar{k}}
&=
\ubf_{k}^{(1)}\ubf_{k}^{(2)} ,
\\[6pt]
\begin{array}{|c|}
(\Lbf^{-(1)}\Lbf^{-(2)})^{\Bar{k}}_{k,k+1,\dots,n}
\\
 \Jbf^{k+1,k+2,\dots , n}_{k,k+1,\dots, n} 
\end{array}
_{\, \Bar{k}k} 
&=
\ubf_{k-1}^{(1)}\ubf_{k-1}^{(2)} ,
\end{split}
 \label{reduct-uu1}
\end{align}
which follow from the relations 
$(\Lbf^{+(1)}\Lbf^{+(2)})_{kj}=(\Lbf^{-(1)}\Lbf^{-(2)})_{jk}=0$
for $k>j$, 
we can rewrite Theorem \ref{sol-zero1} as
\begin{theorem}
\label{sol-zero1-P}
For $ 1 \le i \le j \le n $,
\begin{align}
\widetilde{\Lbf}_{ij}^{+(1)}&=
\left(
\overrightarrow{\prod}_{k=1}^{i-1} 
q^{k+1,k+2,\dots,n}_{k \, \Bar{k}}
({\mathbf M}^{k,k+1,\dots , n}_{\Bar{1},\Bar{2},\dots , \Bar{n},1,2,\dots , n})
\right)
q^{i+1,i+2, \dots,n}_{i  \, \Bar{j}}
({\mathbf M}^{i,i+1,\dots , n}_{\Bar{1},\Bar{2},\dots , \Bar{n},1,2,\dots , n}) ,
 \label{Lp1-th-P} 
\\[6pt]
\widetilde{\Lbf}_{ji}^{-(1)}&=
\left(
\overrightarrow{\prod}_{k=1}^{j-1} 
q^{k+1,k+2,\dots,n}_{k \, \Bar{k}}
({\mathbf M}^{k,k+1,\dots , n}_{\Bar{1},\Bar{2},\dots , \Bar{n},1,2,\dots , n})
\right)
q^{j+1,j+2, \dots,n}_{j \, i}
({\mathbf M}^{j,j+1,\dots , n}_{\Bar{1},\Bar{2},\dots , \Bar{n},1,2,\dots , n}) ,
\label{Lm1-th-P}
\\[6pt]
\widetilde{\Lbf}_{ij}^{+(2)}&=
r^{j+1,j+2, \dots,n}_{i \, \Bar{j}}
({\mathbf M}^{\Bar{1},\Bar{2},\dots , \Bar{n},1,2,\dots , n}_{j,j+1,\dots , n})
\overleftarrow{\prod}_{k=1}^{j-1} 
r^{k+1,k+2,\dots,n}_{k \, \Bar{k}}
({\mathbf M}^{\Bar{1},\Bar{2},\dots , \Bar{n},1,2,\dots , n}_{k,k+1,\dots , n})
,
\label{Lp2-th-P}
\\[6pt]
\widetilde{\Lbf}_{ji}^{-(2)}&=
r^{i+1,i+2, \dots,n}_{\Bar{j}, \Bar{i}}
({\mathbf M}^{\Bar{1},\Bar{2},\dots , \Bar{n},1,2,\dots , n}_{i,i+1,\dots , n})
\overleftarrow{\prod}_{k=1}^{i-1} 
r^{k+1,k+2,\dots,n}_{k \, \Bar{k}}
({\mathbf M}^{\Bar{1},\Bar{2},\dots , \Bar{n},1,2,\dots , n}_{k,k+1,\dots , n})
 \label{Lm2-th-P}
\end{align}
solve the zero-curvature relation \eqref{ZC-an}.
\end{theorem}
%
Similarly, 
we find that Theorem  \ref{sol-zero1-o} can be rewritten in terms of 
quasi-Pl\"{u}cker coordinates of a block matrix:
\begin{align}
\widetilde{\mathbf M}
&=
\begin{pmatrix}
{\mathbf 0} &
   \widetilde{\Lbf}^{+(2)}\widetilde{\Lbf}^{+(1)}  \\
\widetilde{\Lbf}^{-(2)}\widetilde{\Lbf}^{-(1)} & 
  \widetilde{\Lbf}^{+(2)}\widetilde{\Lbf}^{-(1)}
\end{pmatrix}
=
\begin{pmatrix}
{\mathbf 0} &
   \widetilde{\Lbf}^{+(2)}\widetilde{\Lbf}^{+(1)}  \\
\widetilde{\Lbf}^{-(2)}\widetilde{\Lbf}^{-(1)} & 
  \widetilde{\Jbf}
\end{pmatrix}
.
\label{singlmat-qo}
\end{align}
For $\Bar{i_{1}}, \Bar{i_{2}},\dots , \Bar{i_{a}}, 
k_{1},k_{2},\dots, k_{c}, \Bar{j_{1}}, \Bar{j_{2}},\dots , \Bar{j_{b}}, 
l_{1},l_{2},\dots, l_{d} \in \{1,2,\dots, n \}$, we 
define a block matrix as
\begin{align}
\widetilde{\mathbf M}^{\Bar{i_{1}}, \Bar{i_{2}},\dots , \Bar{i_{a}}, 
k_{1},k_{2},\dots, k_{c}}_{\Bar{j_{1}}, \Bar{j_{2}},\dots , \Bar{j_{b}}, 
l_{1},l_{2},\dots, l_{d}}&=
\begin{pmatrix}
{\mathbf 0}
 & (\widetilde{\Lbf}^{+(2)}\widetilde{\Lbf}^{+(1)})^{\Bar{i_{1}}, \Bar{i_{2}},\dots , \Bar{i_{a}}}_ {l_{1},l_{2},\dots, l_{d}} \\[5pt]
(\widetilde{\Lbf}^{-(2)}\widetilde{\Lbf}^{-(1)})^{k_{1},k_{2},\dots, k_{c}}_{
\Bar{j_{1}}, \Bar{j_{2}},\dots , \Bar{j_{b}}}
 & (\widetilde{\Lbf}^{+(2)}\widetilde{\Lbf}^{-(1)})^{k_{1},k_{2},\dots, k_{c}}_ {l_{1},l_{2},\dots, l_{d}}
\end{pmatrix}
.
\label{singlsubmat-qo}
\end{align}
Taking note on the relations: 
\begin{align}
\begin{split}
\begin{array}{|c|}
(\widetilde{\Lbf}^{+(2)}\widetilde{\Lbf}^{+(1)})^{\Bar{k}}_{1,2,\dots,k}
\\
 \widetilde{\Jbf}^{1,2,\dots , k-1}_{1,2,\dots, k} 
\end{array}
_{\, \Bar{k}k} 
&=
\widetilde{\ubf}_{k}^{(2)}\widetilde{\ubf}_{k}^{(1)},
\\[6pt]
\begin{array}{|cc|}
(\widetilde{\Lbf}^{-(2)}\widetilde{\Lbf}^{-(1)})^{1,2,\dots ,k}_{\Bar{k}}
&
 \widetilde{\Jbf}^{1,2,\dots , k}_{1,2,\dots, k-1} 
\end{array}
_{\, k \Bar{k}}
&=
\widetilde{\ubf}_{k-1}^{(2)}\widetilde{\ubf}_{k-1}^{(1)} ,
\end{split}
\label{reduct-uu2}
\end{align}
which follow from the relations 
$(\widetilde{\Lbf}^{+(2)}\widetilde{\Lbf}^{+(1)})_{kj}=
(\widetilde{\Lbf}^{-(2)}\widetilde{\Lbf}^{-(1)})_{jk}=0$
for $k>j$, 
we can rewrite Theorem \ref{sol-zero1-o} as

\begin{theorem}
\label{sol-zero1-oP}
For $ 1 \le i \le j \le n $,
\begin{align}
\Lbf_{ij}^{+(1)}&=
r^{1,2, \dots,j-1}_{\Bar{i} \, j}
(\widetilde{\mathbf M}^{\Bar{1},\Bar{2},\dots , \Bar{n},1,2,\dots , n}_{1,2,\dots , j})
\overleftarrow{\prod}_{k=1}^{j-1} 
r^{1,2,\dots,k-1}_{\Bar{k} \, k}
(\widetilde{\mathbf M}^{\Bar{1},\Bar{2},\dots , \Bar{n},1,2,\dots , n}_{1,2,\dots , k})
 \label{Lp1-th-oP} 
\\[6pt]
\Lbf_{ji}^{-(1)}&=
r^{1,2, \dots,i-1}_{j \, i}
(\widetilde{\mathbf M}^{\Bar{1},\Bar{2},\dots , \Bar{n},1,2,\dots , n}_{1,2,\dots , i})
\overleftarrow{\prod}_{k=1}^{i-1} 
r^{1,2,\dots,k-1}_{\Bar{k} \, k}
(\widetilde{\mathbf M}^{\Bar{1},\Bar{2},\dots , \Bar{n},1,2,\dots , n}_{1,2,\dots , k}) ,
\label{Lm1-th-oP}
\\[6pt]
\Lbf_{ij}^{+(2)}&=
\left(
\overrightarrow{\prod}_{k=1}^{i-1} 
q^{1,2,\dots,k-1}_{\Bar{k} \, k}
(\widetilde{\mathbf M}^{1,2,\dots , k}_{\Bar{1},\Bar{2},\dots , \Bar{n},1,2,\dots , n})
\right)
q^{1,2, \dots,i-1}_{\Bar{i} \, j}
(\widetilde{\mathbf M}^{1,2,\dots , i}_{\Bar{1},\Bar{2},\dots , \Bar{n},1,2,\dots , n})
,
\label{Lp2-th-oP}
\\[6pt]
\Lbf_{ji}^{-(2)}&=
\left(
\overrightarrow{\prod}_{k=1}^{j-1} 
q^{1,2,\dots,k-1}_{\Bar{k} \, k}
(\widetilde{\mathbf M}^{1,2,\dots , k}_{\Bar{1},\Bar{2},\dots , \Bar{n},1,2,\dots , n})
\right)
q^{1,2, \dots,j-1}_{\Bar{j} \, \Bar{i}}
(\widetilde{\mathbf M}^{1,2,\dots , j}_{\Bar{1},\Bar{2},\dots , \Bar{n},1,2,\dots , n})
 \label{Lm2-th-oP}
\end{align}
give the inverse map for \eqref{Lp1-th-P}-\eqref{Lm2-th-P} and solve 
 the zero-curvature relation \eqref{ZC-an}.
\end{theorem}
The zero curvature representation
\footnote{The zero matrices in \eqref{singlmat-q} and \eqref{singlmat-qo} 
play no role in Theorem \ref{sol-zero1-oP} and \ref{sol-zero1-P}. Thus one can replace these  
zero matrices with any matrices of the same size.  For example, 
we could define the matrices ${\mathbf M}$ and $\widetilde{\mathbf M}$ as  
\begin{align}
{\mathbf M}&=
(\Lbf^{+(1)},\Lbf^{-(1)}) \otimes 
\begin{pmatrix}
\Lbf^{-(2)} \\
\Lbf^{+(2)}
\end{pmatrix},
\qquad 
\widetilde{\mathbf M}=
(\widetilde{\Lbf}^{-(2)},\widetilde{\Lbf}^{+(2)}) \otimes 
\begin{pmatrix}
\widetilde{\Lbf}^{+(1)} \\
\widetilde{\Lbf}^{-(1)}
\end{pmatrix}
\label{extraM}
\end{align}
instead of  \eqref{singlmat-q} and \eqref{singlmat-qo}. 
However, \eqref{ZC-an-M} for these produce an extra zero-curvature type relation 
(cf.\ \eqref{ZC-an*}), 
which is incompatible with \eqref{ZC-an}. So we did not use \eqref{extraM} here. 
}
 \eqref{ZC-an} can be rewritten as
\begin{align}
{\mathbf M}&=^{\; t \! \!}\widetilde{\mathbf M}, 
 \label{ZC-an-M}
\end{align}
where the transposition $^{t}$ is taken only for the $(2 \times 2)$-block matrix space. 
Let us introduce matrices
\begin{align}
{\mathbf M}^{*}&=
\begin{pmatrix}
{\mathbf 0} & \Lbf^{+(1)}\Lbf^{+(2)}  \\
\Lbf^{-(1)}\Lbf^{-(2)} & \Lbf^{+(1)}\Lbf^{-(2)}
\end{pmatrix}
, 
\qquad
\widetilde{\mathbf M}^{*}
=
\begin{pmatrix}
{\mathbf 0} &
   \widetilde{\Lbf}^{-(2)}\widetilde{\Lbf}^{-(1)}  \\
\widetilde{\Lbf}^{+(2)}\widetilde{\Lbf}^{+(1)} & 
  \widetilde{\Lbf}^{-(2)}\widetilde{\Lbf}^{+(1)}
\end{pmatrix}
.
\label{singlmat-q*}
\end{align}
Then another zero-curvature relation 
 \eqref{ZC-an*} can be expressed as
\begin{align}
{\mathbf M}^{*}&=^{\; t \! \!}\widetilde{\mathbf M}^{*} .
 \label{ZC-an-M*}
\end{align}
The quantum Yang-Baxter map ${\mathcal R}^{*}$ \eqref{defYB-map-t*} is obtained from 
the inverse of the map 
${\mathcal R}$ \eqref{defYB-map-t} by applying the map 
\eqref{*trans} to Theorem \ref{sol-zero1-oP}, where $\widetilde{\mathbf M}$ 
is replaced by ${\mathbf M}^{*}$.  Similarly, the inverse map ${\mathcal R}^{*-1}$ 
can be derived from the map ${\mathcal R}$ by applying the map 
\eqref{*trans} to Theorem \ref{sol-zero1-P}, where ${\mathbf M}$ 
is replaced by $\widetilde{\mathbf M}^{*}$.

The fact that the quantum Yang-Baxter maps are expressed in terms
 of quasi-Pl\"{u}cker coordinates over matrices 
  \eqref{singlmat-q}, \eqref{singlmat-qo}  and \eqref{singlmat-q*} 
composed of the L-operators may 
 imply an underlying quantum Grassmannian of the system (cf.\ \cite{Lauve04}). 
It is known that L-operators are related to Manin matrices \cite{CFR09,CFRS12}. 
Thus it might be interesting to investigate the matrices 
\eqref{singlmat-q}, \eqref{singlmat-qo}, \eqref{singlmat-q*}, and 
the quantum Yang-Baxter maps in the split of  \cite{CFR09,CFRS12}. 

Finally, let us discuss what will happen in the solution of the zero-curvature relation 
if we chose the gauge of the universal R-matrix and 
the L-operators in subsection \ref{section-Rmatform0} instead of the ones in 
subsection \ref{subsec-twist}. 
The relations corresponding to \eqref{gau3}-\eqref{gau5} now have the form:
\begin{align}
& {\mathbb E}_{ij} =\widetilde{\gbf}^{(2)}_{i} 
\widetilde{\Ebf}^{(2)}_{ji}
(\widetilde{\gbf}_{j}^{(2)})^{-1}, 
\qquad 
{\mathbb F}_{ji}= -\widetilde{\gbf}^{(1)}_{j} 
\widetilde{\Ebf}^{(1)}_{ij}
(\widetilde{\gbf}_{i}^{(1)})^{-1} 
\qquad \text{for} \quad 1 \le i<j \le n, 
\label{dgau3}
\\[5pt]
& {\mathbb H}_{i}=
\widetilde{\gbf}_{i}^{(2)} (\widetilde{\gbf}_{i}^{(1)})^{-1} 
\qquad \text{for} \quad 1 \le i \le n, 
\label{dgau4}
\\[5pt]
&\gbf_{i}^{(1)} \gbf_{i}^{(2)} =
\widetilde{\gbf}_{i}^{(2)}\widetilde{\gbf}_{i}^{(1)},
\qquad 
(\gbf_{i}^{(1)})^{-1} (\gbf_{i}^{(2)})^{-1} =
(\widetilde{\gbf}_{i}^{(2)})^{-1}(\widetilde{\gbf}_{i}^{(1)})^{-1}
\qquad \text{for} \quad 1 \le i \le n, 
\label{dgau5}
\end{align}
where we set $\gbf_{i} = q^{\Ebf_{ii}}$ and consider the Gauss decomposition 
of $\Lbb^{-(1)}\Lbb^{+(2)}={\mathbb E}{\mathbb H}{\mathbb F}$ 
(instead of $\Lbf^{-(1)}\Lbf^{+(2)}$). 
The equations 
\eqref{dgau4} and \eqref{dgau5} produce 
\begin{align}
(\widetilde{\gbf}_{i}^{(2)})^2={\mathbb H}_{i}\gbf_{i}^{(2)}\gbf_{i}^{(1)},
\qquad
(\widetilde{\gbf}_{i}^{(1)})^2={\mathbb H}_{i}^{-1}\gbf_{i}^{(1)}\gbf_{i}^{(2)}.
 \label{dgeq}
\end{align}
The solutions of these equations contain square roots. 
We could not remove these square roots. 
This shows the advantage of choosing the gauge in subsection \ref{subsec-twist}, 
which does not produce square roots. 
%
\section{Quasi-classical limit and classical Yang-Baxter map}
In this section, we will consider the quasi-classical limit of the quantum Yang-Baxter maps  
discussed in the previous section. 
Subsections \ref{subsec-PA}-\ref{subsecCR} correspond to higher rank analogues
 of subsections 3.1, 3.3, 3.5 and 3.8 of \cite{BS15}. 
 The determinant formulas of the classical Yang-Baxter maps in 
subsection \ref{subsec-CYBD} are not mentioned in \cite{BS15}.

\subsection{Poisson algebra}
\label{subsec-PA}
We consider the quasi-classical limit
\begin{align}
q=e^{\pi i \bsf^{2}}, \qquad \bsf \to 0. 
 \label{qc-limit}
\end{align}
Quasi-classical limits of quantum algebras 
were previously considered, for example, in \cite{Babelon88,KMY12,DMV13}.
In the limit \eqref{qc-limit}, the quantum algebra $U_{q}(gl(n))$ 
 reduces to a Poisson algebra 
${\mathcal P}(gl(n)) $. 
Let us make a substitution
\begin{align}
q^{\Ebf_{ii}} \to k_{i}, 
\quad 
\text{and}
\quad 
\Ebf_{ij} \to e_{ij}
\quad 
\text{for}
\quad 
i \ne j, 
\end{align}
and replacement of the 
commutators by Poisson brakets,
\begin{align}
[\; \, , \; ] \to 2 \pi i \bsf^{2} \{\; \, , \; \} , 
\qquad \bsf \to 0,
\end{align}
in the relations 
 \eqref{def-gln}-\eqref{def-eij}.
Then we obtain the following 
relations for the  Posisson algebra ${\mathcal P}(gl(n)) $, 
\begin{align}
\begin{split}
\{k_{l},e_{ij}\} &=\frac{\delta_{il}-\delta_{jl}}{2}e_{ij}k_{l},
\qquad 
\{k_{i},k_{j}\}=0, 
\\[5pt]
\{e_{i,i+1},e_{j+1,j}\}&=\delta_{ij}(k_{i}k_{i+1}^{-1}-k_{i}^{-1}k_{i+1}),
\\[5pt]
\{e_{i,i+1},e_{j,j+1}\}&=\{e_{i+1,i},e_{j+1,j}\}=0 
\quad \text{for} \quad |i-j| \ge 2,
\end{split} 
\label{def-Poisson}
\end{align}
the Serre relations, 
\begin{align}
\begin{split}
\{e_{i,i+1},\{e_{i,i+1},e_{i+1,i+2} \}\}-
\frac{1}{4}e^{2}_{i,i+1}e_{i+1,i+2} &=0,
\\[5pt]
\{e_{i+1,i+2},\{e_{i+1,i+2},e_{i,i+1} \}\}-
\frac{1}{4}e^{2}_{i+1,i+2}e_{i,i+1} &=0,
\\[5pt]
\{e_{i+1,i},\{e_{i+1,i},e_{i+2,i+1} \}\}-
\frac{1}{4}e^{2}_{i+1,i}e_{i+2,i+1} &=0,
\\[5pt]
\{e_{i+2,i+1},\{e_{i+2,i+1},e_{i+1,i} \}\}-
\frac{1}{4}e^{2}_{i+2,i+1}e_{i+1,i} &=0,
\end{split}
 \label{Serre-cl}
\end{align}
and the definition for the other generators:   
for $i,j \in \{1,2,\dots, n\}$ with $j-i \ge 2 $ and $k \in \{i+1,i+2,\dots, j-1 \}$, we define 
\begin{align}
\begin{split}
e_{ij}& =\{e_{ik},e_{kj}\} - \frac{1}{2} e_{kj}e_{ik},
\\[5pt]
e_{ji}& =\{e_{jk},e_{ki}\}+ \frac{1}{2} e_{ki}e_{jk} 
.
\end{split}
\label{def-eijPoisson}
\end{align}
Here \eqref{def-eijPoisson} does not depend on the choice of $k$.
Let $\{ \alpha_{i} \}_{i=1}^{n-1}$ be a positive simple root system of $sl(n)$ with a 
bilinear form $(\alpha_{i}|\alpha_{j})=2\delta_{ij}-\delta_{i+1,j}-\delta_{i,j+1}$. 
The root vectors for the simple roots are denoted as 
$ e_{i,i+1}=e_{\alpha_{i}}$, $ e_{i+1,i}=e_{-\alpha_{i}}$. 
For any root vectors $e_{\pm \alpha}$ and $e_{\pm \beta}$ associated 
with positive roots $\alpha $ and $\beta $, an extended Poisson bracket is defined as
\begin{align}
\{e_{\alpha},e_{\beta } \}_{\mathrm{ex}}=\{e_{\alpha},e_{\beta } \}
+\frac{(\alpha |\beta)}{2} e_{\alpha}e_{\beta }, 
\quad 
\{e_{-\alpha},e_{-\beta } \}_{\mathrm{ex}}=\{e_{-\alpha},e_{-\beta } \}
-\frac{(\alpha |\beta)}{2} e_{-\alpha}e_{-\beta } . 
\end{align}
The general root vectors \eqref{def-eijPoisson}, 
namely $e_{ij}=e_{\alpha_{i}+\alpha_{i+1}+\cdots +\alpha_{j-1}}$ and 
$e_{ji}=e_{-\alpha_{i}-\alpha_{i+1}-\cdots -\alpha_{j-1}}$ for $i<j$, are expressed as 
\begin{align}
e_{ij}& =\{e_{ik},e_{kj}\}_{\mathrm{ex}} \, ,
\qquad 
e_{ji} =\{e_{jk},e_{ki}\}_{\mathrm{ex}} \, , 
\qquad 
i<k<j.
\label{def-eijPoisson-ex}
\end{align}
Then the quasi-classical Serre relations \eqref{Serre-cl} can be rewritten 
in a more standard form as 
\begin{align}
\{e_{\pm \alpha_{i}}, \{e_{\pm \alpha_{i}},e_{\pm \alpha_{i+1} } \}_{\mathrm{ex}} \}_{\mathrm{ex}}=
\{e_{\pm \alpha_{i+1}}, \{e_{\pm \alpha_{i+1}},e_{\pm \alpha_{i} } \}_{\mathrm{ex}} \}_{\mathrm{ex}}=0 .
\end{align}
The quasi-classical limit preserves 
the structure of the Hopf algebra, which has the co-multiplication, co-unit and antipode. 
The co-multiplication $\overline{\Delta}$ is 
an algebra homomorphism from the Poisson algebra to its tensor square 
$\overline{\Delta}: 
{\mathcal P}(gl(n)) \to {\mathcal P}(gl(n)) \otimes {\mathcal P}(gl(n))$, and is 
defined by 
\begin{align}
\begin{split}
\overline{\Delta}(e_{i,i+1})&=e_{i,i+1} \otimes k_{i}k_{i+1}^{-1}
+1 \otimes e_{i,i+1},
\\[5pt]
\overline{\Delta}(e_{i+1,i})&=e_{i+1,i} \otimes 1 + k_{i}^{-1}k_{i+1}
 \otimes e_{i+1,i} \qquad \text{for}
\quad i \in \{ 1,2,\dots, n-1 \}, 
\\[5pt]
\overline{\Delta}(k_{j})&=k_{j} \otimes k_{j}
\qquad \text{for} 
\quad j \in \{ 1,2,\dots, n\},
\end{split}
\end{align}
and the corresponding opposite co-multiplication $\overline{\Delta}^{\prime}$ 
is defined by
\begin{align}
\overline{\Delta}^{\prime} =
\sigma \circ \overline{\Delta},
\end{align}
where $\sigma (a \otimes b )=
b \otimes a $ 
for any $a,b \in {\mathcal P}(gl(n))  $. 
The gauge transformed version of the co-multiplication \eqref{co-pro-gauge} 
reduces to 
\begin{align}
\begin{split}
\overline{\Delta}^{F}(e_{i,i+1})&=e_{i,i+1} \otimes k_{i}
+k^{-1}_{i} \otimes e_{i,i+1},
\\[5pt]
\overline{\Delta}^{F}(e_{i+1,i})&=e_{i+1,i} \otimes k^{-1}_{i+1} + k_{i+1}
 \otimes e_{i+1,i} \qquad \text{for}
\quad i \in \{ 1,2,\dots, n-1 \}, 
\\[5pt]
\overline{\Delta}^{F}(k_{j})&=k_{j} \otimes k_{j}
\qquad \text{for} 
\quad j \in \{ 1,2,\dots, n\}. 
\end{split}
\label{co-pro-gauge-cl}
\end{align}
In the same way as \eqref{counit} and \eqref{antipode1}, 
the co-unit, which is an algebra homomorphism
 from the algebra $ {\mathcal P}(gl(n))  $ to complex numbers, is defined by 
\begin{align}
 \overline{\epsilon} (e_{ij})=0, 
 \qquad
  \overline{\epsilon} (k_{i})=1, 
 \qquad
  \overline{\epsilon}(1)=1, 
 \label{counit-cl}
\end{align}
and the antipode $S$, which is an algebra anti-homomorphism 
on the algebra $ {\mathcal P}(gl(n))  $, is defined by 
\begin{multline}
\overline{S}(k_{i})=k^{-1}_{i}, 
\qquad 
\overline{S}(e_{i,i+1})=-e_{i,i+1} k_{i}^{-1}k_{i+1}, 
\qquad 
\overline{S}(e_{i+1,i})=- k_{i}k_{i+1}^{-1} e_{i+1,i} . 
\label{antipode1-cl}
\end{multline}
The relations \eqref{antipode2} for the classical case is 
the same as the ones for the quantum case: 
\begin{align}
\qquad (\overline{S} \otimes \overline{S}) \circ \overline{\Delta} 
 = \overline{\Delta}^{\prime } \circ \overline{S},
 \qquad 
 \overline{S}(1)=1, \qquad \overline{\epsilon} \circ \overline{S}=\overline{\epsilon}
. 
\label{antipode2-cl}
\end{align}
\subsection{Classical Yang-Baxter map}
The quasi-classical limit of the universal R-matrix becomes singular 
as in \eqref{UR-limit1} and \eqref{UR-limit2}. 
This is not an obstacle to consider the quasi-classical limit of the quantum 
Yang-Baxter maps. In fact, the adjoint action of the universal R-matrix 
on the elements in $\A \otimes \A $ (in \eqref{defYB-map}) is well defined. 
 Then we denote the limit of the quantum Yang-Baxter map by 
\begin{align}
 \Rbar = \lim_{q \to 1} \Rcal . 
\end{align}

Let $x=\{ e_{ij}, k_{l}, k^{-1}_{l} \}$ be the set of generators of the Poisson algebra 
${\mathcal P}(gl(n))$ 
and $x^{(a)}$ be the corresponding components in 
${\mathcal P}(gl(n)) \otimes {\mathcal P}(gl(n))$,
\begin{align}
x^{(1)}=
\{ \xi  \otimes 1 | \xi \in x \}
, \qquad 
x^{(2)}=
\{1\otimes \xi | \xi \in x \}
, 
\qquad 
x=\{ e_{ij}, k_{l}, k^{-1}_{l} \}, 
\quad i \ne j.  
 \label{vari-cl}
\end{align}
In a similar way\footnote{but only through the classical limit, at this stage} 
as the quantum case \eqref{defYB-map}, 
the classical Yang-Baxter map $\Rbar$ is defined in 
${\mathcal P}(gl(n)) \otimes {\mathcal P}(gl(n))$,
\begin{align}
\Rbar : 
(x^{(1)}, x^{(2)})
\mapsto 
(\widetilde{x}^{(1)}, \widetilde{x}^{(2)}), 
\qquad 
(\widetilde{x}^{(1)}, \widetilde{x}^{(2)})
=
\Rbar (x^{(1)}, x^{(2)}). 
\label{defcl-map}
\end{align}
We note that this map preserve the tensor product structure of 
${\mathcal P}(gl(n)) \otimes {\mathcal P}(gl(n))$, and 
 that the elements of both $\tilde{x}^{(a)}$ and  $x^{(a)}$ ($a=1,2$) satisfy  
the same defining relations \eqref{def-Poisson}-\eqref{def-eijPoisson} 
of the Poisson algebra ${\mathcal P}(gl(n))$. 

\subsection{$r$-matrix form of the Poisson algebra}
\label{subsecCR}
Let us take the quasi-classical limit  \eqref{qc-limit} 
of remaining equations in subsections \ref{section-Rmatform0} 
and \ref{subsec-twist}. As opposed to the universal R-matrix, they do not have singularities. 
In the quasi-classical limit, 
L-operators \eqref{Lmp0}, 
\eqref{L-gauge1}, \eqref{L-gauge2}, 
\eqref{L-sp0}  and  \eqref{L-sp} reduce to 
\begin{align}
\begin{split}
l^{-}&
=\sum_{j=1}^{n}E_{jj}\otimes k_{j}^{-1}
 -\sum_{i<j}E_{ji} \otimes   e_{ij}  k_{i}^{-1},
\\[5pt]
l^{+}&=
\sum_{j=1}^{n}E_{jj}\otimes k_{j}
 +\sum_{i<j}E_{ij} \otimes  k_{i}e_{ji},
\end{split}
\\[8pt]
\begin{split}
\ell^{-}&
=\sum_{j=1}^{n}E_{jj}\otimes (k_{1}\cdots k_{j-1})^{2}
 -\sum_{i<j}E_{ji} \otimes 
(k_{1}\cdots k_{i-1})(k_{1}\cdots k_{j-1})e_{ij} ,
\\[5pt]
\ell^{+}&=
\sum_{j=1}^{n}E_{jj}\otimes  (k_{1}\cdots k_{j})^{2}
 +\sum_{i<j}E_{ij} \otimes 
(k_{1}\cdots k_{i})(k_{1}\cdots k_{j})e_{ji}, 
\end{split}
\label{L-clt} 
\end{align} 
and 
\begin{align}
l(\lambda)&=\lambda l^{+}- \lambda^{-1}l^{-}, 
\\[5pt]
\ell(\lambda)&=\lambda \ell^{+}- \lambda^{-1}\ell^{-}. 
\end{align}
For the R-matrices \eqref{Rmp0} and \eqref{Rmp}, we obtain 
\begin{align}
R^{\pm}& =1 \pm i\pi \bsf^2 +2i \pi \bsf^2 r^{\pm} + O(\bsf^{4}),
\\[5pt]
\Rsf^{\pm}& =1 +i\pi \bsf^2 (1 \pm 1) +2i \pi \bsf^2 \rsf^{\pm} + O(\bsf^{4}),
\end{align}
where 
\begin{align}
\begin{split}
r^{+} &= \sum_{i<j}E_{ij} \otimes E_{ji} - 
\frac{1}{2}\sum_{i \ne j}E_{ii} \otimes E_{jj}, 
\\[5pt]
r^{-} &= -\sum_{i>j}E_{ij} \otimes E_{ji} +
\frac{1}{2}\sum_{i \ne j}E_{ii} \otimes E_{jj},
\end{split}
\\[8pt]
\begin{split}
\rsf^{+} &= \sum_{i<j}(E_{ij} \otimes E_{ji} - 
E_{ii} \otimes E_{jj}), 
\\[5pt]
\rsf^{-} &= -\sum_{i>j} (E_{ij} \otimes E_{ji} -
E_{ii} \otimes E_{jj}). 
\end{split}
\end{align}
In the same way as \eqref{Rs0} and \eqref{Rs}, we define 
the classical r-matrices
\begin{align}
r(\lambda)&=\lambda r^{+}- \lambda^{-1}r^{-},  
\\[5pt]
\rsf(\lambda)&=\lambda \rsf^{+}- \lambda^{-1}\rsf^{-} .
\end{align}
These satisfy the classical Yang-Baxter equations \cite{Skl79}
\begin{align}
& [r_{12}(\lambda),r_{13}(\lambda \mu)]+
[r_{12}(\lambda),r_{23}(\mu)]+
[r_{13}(\lambda \mu),r_{23}(\mu)]=0,
\\[5pt]
& [\rsf_{12}(\lambda),\rsf_{13}(\lambda \mu)]+
[\rsf_{12}(\lambda),\rsf_{23}(\mu)]+
[\rsf_{13}(\lambda \mu),\rsf_{23}(\mu)]=0,
\end{align}
where $\lambda, \mu \in \mathbb{C}$. 
The equations corresponding to 
the Yang-Baxter equations \eqref{YBE-sp0} and \eqref{YBE-sp} 
have the following from
\footnote{These equations are derived by assembling the 
quasi-classical limit  of the Yang-Baxter equations \eqref{FRT0} and  \eqref{FRT} rather than 
taking directly quasi-classical limit of \eqref{YBE-sp0} and \eqref{YBE-sp}.}
\begin{align}
\left\{l_{13}(\lambda), l_{23}(\mu)\right\}
& =
\frac{\lambda \mu}{\mu^{2}-\lambda^{2}}
\left[r_{12}(\lambda \mu^{-1}), l_{13}(\lambda)\, l_{23}(\mu) \right], 
 \label{Poisson-rmat-def0}
\\[5pt]
\left\{\ell_{13}(\lambda), \ell_{23}(\mu)\right\}
& =
\frac{\lambda \mu}{\mu^{2}-\lambda^{2}}
\left[\rsf_{12}(\lambda \mu^{-1}), \ell_{13}(\lambda)\, \ell_{23}(\mu) \right].
 \label{Poisson-rmat-def1}
\end{align}
These equations  \eqref{Poisson-rmat-def0} and \eqref{Poisson-rmat-def1}
are r-matrix forms of the defining relations of the Poisson algebra ${\mathcal P}(gl(n))$. 
The quantum zero curvature representations  \eqref{ZC-an0} 
and \eqref{ZC-an} reduce to the classical zero curvature representations: 
\begin{align}
l^{+(1)} l^{+(2)} &= \widetilde{l}^{+(2)}  \widetilde{l}^{+(1)}, 
\qquad 
l^{-(1)} l^{+(2)}= \widetilde{l}^{+(2)}  \widetilde{l}^{-(1)}, 
\qquad 
l^{-(1)} l^{-(2)}= \widetilde{l}^{-(2)}  \widetilde{l}^{-(1)}, 
\label{zcr0}
\\[6pt]
\ell^{+(1)} \ell^{+(2)} &= \widetilde{\ell}^{+(2)}  \widetilde{\ell}^{+(1)}, 
\qquad 
\ell^{-(1)} \ell^{+(2)}= \widetilde{\ell}^{+(2)}  \widetilde{\ell}^{-(1)}, 
\qquad 
\ell^{-(1)} \ell^{-(2)}= \widetilde{\ell}^{-(2)}  \widetilde{\ell}^{-(1)}, 
\label{zcr}
\end{align}
The equations \eqref{zcr0} and  \eqref{zcr}
 define  classical Yang-Baxter maps 
independent of the notion of the universal R-matrix for the
quantum algebra. 
The solution of  \eqref{zcr0} does not produce the rational map 
(it contains square roots) for the variables \eqref{vari-cl}. Then we will consider only the gauge 
transformed version of it, namely, \eqref{zcr} with the following variables, in the next subsection: 
\begin{multline}
x^{(1)}=
\{ \xi^{(1)} :=\xi  \otimes 1 | \xi \in x \}
, \qquad 
x^{(2)}=
\{\xi^{(2)} :=1\otimes \xi | \xi \in x \}
, 
\\[6pt]
x=\{ \ell^{+}_{ij}, \ \ell^{-}_{ji} | 1 \le i \le j \le n \}, 
 \label{vari-cl-t}
\end{multline}
where these are related to the matrix elements
\footnote{$\ell^{\pm}=\sum_{ij }E_{ij} \otimes \ell^{\pm}_{ij} $}
 of the L-operators \eqref{L-clt} 
as 
\begin{align}
\begin{split}
\ell^{+}_{ij}& =(k_{1}k_{2}\cdots k_{i})(k_{1}k_{2}\cdots k_{j})e_{ji}, 
\qquad 
\ell^{-}_{ji}=-(k_{1}k_{2}\cdots k_{i-1})(k_{1}k_{2}\cdots k_{j-1})e_{ij} , 
 \\[3pt]
\ell^{+}_{ji}&=\ell^{-}_{ij}=0 \qquad  \text{for} \quad 1 \le i < j \le n,
\\[6pt]
 \ell^{+}_{ii}& =(k_{1}k_{2}\cdots k_{i})^{2} =: u_{i} , 
  \qquad 
  \ell^{-}_{ii} =(k_{1}k_{2}\cdots k_{i-1})^{2} =u_{i-1} 
  \quad 
  \text{for} \quad 1 \le i  \le n. 
\end{split}
\label{elem-cl}
\end{align}
\subsection{Solution of the zero-curvature representation}
\label{subsec-CYBD}
Some of the formulas in subsection \ref{subsec-CYBD} become 
simple in the quasi-classical limit. 
We denote the sub-matrix 
$(a_{i_{a},j_{b}})_{1 \le a \le m, 1 \le b \le n}$ 
of a matrix $A=(a_{a,b})_{1 \le a,b \le N}$ as 
$A^{i_{1},\dots, i_{m}}_{j_{1},\dots, j_{n}} $, 
where $ 1 \le i_{1} \le \dots \le i_{m} \le N$ and 
$ 1 \le j_{1} \le \dots \le j_{n} \le N$. 
Let us solve the system of equations \eqref{zcr} 
for the matrices  with commutative entries 
$\ell^{\pm (a)}=(\ell^{\pm (a)}_{ij})_{1 \le i,j \le n}$ and 
$\tilde{\ell}^{\pm (a)}=
(\tilde{\ell}^{\pm (a)}_{ij})_{1 \le i,j \le n}$ 
for $a=1,2$ 
under the condition 
$\ell^{+(a)}_{ij}=\widetilde{\ell}^{+(a)}_{ij}=
\widetilde{\ell}^{-(a)}_{ji}=\ell^{-(a)}_{ji}=0$ for $i>j$, 
$\ell^{+(a)}_{ii}=u_{i}^{(a)}$, 
$\widetilde{\ell}^{+(a)}_{ii}=\widetilde{u}_{i}^{(a)}$, 
$\ell^{-(a)}_{ii}=u_{i-1}^{(a)}$, 
$\widetilde{\ell}^{-(a)}_{ii}=\widetilde{u}_{i-1}^{(a)}$ 
and 
  $u_{0}^{(a)}=\widetilde{u}_{0}^{(a)}=1$. 
  These variables come from \eqref{vari-cl-t} and \eqref{elem-cl}. However,  
at this stage,  we do not a priori assume the defining relations of ${\mathcal P}(gl(n))$ explicitly. 
Let us denote the left hand side of 
the second equation in \eqref{ZC-an} as 
$\Jsf=(\Jsf_{a,b})_{1 \le a,b \le n}
:=\ell^{-(1)} \ell^{+(2)}$, where 
\begin{align}
\Jsf_{a,b}=
\sum_{k=1}^{\min\{a,b\}} \ell^{-(1)}_{ak}\ell^{+(2)}_{kb}.
\end{align}  
In the classical limit, the quasi-determinants reduce to ratios of determinants.  
Consequently, Theorems \ref{sol-zero1} and \ref{sol-zero1-o} reduce to
\footnote{We denote the determinant of a matrix $A$ as $|A|$. 
The quantum zero curvature relation \eqref{ZC-an} has 
the same form as the classical one \eqref{zcr}. 
In addition, the matrix elements of the quantum L-operators \eqref{L-Chv} 
have the same form as the classical ones \eqref{elem-cl}, although 
the former are, in general, non-commutative elements and the latter are commutative ones. 
Then we consider that the solution of the classical zero curvature relation
 coincides with the 
classical limit of the solution of the quantum zero curvature relation.}
\begin{theorem}
\label{sol-zero1-cl}
For $1 \le i \le j \le n$, 
\begin{align}
\widetilde{\ell}^{+(1)}_{ij}
&= 
\begin{array}{|cc|}
(\ell^{+(1)}\ell^{+(2)})^{i,\dots, n}_{j}
&
\Jsf^{i,\dots, n}_{i+1,\dots,n}
\end{array}
\, 
u_{n}^{(1)}
\,
\prod_{k=i}^{n} (u_{k}^{(1)}
u_{k}^{(2)} )^{-1}
, 
 \label{lp1-cl-th}
 \\[6pt]
\widetilde{\ell}_{ji}^{-(1)}&=
|\Jsf^{j, \dots, n}_{i,j+1,\dots, n}|
u_{n}^{(1)}\prod_{k=j}^{n} (u_{k}^{(1)}u_{k}^{(2)})^{-1} 
 ,
\label{lm1-cl-th}
\\[6pt]
\widetilde{\ell}_{ij}^{+(2)}&=
\frac{|\Jsf^{i,j+1,\dots, n}_{j,\dots, n}|}
{|\Jsf^{j, \dots,n}_{j,\dots, n}|
|\Jsf^{j+1,\dots, n}_{j+1,\dots, n}|}
\, 
(u_{n}^{(1)})^{-1}
\prod_{k=j}^{n} u_{k}^{(1)}u_{k}^{(2)} 
,
\label{lp2-cl-th}
\\[6pt]
\widetilde{\ell}^{-(2)}_{ji}
&= 
\frac{
\begin{array}{|c|}
(\ell^{-(1)}\ell^{-(2)})^{j}_{i,\dots,n}
\\[3pt]
\Jsf^{i+1,\dots, n}_{i,\dots, n}
\end{array}
}
{|\Jsf^{i+1,\dots, n}_{i+1,\dots, n}|
 |\Jsf^{i,\dots, n}_{i,\dots, n}|}
\, 
(u_{n}^{(1)})^{-1}
\prod_{k=i}^{n} u_{k}^{(1)}u_{k}^{(2)} 
 \label{lm2-cl-th} 
\end{align}
solve the zero curvature relation \eqref{zcr}.
\end{theorem}
\begin{theorem}
\label{sol-zero1-o-cl}
For $ 1 \le i \le j \le n$, 
\begin{align}
\ell^{+(1)}_{ij}
&= 
\frac{
\begin{array}{|c|}
\widetilde{\Jsf}^{1,\dots, j-1}_{1,\dots, j}
\\[3pt]
(\widetilde{\ell}^{+(2)}\widetilde{\ell}^{+(1)})^{i}_{1,\dots,j}
\end{array}
}
{|\widetilde{\Jsf}^{1,\dots, j-1}_{1,\dots, j-1}|
 |\widetilde{\Jsf}^{1,\dots, j}_{1,\dots, j}|}
\, 
\prod_{k=1}^{j-1} \widetilde{u}_{k}^{(2)}\widetilde{u}_{k}^{(1)} 
, \qquad 1 \le i \le j \le n, 
  \label{lp1-in-cl-th}
\\[6pt]
\ell_{ji}^{-(1)}&=
\frac{|\widetilde{\Jsf}^{1,\dots, i-1,j}_{1,\dots, i-1,i}|}
{|\widetilde{\Jsf}^{1,\dots, i-1}_{1,\dots, i-1}|
|\widetilde{\Jsf}^{1,\dots, i}_{1,\dots, i}|}
\, 
\prod_{k=1}^{i-1} \widetilde{u}_{k}^{(2)}\widetilde{u}_{k}^{(1)} 
, 
\label{lm1-in-cl-th}
\\[6pt]
\ell_{ij}^{+(2)}&=
|\widetilde{\Jsf}^{1,\dots, i-1,i}_{1,\dots, i-1,j}|
\prod_{k=1}^{i-1} (\widetilde{u}_{k}^{(2)}\widetilde{u}_{k}^{(1)})^{-1} 
 , \qquad 1 \le i \le j \le n,
\label{lp2-in-cl-th}
\\[6pt]
\ell^{-(2)}_{ji}
&= 
\begin{array}{|cc|}
\widetilde{\Jsf}^{1,\dots, j}_{1,\dots,j-1}
&
(\widetilde{\ell}^{-(2)}\widetilde{\ell}^{-(1)})^{1,\dots, j}_{i}
\end{array}
\, 
\prod_{k=1}^{j-1} (\widetilde{u}_{k}^{(2)}
\widetilde{u}_{k}^{(1)} )^{-1}
, \qquad 1 \le i \le j \le n.
 \label{lm2-in-cl-th}
\end{align}
give the inverse transformation of Theorem \ref{sol-zero1-cl} and 
solve the zero curvature relation \eqref{zcr}.
\end{theorem}
The Gauss decomposition formulas for matrices with non-commutative entries, 
which played a key role in the previous section, reduce to 
the ones for matrices with commutative entries. Namely, 
\eqref{gauss-h1}-\eqref{gauss-f1}, \eqref{Einv} and \eqref{Finv} reduce to
\begin{align}
{\mathbb H}_{i}&=
\frac{|\Jsf^{i,\dots, n}_{i,\dots, n}|}
{|\Jsf^{i+1,\dots, n}_{i+1,\dots, n}|} \qquad \text{for} \quad 1 \le i \le n,
\nonumber 
\\[6pt]
{\mathbb E}_{ij}&=
\frac{|\Jsf^{i,j+1,\dots, n}_{j,j+1,\dots, n}|}
{|\Jsf^{j,\dots, n}_{j,\dots, n}|},
\qquad 
{\mathbb F}_{ji}=
\frac{|\Jsf^{j,j+1,\dots, n}_{i,j+1,\dots, n}|}
{|\Jsf^{j,\dots, n}_{j,\dots, n}|}, 
\nonumber 
\\[6pt]
({\mathbb E}^{-1})_{ij}&= (-1)^{j-i}
\frac{|\Jsf^{i,\dots, \hat{j},\dots, n}_{i+1,\dots, n}|}
{|\Jsf^{i+1,\dots, n}_{i+1,\dots, n}|}, 
\quad 
({\mathbb F}^{-1})_{ji}= (-1)^{j-i}
\frac{|\Jsf^{i+1,\dots, n}_{i,\dots,\widehat{j}, \dots,n}|}
{|\Jsf^{i+1,\dots, n}_{i+1,\dots, n}|}
\quad \text{for} \quad 1 \le i \le j \le n, 
\nonumber 
\\[5pt]
({\mathbb E}^{-1})_{ij}&=({\mathbb F}^{-1})_{ji}=0 
\qquad \text{for} \qquad 1 \le j < i \le n, 
\label{Gau-mat1}
\end{align}
where $\hat{i}$ means that $i$ is removed from $1,\dots, j$; 
and 
\eqref{gauss-h2}-\eqref{gauss-f2}, \eqref{Einv-o} and \eqref{Finv-o} reduce to
\begin{align}
\widetilde{\mathbb H}_{i}&=
\frac{|\widetilde{\Jsf}^{1,\dots, i}_{1,\dots, i}|}
{|\widetilde{\Jsf}^{1,\dots, i-1}_{1,\dots, i-1}|} \qquad \text{for} \qquad 1 \le i \le n,
\nonumber 
\\[6pt]
\widetilde{\mathbb E}_{ij}&=
\frac{|\widetilde{\Jsf}^{1,\dots, i-1,i}_{1,\dots, i-1,j}|}
{|\widetilde{\Jsf}^{1,\dots, i}_{1,\dots, i}|},
\qquad 
\widetilde{\mathbb F}_{ji}=
\frac{|\widetilde{\Jsf}^{1,\dots, i-1,j}_{1,\dots, i-1,i}|}
{|\widetilde{\Jsf}^{1,\dots, i}_{1,\dots, i}|}, 
\nonumber 
\\[6pt]
(\widetilde{\mathbb E}^{-1})_{ij}&= (-1)^{j-i}
\frac{|\widetilde{\Jsf}^{1,\dots, j-1}_{1,\dots,\widehat{i}, \dots,j}|}
{|\widetilde{\Jsf}^{1,\dots, j-1}_{1,\dots, j-1}|}, 
\quad 
(\widetilde{\mathbb F}^{-1})_{ji}= (-1)^{j-i}
\frac{|\widetilde{\Jsf}^{1,\dots, \hat{i},\dots, j}_{1,\dots, j-1}|}
{|\widetilde{\Jsf}^{1,\dots, j-1}_{1,\dots, j-1}|}
\quad \text{for} \quad 1 \le i \le j \le n,
\nonumber 
\\[6pt]
(\widetilde{\mathbb E}^{-1})_{ij}&=
(\widetilde{\mathbb F}^{-1})_{ji}=0
\qquad \text{for} \qquad 1 \le j < i \le n.
 \label{Gau-mat2}
\end{align}
One can prove \eqref{Gau-mat1} and \eqref{Gau-mat2}, and based on these, 
Theorem \ref{sol-zero1-cl} and Theorem \ref{sol-zero1-o-cl} 
independent of the quasi-determinants. For this, one may use 
properties of minor determinants of triangular matrices and a 
multiplicative formula for minor determinants of the form  
$\left|(AB)^{i_{1},\dots, i_{r} }_{j_{1},\dots, j_{r} }\right|=\sum_{k_{1}<\dots <k_{r}}
\left|A^{i_{1},\dots, i_{r} }_{k_{1},\dots, k_{r} }\right|
\left|B^{k_{1},\dots, k_{r}}_{j_{1},\dots, j_{r} }\right| $
 (cf.\ \cite{Noumi}). 


Note that 
the factor $|\Jsf^{j,\dots, n}_{j,\dots, n}|$ becomes simple at $j=1$: 
$|\Jsf^{1,\dots, n}_{1,\dots, n}| = |\ell^{-(1)} |  |\ell^{+(2)} | =
 u_{n}^{(2)} \prod_{k=1}^{n-1} u_{k}^{(1)}u_{k}^{(2)} $.


We remark that factors repeatedly appear in the above formulas can also 
be written in terms of  minor determinants of products of quasi-classical L-operators (they are 
co-multiplications of quasi-classical L-operators): 
\begin{align}
| (\ell^{+(1)}\ell^{+(2)})^{1,\dots , i}_{1,\dots , i} |=
| (\ell^{-(1)}\ell^{-(2)})^{1,\dots , i+1}_{1,\dots , i+1} | = \prod_{k=1}^{i} u_{k}^{(1)}u_{k}^{(2)} 
, \qquad 
| (\ell^{+(1)}\ell^{+(2)})^{i,\dots , n}_{i,\dots , n} |
 = \prod_{k=i}^{n} u_{k}^{(1)}u_{k}^{(2)} .
\end{align}
The quasi-classical analogues of  the relations 
\eqref{reduct-uu1} and \eqref{reduct-uu2} also hold. 
The quasi-Pl\"{u}cker coordinates reduce to ratios of Pl\"{u}cker coordinates \eqref{qP-cl}. 
Then quasi-classical analogues of Theorems \ref{sol-zero1-P} and \ref{sol-zero1-oP} hold. 
Thus all the formulas can be written in terms of ratios of minor determinants over the 
 matrices $\mathsf{M}$ and $\widetilde{\mathsf M}$, 
 which are quasi-classical counterparts of \eqref{singlmat-q} 
and \eqref{singlmat-qo}:
\begin{align}
\mathsf{M}&=
\begin{pmatrix}
{\mathbf 0} & \ell^{-(1)}\ell^{-(2)}  \\
\ell^{+(1)}\ell^{+(2)} & \ell^{-(1)}\ell^{+(2)}
\end{pmatrix}
=
\begin{pmatrix}
{\mathbf 0} & \ell^{-(1)}\ell^{-(2)}  \\
\ell^{+(1)}\ell^{+(2)} & \Jsf
\end{pmatrix}
,
\\[6pt]
\widetilde{\mathsf M}&=
\begin{pmatrix}
{\mathbf 0} &
   \widetilde{\ell}^{+(2)}\widetilde{\ell}^{+(1)}  \\
\widetilde{\ell}^{-(2)}\widetilde{\ell}^{-(1)} & 
  \widetilde{\ell}^{+(2)}\widetilde{\ell}^{-(1)}
\end{pmatrix}
=
\begin{pmatrix}
{\mathbf 0} &
   \widetilde{\ell}^{+(2)}\widetilde{\ell}^{+(1)}  \\
\widetilde{\ell}^{-(2)}\widetilde{\ell}^{-(1)} & 
  \widetilde{\Jsf}
\end{pmatrix}
.
\label{singlmat-cl}
\end{align}
The map ${\mathcal R}^{*}$ \eqref{defYB-map-t*} and its inverse also have 
natural quasi-classical counterparts, which are easily obtained. 

Let us give examples for  $n=3$ case. 
For simplicity, we impose the ${\mathcal P}(sl(n))$-type condition 
$u^{(1)}_{n}=u^{(2)}_{n}=\widetilde{u}^{(1)}_{n}=\widetilde{u}^{(2)}_{n}=1 $ 
for $n=3$. 
The formula  is composed of  matrices:
\begin{align}
 \ell^{-(a)} &=\begin{pmatrix}
1 & 0 & 0 \\
\ell^{-(a)}_{21} & u^{(a)}_{1} & 0 \\ 
\ell^{-(a)}_{31} & \ell^{-(a)}_{32} & u^{(a)}_{2} 
\end{pmatrix},
\qquad 
 \ell^{+(a)} =\begin{pmatrix}
u^{(a)}_{1}  & \ell^{+(a)}_{12} & \ell^{+(a)}_{13} \\
0 & u^{(a)}_{2} & \ell^{+(a)}_{23} \\ 
0 & 0 & 1
\end{pmatrix},
\qquad 
a=1,2,
\\[5pt]
 \ell^{-(1)}\ell^{+(2)} &=\Jsf =
\begin{pmatrix}
\Jsf_{11} & \Jsf_{12} & \Jsf_{13} \\
\Jsf_{21} & \Jsf_{22} & \Jsf_{23} \\ 
\Jsf_{31} & \Jsf_{32} & \Jsf_{33} 
\end{pmatrix}
\nonumber 
\\[6pt]
&
=
\begin{pmatrix}
u^{(2)}_{1} & \ell^{+(2)}_{12} & \ell^{+(2)}_{13} \\
\ell^{-(1)}_{21} u^{(2)}_{1} & \ell^{-(1)}_{21}\ell^{+(2)}_{12} + u^{(1)}_{1}u^{(2)}_{2} 
   & \ell^{-(1)}_{21}\ell^{+(2)}_{13} + u^{(1)}_{1} \ell^{+(2)}_{23} \\
\ell^{-(1)}_{31} u^{(2)}_{1} & \ell^{-(1)}_{31}\ell^{+(2)}_{12} + \ell^{-(1)}_{32}u^{(2)}_{2} 
   & \ell^{-(1)}_{31}\ell^{+(2)}_{13} + \ell^{-(1)}_{32} \ell^{+(2)}_{23} +u^{(1)}_{2}
\end{pmatrix}
.
\end{align}
Then Theorem \ref{sol-zero1-cl} for $n=3$ reduces to
\begin{align}
\widetilde{u}^{(1)}_{1} &= 
\begin{array}{|cc|}
\Jsf_{22} & \Jsf_{23} \\
\Jsf_{32} & \Jsf_{33}
\end{array}
\, 
(u^{(1)}_{2}u^{(2)}_{2})^{-1} ,
\qquad 
\widetilde{u}^{(1)}_{2} =  \Jsf_{33} ,
\\[6pt]
\widetilde{\ell}^{+(1)}_{12} &=
\begin{array}{|ccc|}
(\ell^{+(1)}\ell^{+(2)})_{12} & \Jsf_{12}  & \Jsf_{13} \\
u^{(1)}_{2}u^{(2)}_{2} & \Jsf_{22} & \Jsf_{23} \\
0 & \Jsf_{32} &  \Jsf_{33}
\end{array}
(u^{(1)}_{1} u^{(2)}_{1} u^{(1)}_{2} u^{(2)}_{2} )^{-1} , 
\\[6pt]
\widetilde{\ell}^{+(1)}_{23} &=
\begin{array}{|cc|} 
(\ell^{+(1)}\ell^{+(2)})_{23} & \Jsf_{23} \\
1 & \Jsf_{33}
\end{array}
\, 
(u^{(1)}_{2} u^{(2)}_{2} )^{-1} , 
\\[6pt]
\widetilde{\ell}^{+(1)}_{13} &=
\begin{array}{|ccc|}
(\ell^{+(1)}\ell^{+(2)})_{13} & \Jsf_{12}  & \Jsf_{13} \\
(\ell^{+(1)}\ell^{+(2)})_{23} & \Jsf_{22} & \Jsf_{23} \\
1 & \Jsf_{32} &  \Jsf_{33}
\end{array}
\, 
(u^{(1)}_{1} u^{(2)}_{1} u^{(1)}_{2} u^{(2)}_{2} )^{-1} , 
\\[6pt]
\widetilde{\ell}^{-(1)}_{21} &= 
\begin{array}{|cc|}
\Jsf_{21} & \Jsf_{23} \\
\Jsf_{31} & \Jsf_{33}
\end{array}
\, 
(u^{(1)}_{2}u^{(2)}_{2})^{-1} ,
\qquad 
\widetilde{\ell}^{-(1)}_{31} = \Jsf_{31},
\qquad 
\widetilde{\ell}^{-(1)}_{32} = \Jsf_{32},
\\[6pt]
\widetilde{u}^{(2)}_{1} &= 
\frac{u^{(1)}_{1}u^{(2)}_{1}u^{(1)}_{2}u^{(2)}_{2}
}{
\begin{array}{|cc|}
\Jsf_{22} & \Jsf_{23} \\
\Jsf_{32} & \Jsf_{33}
\end{array} 
} ,
\qquad 
\widetilde{u}^{(2)}_{2} = \frac{u^{(1)}_{2}u^{(2)}_{2}}{ \Jsf_{33}} ,
\\[6pt]
\widetilde{\ell}^{+(2)}_{12} &=
\frac{ 
\begin{array}{|cc|}
\Jsf_{12} & \Jsf_{13} \\
\Jsf_{32} & \Jsf_{33}
\end{array}
\, 
u^{(1)}_{2}u^{(2)}_{2}
}{ 
\begin{array}{|cc|}
\Jsf_{22} & \Jsf_{23} \\
\Jsf_{32} & \Jsf_{33}
\end{array}
\, 
\Jsf_{33}
},
\qquad 
\widetilde{\ell}^{+(2)}_{13} =\frac{\Jsf_{13}}{\Jsf_{33}}, 
\qquad 
\widetilde{\ell}^{+(2)}_{23} =\frac{\Jsf_{23}}{\Jsf_{33}}, 
\\[6pt]
\widetilde{\ell}^{-(2)}_{21} &=
\frac{ 
\begin{array}{|ccc|}
(\ell^{-(1)}\ell^{-(2)})_{21} &  u^{(1)}_{1}u^{(2)}_{1}& 0 \\
\Jsf_{21} & \Jsf_{22} & \Jsf_{23} \\
\Jsf_{31} & \Jsf_{32} &  \Jsf_{33}
\end{array}
}{ 
\begin{array}{|cc|}
\Jsf_{22} & \Jsf_{23} \\
\Jsf_{32} & \Jsf_{33}
\end{array}
},
\\[6pt]
\widetilde{\ell}^{-(2)}_{31} & =
\frac{ 
\begin{array}{|ccc|}
(\ell^{-(1)}\ell^{-(2)})_{31} & (\ell^{-(1)}\ell^{-(2)})_{32} & u^{(1)}_{2}u^{(2)}_{2} \\
\Jsf_{21} & \Jsf_{22} & \Jsf_{23} \\
\Jsf_{31} & \Jsf_{32} &  \Jsf_{33}
\end{array}
}{ 
\begin{array}{|cc|}
\Jsf_{22} & \Jsf_{23} \\
\Jsf_{32} & \Jsf_{33}
\end{array}
},
\\[6pt]
\widetilde{\ell}^{-(2)}_{32} &=
\frac{ 
\begin{array}{|cc|}
 (\ell^{-(1)}\ell^{-(2)})_{32} & u^{(1)}_{2}u^{(2)}_{2} \\
 \Jsf_{32} &  \Jsf_{33}
\end{array}
\,
u^{(1)}_{2}u^{(2)}_{2}
}{ 
\Jsf_{33} 
\,
\begin{array}{|cc|}
\Jsf_{22} & \Jsf_{23} \\
\Jsf_{32} & \Jsf_{33}
\end{array}
} ,
\end{align}
where the relation $|\Jsf^{123}_{123} | =u^{(1)}_{1}u^{(2)}_{1}u^{(1)}_{2}u^{(2)}_{2}$, 
which follows from 
$u^{(1)}_{3}=u^{(2)}_{3}=\widetilde{u}^{(1)}_{3}=\widetilde{u}^{(2)}_{3}=1 $, is used. 
\section{Concluding remarks}
In this paper, we studied Yang-Baxter maps from 
the point of view of the quantum group theory. 
Solving the zero curvature relation, we obtained  
quasi-determinant expressions of quantum Yang-Baxter maps  
 associated with $U_{q}(gl(n))$. In the quasi-classical limit, 
these reduce to new determinant expressions of the 
classical Yang-Baxter maps.  

In this paper, we focused our discussions only on algebraic structures 
of the Yang-Baxter maps. The next step would be to investigate the maps 
for particular representations or realizations of the algebra. 
In \cite{BS15}, they considered a Heisenberg-Weyl realization of  $U_{q}(sl(2))$, and 
obtained discrete Liouville equations in the quasi-classical limit. 
We expected at first that a natural higher rank analogue of their results might produce 
discrete Toda field equations. However, this is not clear, at the moment. 
Another interesting question would be how or if our results for the quantum case 
are related to 
the non-Abelian Hirota-Miwa equations whose solutions are written 
in terms of quasi-determinants \cite{Nimmo06,GNO07}. 

\section*{Acknowledgments}
The author would like to thank 
Vladimir Bazhanov for encouraging him to consider higher rank 
 generalization of their results on $U_{q}(sl(2))$ case \cite{BS15}, and 
Vladimir Bazhanov, 
Sergey Khoroshkin and Sergey Sergeev 
for collaboration at the earlier stage of the present work. 
What the author learned from them are cited as private communications \cite{Sergeev15,BK13}. 
The parts inherited from \cite{BS15} are explained at the beginning of sections 2 and 3. 
This work was started when he was at 
Department of Theoretical Physics, RSPE, 
Australian National University 
(where he was supported by the Australian Research Council) in 2013, 
and was developed at 
School of Mathematics and Statistics, 
The University of Melbourne 
(where he was supported by the Australian Research Council), 
Fakult\"at f\"ur Mathematik und Naturwissenschaften, 
Bergische Universit\"at Wuppertal (where he was supported by the university), and 
Laboratoire de Math\'ematiques et Physique Th\'eorique CNRS/UMR 7350,
 F\'ed\'eration Denis Poisson FR2964,
Universit\'e de Tours (where he was supported by CNRS), 
and has been finalized at 
 LPTENS (where he is supported by 
 the European Research Council 
(Programme ``Ideas'' ERC-2012-AdG 320769 AdS-CFT-solvable)).  
He thanks Vladimir Bazhanov, Omar Foda, Hermann Boos, Pascal Baseilhac and Vladimir Kazakov 
for support at these institutions. 
A part of this work was previously presented at a seminar at 
Bergische Universit\"at Wuppertal in July 2016. 
He also thanks an anonymous referee for useful comments. 
\appendix
\section{Properties of the quantum Yang-Baxter map}
Properties of the quantum Yang-Baxter map were already explained in 
subsection 2.4 of \cite{BS15}. However, 
proofs of some of the relations were omitted in \cite{BS15}.
The purpose of this section is to supplement their discussions by giving details.  

The co-multiplication for any element $\xbf \in \Xbf $ has the form
\begin{align}
\Delta (\xbf)=\sum_{i} \ybf^{(1)}_{i} \zbf^{(2)}_{i},  
 \label{co-prod-gen}
\end{align}
where $\ybf^{(1)}_{i} $ (resp.\ $\zbf^{(2)}_{i}$) is written in terms of only elements of 
$\Xbf^{(1)}$ (resp.\ $\Xbf^{(2)}$). 
Taking note on this fact, we introduce 
the set-theoretic multiplication $\delta$ 
which acts on two sets of generating elements
\footnote{Note that the map $\delta $ 
 is well defined only for the pair of the set $(\Xbf^{(1)}, \Xbf^{(2)})$. 
On the other hand, the map $\R$ can also be defined for each element 
$\xbf^{(a)} \in \Xbf^{(a)}$, $a=1,2$ by 
$\R(\xbf^{(a)})=\Rbb \xbf^{(a)} \Rbb^{-1}$,  
as well as for the pair of the set $(\Xbf^{(1)}, \Xbf^{(2)})$.}
\begin{align}
\delta :  \quad (\Xbf^{(1)}, \Xbf^{(2)}) \mapsto \Xbf^{\prime} 
=\{ \Delta(\xbf) | \xbf \in \Xbf \},
\end{align}
and denote it as 
\begin{align}
 \Xbf^{\prime}= \delta (\Xbf^{(1)}, \Xbf^{(1)})  . 
\end{align}
The co-multiplication preserves the defining relations of the algebra $\A$. 
Thus the algebra generated by the elements in the set 
$\delta ({\mathbf X}^{(1)}, {\mathbf X}^{(2)}) $
is isomorphic to the original algebra $\A$.  
The map $\delta $ has the following property
\begin{align}
\Rbb \delta({\mathbf X}^{(1)}, {\mathbf X}^{(2)} ) \Rbb^{-1}
=  \delta( \Rbb {\mathbf X}^{(1)} \Rbb^{-1}, \Rbb {\mathbf X}^{(2)}\Rbb^{-1} ) .
\label{delta-R}
\end{align}
This follows from the relation 
\begin{align}
\Rbb \Delta (\xbf)\Rbb^{-1}= \sum_{i} (\Rbb \ybf^{(1)}_{i} \Rbb^{-1}) (\Rbb \zbf^{(2)}_{i} \Rbb^{-1}),  
\end{align}
for \eqref{co-prod-gen}. 

Once the map \eqref{defYB-map} is given, 
 it can be interpreted as 
a function of  $ ({\mathbf X}^{(1)}, {\mathbf X}^{(2)}) $. 
Below, we will write this 
in a functional form
\begin{align}
(\widetilde{\mathbf X}^{(1)}, \widetilde{\mathbf X}^{(2)})=
{\mathcal R}
({\mathbf X}^{(1)}, {\mathbf X}^{(2)}) . 
 \label{YB-map-fun1}
\end{align}
Note that we can define a map by 
 replacing  $({\mathbf X}^{(1)}, {\mathbf X}^{(2)})$ with 
 a pair of two copies of the set $\Xbf $ if 
 any two elements, each of which is taken from a different copy, 
are assumed to be commutative.  The map \eqref{YB-map-fun1} 
has an important property: 
\begin{lemma}
For any algebra homomorphism $f$ on $\A \otimes \A$, we have 
\begin{align}
f({\mathcal R}({\mathbf X}^{(1)}, {\mathbf X}^{(2)}))
=
{\mathcal R} (f({\mathbf X}^{(1)}), f({\mathbf X}^{(2)}) )  ,
\label{homo-R}
\end{align}
where we define
$f((\widetilde{\mathbf X}^{(1)}, \widetilde{\mathbf X}^{(2)}))=
(f(\widetilde{\mathbf X}^{(1)}), f(\widetilde{\mathbf X}^{(2)}))$ and 
 $f({\mathbf X}^{(a)}) =\{ f(\xi) | \xi \in {\mathbf X}^{(a)} \}$, $a=1,2$.
\end{lemma}

Next, we consider the tensor product 
 ${\mathcal A}^{\otimes L}$ 
of the algebra $ {\mathcal A}$ for $L \ge 2$, and introduce the notation:  
\begin{align}
{\mathbf X}^{(a)}=
\{ 1^{\otimes (a-1)} \otimes  {\mathbf x} \otimes 1^{\otimes (L-a)} 
 | {\mathbf x} \in {\mathbf X} \} ,
\qquad a=1,2,\dots , L. 
\end{align}
In the same way as \eqref{defYB-map}, we define 
a map $\R_{ij}$ by the adjoint action of the 
universal R-matrix $\Rbb_{ij}$. 
\begin{multline}
\R_{ij} : 
({\mathbf X}^{(1)}, {\mathbf X}^{(2)} , \dots, {\mathbf X}^{(L)})
\mapsto 
(\widetilde{\mathbf X}^{(1)}, \widetilde{\mathbf X}^{(2)} , \dots , \widetilde{\mathbf X}^{(L)}), 
\\
\widetilde{\mathbf X}^{(a)}=
\Rbb_{ij} \Xbf^{(a)}
\Rbb_{ij}^{-1}
=
\left\{ 
\Rbb_{ij} {\mathbf x}^{(a)}
\Rbb_{ij}^{-1} | {\mathbf x}^{(a)} \in {\mathbf X}^{(a)} 
\right\}, 
\qquad a=1,2,\dots , L. 
\label{defYB-map2}
\end{multline}
In the same way as \eqref{YB-map-fun1}, 
this defines a functional form of the map 
\begin{align}
(\widetilde{\mathbf X}^{(1)}, \widetilde{\mathbf X}^{(2)}, \dots, \widetilde{\mathbf X}^{(L)})=
{\mathcal R}_{ij}
({\mathbf X}^{(1)}, {\mathbf X}^{(2)}, \dots, {\mathbf X}^{(L)}) . 
 \label{YB-map-fun2}
\end{align}
The function map \eqref{YB-map-fun1} is related to $\R_{12}$ as 
\begin{align}
{\mathcal R}_{12}
({\mathbf X}^{(1)}, {\mathbf X}^{(2)},  {\mathbf X}^{(3)}, \dots, {\mathbf X}^{(L)})
=
(\R({\mathbf X}^{(1)}, {\mathbf X}^{(2)}),  {\mathbf X}^{(3)}, \dots, {\mathbf X}^{(L)})  
\end{align}
under the interpretation $((\cdot , \cdot ), \cdot , \dots, \cdot )=
(\cdot , \cdot , \cdot , \dots, \cdot )$. 
We also define the permutation map $\sigma_{ij}$ for $i \le j$, which exchanges the $i$-th and
 the $j$-th 
components
\footnote{For $i=j$, this becomes an identity map $\sigma_{ii}=1$.},  as  
\begin{align}
({\mathbf X}^{(1)}, \dots, 
{\mathbf X}^{(j)}, \dots, {\mathbf X}^{(i)}, \dots, 
{\mathbf X}^{(L)})=
\sigma_{ij}
({\mathbf X}^{(1)}, \dots, {\mathbf X}^{(i)}, \dots, {\mathbf X}^{(j)}, \dots, {\mathbf X}^{(L)}) . 
 \label{permu}
\end{align}
Then \eqref{YB-map-fun2} can also be represented as 
\begin{align}
{\mathcal R}_{ij} = 
  \sigma_{1i} \circ \sigma_{2j} \circ {\mathcal R}_{12} \circ \sigma_{1i} \circ \sigma_{2j} . 
\end{align}
%
As remarked in \cite{BS15},  a successive application of the adjoint action 
of the universal R-matrix is equivalent to  the corresponding set-theoretic map 
 in the reverse order. 
\begin{align}
\Rbb_{i_{1}j_{1}} \Rbb_{i_{2}j_{2}} 
({\mathbf X}^{(1)}, {\mathbf X}^{(2)}, \dots, {\mathbf X}^{(L)})
\Rbb_{i_{2}j_{2}}^{-1}  \Rbb_{i_{1}j_{1}}^{-1}
=
\R_{i_{2}j_{2}} (\R_{i_{1}j_{1}}  ({\mathbf X}^{(1)}, {\mathbf X}^{(2)}, \dots, {\mathbf X}^{(L)})). 
\end{align}
Then the functional Yang-Baxter equation (for $L=3$) follows from \eqref{YBE}.
\begin{align}
\R_{23} \circ \R_{13} \circ \R_{12}= 
\R_{12} \circ \R_{13} \circ \R_{23}. 
\end{align}
We also define 
\begin{align}
\begin{split}
\delta_{12}({\mathbf X}^{(1)}, {\mathbf X}^{(2)}, {\mathbf X}^{(3)})
&=
(\delta({\mathbf X}^{(1)}, {\mathbf X}^{(2)}), {\mathbf X}^{(3)}), 
\\[6pt]
\delta_{23}({\mathbf X}^{(1)}, {\mathbf X}^{(2)}, {\mathbf X}^{(3)})
&=
({\mathbf X}^{(1)},\delta({\mathbf X}^{(2)}, {\mathbf X}^{(3)})), 
\end{split}
\end{align}
\begin{align}
(\epsilon \otimes 1)({\mathbf X}^{(1)}, {\mathbf X}^{(2)}) &=
(\{ \epsilon (\xbf) \otimes 1  | \xbf \in \Xbf  \},  \{ \epsilon (1) \otimes \xbf  | \xbf \in \Xbf  \} ), 
\\[6pt] 
(1 \otimes \epsilon )({\mathbf X}^{(1)}, {\mathbf X}^{(2)}) &=
( \{ \xbf \otimes \epsilon (1)  | \xbf \in \Xbf  \}, \{ 1 \otimes \epsilon (\xbf)   | \xbf \in \Xbf  \} ), 
\\[6pt]
(S \otimes S)({\mathbf X}^{(1)}, {\mathbf X}^{(2)}) &=
(\{ S (\xbf) \otimes S(1)  | \xbf \in \Xbf  \},  
\{ S(1) \otimes S (\xbf)   | \xbf \in \Xbf  \}  ). 
\end{align}
The following propositions were written in \cite{BS15} without proof. 
\begin{proposition}
The following relations hold 
\begin{align} 
 \delta \circ \sigma  & =\delta \circ \R ,  
 \label{URmap1}
\\[6pt]
 \R \circ \delta_{12} &= \delta_{12} \circ \R_{23} \circ \R_{13} , 
 \label{URmap2}
\\[6pt]
\R \circ \delta_{23} &= \delta_{23} \circ \R_{12} \circ \R_{13} , 
 \label{URmap3}
\end{align}
\end{proposition}
{\em Proof:}
We will prove 
\eqref{URmap1} on $ ({\mathbf X}^{(1)}, {\mathbf X}^{(2)})$: 
\begin{align}
 \delta \circ \sigma   ({\mathbf X}^{(1)}, {\mathbf X}^{(2)} ) 
&= \delta ( {\mathbf X}^{(2)}, {\mathbf X}^{(1)}) 
\nonumber 
\\[6pt]
&= \{ \Delta^{\prime}(\xbf) |\xbf \in \Xbf \}
\nonumber 
\\[6pt]
&= \{ \Rbb \Delta(\xbf) \Rbb^{-1} |\xbf \in \Xbf \}
 \qquad [\text{by \eqref{UR-def}}] 
\nonumber 
\\[6pt]
&= \Rbb \delta({\mathbf X}^{(1)}, {\mathbf X}^{(2)} ) \Rbb^{-1}
\nonumber 
\\[6pt]
&=  \delta( \Rbb {\mathbf X}^{(1)} \Rbb^{-1}, \Rbb {\mathbf X}^{(2)}\Rbb^{-1} )  
\qquad [\text{by \eqref{delta-R}}] 
\nonumber 
\\[6pt]
&
=\delta \circ \R ({\mathbf X}^{(1)}, {\mathbf X}^{(2)} ). 
\end{align}
We will prove 
\eqref{URmap2} on $ ({\mathbf X}^{(1)}, {\mathbf X}^{(2)}, {\mathbf X}^{(3)})$: 
\begin{align}
& \R \circ \delta_{12}  ({\mathbf X}^{(1)}, {\mathbf X}^{(2)}, {\mathbf X}^{(3)}) 
 =\R (\delta ({\mathbf X}^{(1)}, {\mathbf X}^{(2)}), {\mathbf X}^{(3)}) 
\nonumber 
\\[6pt]
& = \R (\{ \Delta (\xbf)\otimes 1| \xbf \in \Xbf \}, \{ \Delta (1)\otimes \xbf | \xbf \in \Xbf \}) 
\qquad [\text{use $\Delta(1)=1\otimes 1$}]  
\nonumber 
\\[6pt]
&= (\{ (\Delta \otimes 1) (\Rbb (\xbf \otimes 1) \Rbb^{-1}) | \xbf \in \Xbf \},
 \{ (\Delta \otimes 1) (\Rbb (1 \otimes \xbf ) \Rbb^{-1}) | \xbf \in \Xbf \}) 
 \nonumber 
\\
&
\hspace{240pt} [\text{by \eqref{homo-R} for $f=\Delta \otimes 1 $}]
\nonumber 
\\[6pt]
&=\Rbb_{13}\Rbb_{23}  (\{ \Delta (\xbf) \otimes 1)  | \xbf \in \Xbf \},
 \{ 1 \otimes 1 \otimes \xbf  | \xbf \in \Xbf \}) 
\Rbb_{23}^{-1} \Rbb_{13}^{-1}
\quad [\text{by \eqref{UR-def}}]
\nonumber 
\\[6pt]
&=\Rbb_{13}\Rbb_{23}  (\delta( {\mathbf X}^{(1)}, {\mathbf X}^{(2)} ),
  {\mathbf X}^{(3)} ) 
\Rbb_{23}^{-1} \Rbb_{13}^{-1}
\nonumber 
\\[6pt]
&= 
(\delta(  \Rbb_{13}\Rbb_{23} {\mathbf X}^{(1)}\Rbb_{23}^{-1} \Rbb_{13}^{-1}, 
\Rbb_{13}\Rbb_{23} {\mathbf X}^{(2)} \Rbb_{23}^{-1} \Rbb_{13}^{-1} ) ,
\Rbb_{13}\Rbb_{23}  {\mathbf X}^{(3)} \Rbb_{23}^{-1} \Rbb_{13}^{-1}) 
\quad  [\text{by \eqref{delta-R}}] 
\nonumber 
\\[6pt]
&= \delta_{12}
( \Rbb_{13}\Rbb_{23}  ({\mathbf X}^{(1)}, {\mathbf X}^{(2)}, {\mathbf X}^{(3)}) \Rbb_{23}^{-1} \Rbb_{13}^{-1})
\nonumber 
\\[6pt]
&=\delta_{12} \circ \R_{23} \circ \R_{13}
  ({\mathbf X}^{(1)}, {\mathbf X}^{(2)}, {\mathbf X}^{(3)}) 
\end{align}
The proof of \eqref{URmap3} is similar.   \qed
\begin{proposition}
The following relations hold 
\begin{align} 
\R \circ (\epsilon \otimes 1) & =\epsilon \otimes 1 ,
 \label{URmap4} 
\\[6pt]
\R \circ (1 \otimes \epsilon ) &=  1 \otimes \epsilon , 
 \label{URmap5}
\\[6pt]
\R \circ (S \otimes S) &=  (S \otimes S) \circ \R^{-1} , 
\label{R-Rinv}
\end{align}
\end{proposition}
{\em Proof:} Let us prove \eqref{URmap4} on  
$ ({\mathbf X}^{(1)}, {\mathbf X}^{(2)})$. 
\begin{align}
& \R \circ (\epsilon \otimes 1)({\mathbf X}^{(1)}, {\mathbf X}^{(2)})
=\R(  (\epsilon \otimes 1)({\mathbf X}^{(1)}), (\epsilon \otimes 1)({\mathbf X}^{(2)}))
\nonumber \\[6pt]
&=( (\epsilon \otimes 1)(\Rbb {\mathbf X}^{(1)} \Rbb^{-1}), 
 (\epsilon \otimes 1)(\Rbb {\mathbf X}^{(2)} \Rbb^{-1})) 
\quad [\text{by \eqref{homo-R} for $f=\epsilon \otimes 1 $}]
\nonumber
\\[6pt]
&=( (\epsilon \otimes 1)(\Rbb)  (\epsilon \otimes 1)({\mathbf X}^{(1)}) 
 (\epsilon \otimes 1)(\Rbb^{-1}), 
 (\epsilon \otimes 1)(\Rbb)  (\epsilon \otimes 1)({\mathbf X}^{(2)})  (\epsilon \otimes 1)(\Rbb^{-1})))
\nonumber
\\[6pt]
&=((\epsilon \otimes 1)({\mathbf X}^{(1)}) ,  (\epsilon \otimes 1)({\mathbf X}^{(2)})) 
\qquad [\text{by \eqref{UR-relations}}]
\nonumber 
\\[6pt]
&=(\epsilon \otimes 1)({\mathbf X}^{(1)} , {\mathbf X}^{(2)}) . 
\end{align}
The proof of \eqref{URmap5} is similar. 
 \eqref{R-Rinv} is proven based on the fact that $S$ is an anti-homomorphism on $\A$.
\begin{align}
\R \circ (S \otimes S)({\mathbf X}^{(1)}, {\mathbf X}^{(2)})
&=
(\Rbb ((S \otimes S){\mathbf X}^{(1)}) \Rbb^{-1}, 
\Rbb ((S \otimes S){\mathbf X}^{(2)}) \Rbb^{-1})
\nonumber \\[6pt]
&=
( ((S \otimes S)\Rbb )
 ((S \otimes S){\mathbf X}^{(1)})
  ((S \otimes S)\Rbb^{-1}) ,  
 \nonumber  \\
  & \ \ 
 ((S \otimes S)\Rbb )
 ((S \otimes S){\mathbf X}^{(2)})
  ((S \otimes S)\Rbb^{-1}) ) 
\quad [\text{by \eqref{UR-relations2}}] 
\nonumber  \\[6pt]
&=
( (S \otimes S)(\Rbb^{-1}{\mathbf X}^{(1)}\Rbb) ,  
(S \otimes S)(\Rbb^{-1}{\mathbf X}^{(2)} \Rbb) ) 
\nonumber  \\[6pt]
&=
 (S \otimes S) \circ \R^{-1} ({\mathbf X}^{(1)}, {\mathbf X}^{(2)}).
\end{align}
\qed

Most of the relations discussed in this section have natural counterparts in 
the quasi-classical limit. 
They are basically described in section 3.4 of \cite{BS15}. 
For the higher rank case, one has to replace eq. (3.21) in \cite{BS15} with 
\begin{align}
x^{(a)}=\{k_l^{(a)},(k_l^{(a)})^{-1},e_{ij}^{(a)} | i \ne j \},\qquad a=1,2,3,\ldots 
\end{align}
and eq. (3.24) in \cite{BS15} with 
\begin{align}
k'_{i}=k_{i}^{(1)}\,k_{i}^{(2)},
\quad 
e'_{i,i+1}=e_{i,i+1}^{(1)}\,k_{i}^{(2)}(k_{i+1}^{(2)})^{-1}+ e^{(2)}_{i,i+1},
\quad e'_{i+1,i}=
e_{i+1,i}^{(1)} + (k_{i}^{(1)})^{-1} k_{i+1}^{(1)} e^{(2)}_{i+1,i} ,
\end{align}
which is a classical counterpart of \eqref{co-pro}. 
\section{Quasi-classical expansion of the universal R-matrix} 
In the limit \eqref{qc-limit}, the universal R-matrices 
\eqref{UR-exp} and 
\eqref{UR-expi-ren} 
become 
singular, as shown  below
\footnote{This section corresponds to a higher rank analogue of section 3.2 of \cite{BS15}. 
The author learned the procedures to derive \eqref{UR-limit1} for $U_{q}(sl(3))$ case
 from V.\ Bazhanov and S.\ Khoroshkin \cite{BK13}.}.  
The following function
\begin{align}
f(x)=\exp_{q^{-2}}\left((q-q^{-1})^{-1} x\right)
=\prod_{j=0}^{\infty}(1-q^{-2j-1}x)^{-1}, 
\end{align}
satisfies $f(xq^{2})=f(x)/(1-qx)$. Then, in the limit 
\eqref{qc-limit}, 
\begin{align}
\frac{\mathrm{d}}{\mathrm{d}x} \log f(x)
=-\frac{\log(1-x)}{2\pi i {\mathbf b}^2x}
+\frac{1}{2\pi i(1-x)} 
+O({\mathbf b}^2).
\end{align}
holds. 
Integrating this with respect to $x$, we obtain 
\begin{align}
f(x)=(1-x)^{-\frac{1}{2}}
\exp\left( \frac{\mathrm{Li}_{2}(x)}{2\pi i {\mathsf b}^2} \right)(1 + O({\mathsf b^2})) ,
\label{expan-exp} 
\end{align}
where we introduce the dilogarithmic function 
\begin{align}
\mathrm{Li}_{2}(x)=-\int_{0}^{x}\frac{\log(1-t)}{t}
\mathrm{d} t.
\end{align}
Applying \eqref{expan-exp} 
to \eqref{UR-exp} and \eqref{UR-expi-ren}, we arrive at  
\begin{multline}
\Rbb=
\prod_{i<j}(1-e_{ij}\otimes e_{ji})^{-\frac{1}{2}}
\\
\times
\exp
\left(\frac{1}{i\pi {\mathsf b}^2}
\left(
\sum_{i=1}^{n}\log k_{i} \otimes \log k_{i}
+
\frac{1}{2}
\sum_{i<j}
 \mathrm{Li}_{2}(e_{ij} \otimes e_{ji})
\right)
\right)
(1 + O({\mathsf b^2})) ,  \label{UR-limit1}
\end{multline}
and 
\begin{multline}
\Rbf=
\prod_{i<j}(1-e_{ij}\otimes e_{ji})^{-\frac{1}{2}}
\\
\times
\exp
\left(\frac{1}{i\pi {\mathsf b}^2}
\left( 2
\sum_{i \ge j }\log k_{i} \otimes \log k_{j}
+
\frac{1}{2}
\sum_{i<j}
 \mathrm{Li}_{2}(e_{ij} \otimes e_{ji})
\right)
\right)
(1 + O({\mathsf b^2})) . \label{UR-limit2}
\end{multline}
In order to get a finite result, 
we may have to renormalize 
the universal R-matrices as 
\begin{align}
\lim_{{\mathsf b} \to 0}
\Rbb^{\mathsf b^2}&=
\exp
\left(\frac{1}{i\pi }
\left(
\sum_{i=1}^{n}\log k_{i} \otimes \log k_{i}
+
\frac{1}{2}
\sum_{i<j}
 \mathrm{Li}_{2}(e_{ij} \otimes e_{ji})
\right)
\right), 
\label{limitUR1}
\\[5pt]
\lim_{{\mathsf b} \to 0}
\Rbf^{\mathsf b^2}&=
\exp
\left(\frac{1}{i\pi }
\left( 2
\sum_{i \ge j}\log k_{i} \otimes \log k_{j}
+
\frac{1}{2}
\sum_{i<j}
 \mathrm{Li}_{2}(e_{ij} \otimes e_{ji})
\right)
\right). 
\label{limitUR2}
\end{align}


\end{document}